\documentclass[12pt]{article}

\usepackage{authblk}
\usepackage{graphicx,amsmath,hyperref}
\usepackage{mciteplus}

\allowdisplaybreaks[3]
\def\hybrid{\topmargin 0pt      \oddsidemargin 0pt
        \headheight 0pt \headsep 0pt
        \voffset=-0.5cm
        \textwidth 6.25in       %
        \textheight 9.5in       %
        \marginparwidth 0.0in
        \parskip 5pt plus 1pt   \jot = 1.5ex}
\catcode`\@=11
\def\marginnote#1{}

\newcount\hour
\newcount\minute
\newtoks\amorpm
\hour=\time\divide\hour by60
\minute=\time{\multiply\hour by60 \global\advance\minute by-\hour}
\edef\standardtime{{\ifnum\hour<12 \global\amorpm={am}%
        \else\global\amorpm={pm}\advance\hour by-12 \fi
        \ifnum\hour=0 \hour=12 \fi
        \number\hour:\ifnum\minute<10 0\fi\number\minute\the\amorpm}}
\edef\militarytime{\number\hour:\ifnum\minute<10 0\fi\number\minute}

\def\draftlabel#1{{\@bsphack\if@filesw {\let\thepage\relax
   \xdef\@gtempa{\write\@auxout{\string
      \newlabel{#1}{{\@currentlabel}{\thepage}}}}}\@gtempa
   \if@nobreak \ifvmode\nobreak\fi\fi\fi\@esphack}
        \gdef\@eqnlabel{#1}}
\def\@eqnlabel{}
\def\@vacuum{}
\def\draftmarginnote#1{\marginpar{\raggedright\scriptsize\tt#1}}
\def\draftlabel#1{{\@bsphack\if@filesw {\let\thepage\relax
   \xdef\@gtempa{\write\@auxout{\string
      \newlabel{#1}{{\@currentlabel}{\thepage}}}}}\@gtempa
   \if@nobreak \ifvmode\nobreak\fi\fi\fi\@esphack}
        \gdef\@eqnlabel{#1}}
\def\@eqnlabel{}
\def\@vacuum{}
\def\draftmarginnote#1{\marginpar{\raggedright\scriptsize\tt#1}}

\def\draft{\oddsidemargin -.5truein
        \def\@oddfoot{\sl preliminary draft \hfil
        \rm\thepage\hfil\sl\today\quad\militarytime}
        \let\@evenfoot\@oddfoot \overfullrule 3pt
        \let\label=\draftlabel
        \let\marginnote=\draftmarginnote
   \def\@eqnnum{(\theequation)\rlap{\kern\marginparsep\tt\@eqnlabel}%
\global\let\@eqnlabel\@vacuum}  }

\def\numberbysection{\@addtoreset{equation}{section}
        \def\theequation{\thesection.\arabic{equation}}}

\def\underline#1{\relax\ifmmode\@@underline#1\else
        $\@@underline{\hbox{#1}}$\relax\fi}

\def\titlepage{\@restonecolfalse\if@twocolumn\@restonecoltrue\onecolumn
     \else \newpage \fi \thispagestyle{empty}\c@page\z@
        \def\thefootnote{\fnsymbol{footnote}} }

\def\endtitlepage{\if@restonecol\twocolumn \else  \fi
        \def\thefootnote{\arabic{footnote}}
        \setcounter{footnote}{0}}  %
\relax

\numberbysection
\hybrid

\newtheorem{prop}{Proposition}[section]

\newtheorem{cor}[prop]{Corollary}
\newtheorem{lemma}{Lemma}[section]
\newfont{\Bbb}{msbm10 scaled 1\@ptsize00}
\newfont{\Bbbb}{msbm7 scaled 1\@ptsize00}
\newcommand{\CC}{\mbox{\Bbb C}}
\newcommand{\CCC}{\mbox{\Bbbb C}}

\newcommand{\DDD}{\raise-1pt\hbox{$\mbox{\Bbbb D}$}}

\newcommand{\II}{\mbox{\Bbb I}}
\newcommand{\III}{\raise-1pt\hbox{$\mbox{\Bbbb I}$}}

\newcommand{\PP}{\mbox{\Bbb P}}        %

\newcommand{\RR}{\mbox{\Bbb R}}

\newcommand{\UUU}{\raise-1pt\hbox{$\mbox{\Bbbb U}$}}

\newcommand{\ZZ}{\mbox{\Bbb Z}}
\newcommand{\z}{\raise-1pt\hbox{$\mbox{\Bbbb Z}$}}
\newcommand{\omhz}[1][1]{\ensuremath{\left((1-h z_{#1})(1-h \zeta_{#1})\right)}}

\def\beq{\begin{equation}}
\def\eeq{\end{equation}}
\def\p{\partial}

\def\diag{{\rm diag~}}

\DeclareMathOperator{\tr}{tr}
\DeclareMathOperator*{\res}{res}

\begin{document}

\begin{titlepage}

\title{The master $T$-operator for the Gaudin model and the KP hierarchy}

\author[1,2]{Alexander Alexandrov \thanks{E-mail:  {\tt alexandrovsash at gmail.com}}}
\author[3]{Sebastien Leurent \thanks{E-mail:  {\tt sebastien.leurent
      at normalesup.org}}\thanks{present address: Institut de
    math\'ematiques de Bourgogne, 9 avenue Alain Savary, 21000 Dijon, FRANCE}}
\author[4]{Zengo Tsuboi \thanks{E-mail:  {\tt ztsuboi at yahoo.co.jp}
}\thanks{present address: Department of Mathematics and Statistics, 
The University of Melbourne, Royal Parade, Parkville, Victoria 3010, Australia}}
\author[2,5,6]{Anton Zabrodin \thanks{E-mail:  {\tt zabrodin at itep.ru}}}

\affil[1]{\small{Mathematics Institute and Freiburg Institute for Advanced Studies (FRIAS), University of Freiburg, Germany}} 
\affil[2]{ITEP, 25 B.Cheremushkinskaya, Moscow 117218, Russia}
\affil[3]{Imperial College, London SW7 2AZ, United Kingdom}
\affil[4]{Department of Theoretical Physics, 
Research School of Physics and Engineering, 
Australian National University, Canberra, ACT 0200, Australia.}
\affil[5]{Institute of Biochemical Physics, 4 Kosygina st.,
Moscow 119334, Russia}
\affil[6]{National Research University Higher School of Economics,
International Laboratory of Representation Theory and Mathematical Physics,
20 Myasnitskaya Ulitsa, Moscow 101000, Russia}

\date{June 2013}
\maketitle

\vspace{-12.5cm} 
\centerline{ \hfill ITEP-TH-10/13}\centerline{\hfill Imperial/TP/13/SL/03} %
\vspace{12.5cm}

\vspace{-10mm}

\begin{abstract}

Following the approach of
\cite{AKLTZ11}, we construct the master $T$-operator
for the quantum Gaudin model with twisted 
boundary conditions and show that it satisfies
the bilinear identity and Hirota equations 
for the classical KP hierarchy.
We also characterize the class
of solutions to the KP hierarchy
that correspond to eigenvalues of the master $T$-operator
and study dynamics of their zeros as functions of the 
spectral parameter. This implies a remarkable 
connection between 
the quantum Gaudin model and the classical Calogero-Moser
system of particles.

\end{abstract}

\end{titlepage}

\vspace{5mm}

Keywords: classical Calogero-Moser model, 
quantum-classical correspondence, 
quantum Gaudin model, KP hierarchy, 
quantum Jacobi-Trudi and Giambelli formulas, 
master T-operator, matrix derivative, 
tau-function
 \\[3pt]
Journal Reference: Nucl.\ Phys.\ B 883 (2014) 173-223 
\\
DOI: 10.1016/j.nuclphysb.2014.03.008
\vspace{5mm}
\newpage
\tableofcontents

\vspace{5mm}

\section{Introduction}

In this paper we discuss a correspondence between the following
integrable systems:
\begin{itemize}
\item[(i)] The quantum Gaudin model,
\item[(ii)] The classical Kadomtsev-Petviashvili (KP) hierarchy,
\item[(iii)] The classical Calogero-Moser (CM) system of particles.
\end{itemize}
The link (i)-(ii) is a limiting case of
the correspondence between quantum spin chains
with the Yangian $Y(gl(N))$-invariant
rational $R$-matrices and the classical modified
KP hierarchy based on the construction of the master $T$-operator
\cite{AKLTZ11,zabrodin2013master,*Zabrodin2012gx}. 
The link (ii)-(iii) is a well-known story about
dynamics of poles of rational solutions to soliton equations
started by Airault, McKean and Moser \cite{AMM} for the KdV equation, 
developed by Krichever \cite{Kricheverrat1,*Kricheverrat2} 
for the KP equation 
and extended to the whole KP hierarchy
by Shiota \cite{Shiota}. The composition of (i)-(ii) and
(ii)-(iii) implies the connection between the quantum Gaudin model and the
classical CM model which is a limiting
case of the connection between quantum spin chains and classical
Ruijsenaars systems found in \cite{AKLTZ11}.
The link (i)-(iii) was also earlier established in \cite{MTV1,*MTV2}
from a different
reasoning.

The master $T$-operator was
introduced in \cite{AKLTZ11}\footnote{A preliminary form of the master
$T$-operator
was previously discussed in \cite{KLT10}.}.
It is a generating function
for commuting transfer matrices of integrable vertex models
and associated quantum spin chains which unifies the transfer matrices
on all levels of the nested Bethe ansatz and Baxter's
$Q$-operators in one commuting family.

It was also proven in \cite{AKLTZ11} that the master $T$-operator,
as a function of infinitely many auxiliary parameters (``times''),
one of which being the quantum spectral parameter,
satisfies the same hierarchy of bilinear Hirota equations
as the classical tau-function does. This means that any eigenvalue
of the master $T$-operator is the tau-function of a classical
integrable hierarchy.
For finite spin chains with
$Y(gl(N))$-invariant $R$-matrices this tau-function is polynomial
in the quantum spectral parameter. The close connection of the spin chain
spectral problem with
integrable many-body systems of classical mechanics comes from
dynamics of zeros of the polynomial tau-functions.

In this paper we obtain similar results for the 
quantum $gl(N)$ Gaudin model \cite{Gaudin} with twisted 
boundary conditions and spins in the vector representation.
It is known that
this model can be treated as a certain limit of the integrable spin chain.
However, the construction of higher integrals of motion emerging
from the limiting procedure is not obvious.
This makes the master $T$-operator for the Gaudin model a meaningful
object interesting by itself.

The results of this paper can be outlined as follows.

\begin{itemize}
\item
We construct commuting integrals of motion for the $gl(N)$ Gaudin model,
with twisted 
boundary conditions and 
vector representations at the marked points in the quantum space,
corresponding to arbitrary representations in the auxiliary 
space.
They are presented in an explicit form using the matrix derivative 
operation. We also find functional relations satisfied by them.
\item
The master $T$-operator
for the $gl(N)$ Gaudin model
is the most general generating function
for the commuting integrals of motion. 
It depends on an infinite number
of auxiliary ``time variables'' ${\bf t}=\{t_1, t_2, t_3 , \ldots \}$,
where $t_1$ can be identified with the spectral parameter $x$.
The master $T$-operator is constructed 
explicitly using the matrix derivative.
\item
We show that the master $T$-operator satisfies the bilinear identity
for the classical KP hierarchy and hence
any of its eigenvalues is a KP tau-function. 
Here is a short dictionary of the Gaudin-KP correspondence:

\vspace{1mm}

\begin{center}
\begin{tabular}{ccc}
\bf Gaudin & & \bf KP hierarchy \\[10pt]
{\it master $T$-operator} & $\longleftrightarrow$ & 
{\it $\tau$-function}\\[5pt]
{\it spectral parameter} & $\longleftrightarrow$ & 
{\it the $t_1$-variable}\\[5pt]
\it higher transfer matrices& $\longleftrightarrow$ & 
\it Pl\"ucker coordinates\\[1pt]
\end{tabular}
\end{center}

\noindent
Moreover, from the explicit
form of the $R$-matrix and the Yang-Baxter equation it follows that
this tau-function is a polynomial in $x=t_1$. Therefore, according to
\cite{Kricheverrat1,*Kricheverrat2,Shiota}, the dynamics of its 
roots in $t_i$ with $i>1$ is
given by equations of motion of the classical CM system
of particles. 
\item
The Gaudin-Calogero correspondence implies that
the marked points $x_i$ in the Gaudin model (the inhomogeneities
at the sites in the spin chain language) should be identified with
initial coordinates of the CM particles while eigenvalues
of the Gaudin Hamiltonians are their initial momenta. 
Eigenvalues of the Lax matrix for the CM model coincide with
eigenvalues of the twist matrix (with certain multiplicities).
Therefore, with fixed integrals of motion
(action variables) in the CM model determined by invariants of the
twist matrix, there are finite number of solutions for their values
which correspond to different eigenstates of the Gaudin model.
In other words, the eigenstates of the Gaudin Hamiltonians 
are in one-to-one correspondence with (a finite number of) 
intersection points
of two Lagrangian submanifolds in the phase space of the CM model.
In short, the dictionary of the Gaudin-Calogero correspondence is
as follows:

\vspace{1mm}

\begin{center}
\begin{tabular}{ccc}
\bf Gaudin & & \bf Calogero-Moser \\[10pt]
{\it marked points (inhomogeneities)} & $\longleftrightarrow$ & 
{\it initial 
coordinates }\\[5pt]
\it eigenvalues of Hamiltonians & $\longleftrightarrow$ & 
\it initial momenta \\[5pt]
\it twist parameters  & $\longleftrightarrow$ & 
\it integrals of motion \\[1pt]
\end{tabular}
\end{center}
\end{itemize}

Here is a more detailed summary of the Gaudin-Calogero 
correspondence. 
Let $\Phi$ be the $2n$-dimensional phase space of the classical CM model
with canonical coordinates $\{ p_1, \ldots , p_n, \, x_1, \ldots , x_n\}$.
As is known, the model is integrable and the 
complete set of integrals of motion in involution
is given by ${\cal H}_j =\mbox{tr}\, Y^j$ ($j=1, \ldots , n$),
where $\displaystyle{Y_{ik}=-p_i \delta_{ik}-
\frac{1-\delta_{ik}}{x_i-x_k}}$ is the Lax matrix.
In particular, $-{\cal H}_1=\sum_i p_i$ is the total momentum and 
$$
{\cal H}_2 =\sum_{i=1}^n p_i^2 -\sum_{i<j}^n\frac{2}{(x_i-x_j)^2}
$$
is the CM Hamiltonian.
Let ${\cal L}_1\subset \Phi$ be the Lagrangian hyperplane defined by
fixing the $x_i$'s and let ${\cal L}_2\subset \Phi$ be the Lagrangian
submanifold defined as the level set of all the integrals of motion
${\cal H}_j=\sum_{a=1}^N m_a k_a^j$ with some fixed non-negative 
integer $m_a\in \ZZ$ and real $k_a$'s. Their intersection is a union of
a finite number of points: $\displaystyle{{\cal L}_1\cap {\cal L}_2=
\cup_{\alpha \in I}\psi_{\alpha}}$, with $\psi_{\alpha}=(
p_1^{(\alpha )}, \ldots p_n^{(\alpha )}, \, x_1, \ldots , x_n)\in \Phi$.
The claim is that the values $-p_i^{(\alpha )}$ coincide with the spectrum
of quantum Hamiltonians 
$$
H_i =\sum_{a=1}^N k_a e_{aa}^{(i)}+ \sum_{j\neq i}^n \sum_{a,b=1}^N
\frac{e_{ab}^{(i)}e_{ba}^{(j)}}{x_i-x_j}
$$
of the twisted $GL(N)$ Gaudin model with the 
twist matrix $\mbox{diag}\, (k_1, \ldots , k_N)$ in the sector where
the operators $M_a =\sum_{l=1}^{n}e_{aa}^{(l)}$ (higher rank analogues of the 
spin projection operator) have eigenvalues $m_a$.
Here $(e_{ab})_{a'b'}=\delta_{aa'}\delta_{bb'}$ is the basis in the 
space of $N\! \times \! N$ matrices and $e_{ab}^{(i)}=
\II ^{\otimes (i-1)}\otimes e_{ab}\otimes \II ^{\otimes (n-i)}$.

Most of these results can be obtained by a limiting procedure
from the corresponding results for quantum spin chains proved in 
\cite{AKLTZ11}. The limit from spin chains to the Gaudin model is rather
tricky, especially in the construction 
of higher integrals of motion, but in some cases 
it appears to be easier 
than the direct approach. In particular, the direct proof
of the KP hierarchy for the master $T$-operator is
more complicated for the Gaudin model
because of a high degree degeneration of the latter. 
(Roughly speaking, it is technically easier to deal with a number of simple
poles than with a multiple pole of high degree.)
At the same time, along with the proof through the limit
from spin chains (section \ref{sec-limitproof} and
Appendix \ref{sec:proof-means-limit}), we also
give two direct proofs: one based on combinatorics of the
symmetric group (Appendix \ref{sec:alternative-proof}) and 
another one which exploits the matrix derivatives and 
matrix integrals technique
(Appendix \ref{Hirpr}).

\section{The quantum Gaudin model as a limit of inhomogeneous spin chain}

Consider generalized quantum integrable spin chains with
$Y(gl(N))$-invariant $R$-matrix
\begin{equation}\label{R1}
R(x)=\II \otimes \II +\frac{\eta}{x}
\sum_{a,b=1}^{N}e_{ab}\otimes e_{ba}.
\end{equation}
Here $x$ is the spectral parameter, $\eta$ is an auxiliary parameter,
$\II$ is the unity matrix.
By $e_{ab}$ we denote the
basis in the space of $N\! \times \! N$ matrices
such that $e_{ab}$ has only
one non-zero element (equal to 1) at the place $ab$:
$(e_{ab})_{cd}=\delta_{ac}\delta_{bd}$. Note that
$\PP=\sum_{ab}e_{ab} \otimes e_{ba}$ is the permutation matrix
in the space $\CC^N \otimes \CC^N$.
A more general $Y(gl(N))$-invariant $R$-matrix is
\begin{equation}\label{R2}
R_{\lambda}(x)=\II \otimes \II +\frac{\eta}{x}\sum_{a,b=1}^{N}
e_{ab}\otimes \pi_{\lambda}({\sf e}_{ba})
\end{equation}
which acts in the tensor product of the vector representation
space $\CC^N$ and an arbitrary finite-dimensional irreducible
representation $\pi_{\lambda}$ of the algebra $U(gl(N))$
with the highest weight $\lambda$.
We identify $\lambda$ with the Young diagram $\lambda =
(\lambda_1, \lambda_2 , \ldots , \lambda_{\ell})$ with
$\ell=\ell(\lambda )$ non-zero rows, where
$\lambda_i \in \ZZ_{+}$,
$\lambda_1 \geq \lambda_2 \geq \ldots \geq \lambda_{\ell}>0$.
By ${\sf e}_{ab}$ we denote the generators of the algebra
$U(gl(N))$ with the commutation relations
${\sf e}_{ab}{\sf e}_{a'b'}-{\sf e}_{a'b'}{\sf e}_{ab}=
\delta_{a'b}{\sf e}_{ab'}-\delta_{ab'}{\sf e}_{a'b}$.
In this notation we have $e_{ab}=\pi_{(1)}({\sf e}_{ab})$,
where $\pi_{(1)}$ is the $N$-dimensional vector representation
corresponding to the 1-box diagram $\lambda =(1)$.

Fix a matrix $g\in GL(N)$ called
the twist matrix.
For our purpose it is enough to consider diagonal twist matrices.
A family of commuting operators (quantum transfer matrices or $T$-operators)
can be constructed as
\begin{equation}\label{R3}
{\sf T}_{\lambda}(x)=\mbox{tr}_{\pi_{\lambda}}\Bigl (
R_{\lambda}^{10}(x-x_1)R_{\lambda}^{20}(x-x_2)
 \ldots  R_{\lambda}^{n0}(x-x_n)
( \II^{\otimes n} \otimes \pi_{\lambda}(g) )\Bigr ),
\end{equation}
where $x_i$ are arbitrary complex parameters.
The trace is taken in the auxiliary
space $V_{\lambda}$ where the representation $\pi_{\lambda}$ is realized.
By $R_{\lambda}^{j0}(x)$ we denote
the $R$-matrix \eqref{R2} acting  in the tensor product of
the $j$-th local space $\CC^N$ of the chain and the space $V_{\lambda}$
(labeled by $0$)\footnote{
Let us denote \eqref{R2} symbolically as
$R_{\lambda}(x)=\sum_{i}a_{i}\otimes b_{i} $. Then $R_{\lambda}^{j0}(x)$ is realized as
$R_{\lambda}^{j0}(x)=\sum_{i} \II^{\otimes (j-1)} \otimes a_{i}
 \otimes \II^{\otimes (n-j)} \otimes b_{i}$,
 where $j=1,2,\dots , n$.
Here the operator $b_{i}$ acts in the auxiliary space
$V_{\lambda}$ labeled by $0$.
}.
The $T$-operators
act in the physical Hilbert space of the model ${\cal V}=(\CC^N)^{\otimes n}$.
It follows from the Yang-Baxter equation that the $T$-operators
with the same $\eta$, $g$, $x_i$ commute for all $x, \lambda$ and can be
simultaneously diagonalized.

The normalization used above is such that
${\sf T}_{\emptyset}(x)=\II^{\otimes n}$.
It is convenient for the limit
to the Gaudin model. Another useful normalization
is
\begin{equation}\label{R4}
T_{\lambda}(x)=\prod_{j=1}^{n}(x-x_j)\,\cdot
{\sf T}_{\lambda}(x).
\end{equation}
In this normalization all $T_{\lambda}(x)$ and all their eigenvalues
are polynomials in $x$ of degree $n$.

At $n=0$ the transfer matrix
(\ref{R3}) is just the character:
$
{\sf T}_{\lambda}(x)=\chi_{\lambda}(g)
$,
where $\chi_{\lambda}(g)=\mbox{tr}_{\pi_{\lambda}}g$ is
the character of the representation $\pi_{\lambda}$ given by the
Schur polynomials $s_{\lambda}({\bf y})$
of the variables ${\bf y}=\{y_1, y_2, \ldots \}$,
$y_k=\frac{1}{k}\, \mbox{tr}\, g^k$ (the Jacobi-Trudi formula):
\begin{align}
\chi_{\lambda}(g)=s_{\lambda}({\bf y})=
\det_{i,j=1, \ldots , \ell (\lambda )}h_{\lambda_i -i+j}({\bf y}),
 \label{JT-det}
\end{align}
with the complete symmetric polynomials
$h_{k}({\bf y})=s_{(k)}({\bf y})$ defined by
\begin{align}
\exp \Bigl (\xi({\bf y},z) \Bigr )=
\sum_{k=0}^{\infty}h_{k}({\bf y})z^{k},
\label{Schur-p}
\end{align}
where $\xi({\bf y},z) := \sum_{k\geq 1}y_kz^k$.
It is convenient to set $h_{k}=0$ at $k<0$.
A ``dual'' form of \eqref{JT-det} is
\begin{align}
\chi_{\lambda}(g)=
s_{\lambda}({\bf y})=
\det_{i,j=1, \ldots ,  \lambda_{1}}e_{\lambda^{\prime}_i -i+j}({\bf y}),
 \label{JT-det2}
\end{align}
where the elementary symmetric polynomials
$e_{k}({\bf y})=(-1)^{k}h_k(-{\bf y})=s_{(1^k)}({\bf y})$ are defined by
\begin{align}
\exp \Bigl (-\xi({\bf y},z) \Bigr )=
\sum_{k=0}^{\infty}(-1)^{k} e_k({\bf y})z^k.
\label{Schur-ele}
\end{align}
Here and below $\lambda^{\prime}$ is the transposed
(reflected about the main diagonal) of the diagram $\lambda $.
Let $p_1, \ldots , p_N$ be eigenvalues of $g\in GL(N)$ realized
as an element of $\mbox{End} \, (\CC^N)$. Then
$y_k = \frac{1}{k}\, (p_1^k +\ldots +p_N^k)$ and
\beq\label{char-det}
\chi_{\lambda}(g)= \frac{\det_{1\leq i,j\leq N}
\bigl (p_j^{\lambda _i +N-i}\bigr )}{\det_{1\leq i,j\leq N}
\bigl (p_j^{N-i}\bigr )}
\eeq
(see \cite{Macdonald}). This formula implies that
$\chi_{\emptyset}(g)=s_{\emptyset}({\bf y})=1$.

A different but equivalent
construction of quantum transfer matrices was suggested in
\cite{Kazakov2007na}. It uses the special derivative operator
on the group $GL(N)$ called there the co-derivative operator.
Let $g$ be an element of the Lie group $GL(N)$ and $f$ be any
function of $g$.
The (left) co-derivative is defined as\footnote{
Originally, the co-derivative was defined \cite{Kazakov2007na} as
\[
{\sf D}f(g)=\sum_{ab}e_{ab}\frac{\p}{\p \varepsilon_{ab}}\,
f(e^{\varepsilon {\sf e}}g)\Bigr |_{\varepsilon =0}\,, \quad \quad
\varepsilon {\sf e} \equiv \sum_{cd}\varepsilon_{cd}{\sf e}_{dc}.
\]
This definition is equivalent to the one given here.}
\begin{equation}\label{D1}
{\sf D}f(g)=\frac{\p}{\p \varepsilon} \sum_{ab}e_{ab}
f(e^{\varepsilon {\sf e}_{ba}}g)\Bigr |_{\varepsilon =0}.
\end{equation}
The result of the action of ${\sf D}$ to a scalar function is a linear
operator in $\CC^N$, acting by ${\sf D}$ twice we get an operator
in $\CC^N \otimes \CC^N$ and so on.
In practice, we will specify in which space the operator
is acting by adding a suffix, in the following way.
Let $V_i\cong \CC^N$ be several copies of $\CC^N$, then,
applying ${\sf D}$ to a scalar function $k$ times we get
an operator in
$V_1\otimes \ldots \otimes V_k$:
$$
{\sf D}_k \ldots {\sf D}_1 f(g) \in \mbox{End}\,
(V_1\otimes \ldots \otimes V_k)
$$
In particular, applying the second ${\sf D}$ to (\ref{D1}),
we can write
\begin{equation}\label{DD}
{\sf D}_2 {\sf D}_1 f(g)=
\frac{\p}{\p \varepsilon_{2}}
\frac{\p}{\p \varepsilon_{1}}
\sum_{a_2b_2}\sum_{a_1b_1}
e^{(2)}_{a_2b_2}e^{(1)}_{a_1b_1}
f\bigl (e^{\varepsilon_1 {\sf e}_{b_1 a_1}}
e^{\varepsilon_2 {\sf e}_{b_2 a_2}}g \bigr )
\Bigr |_{\varepsilon_1=\varepsilon_{2}=0}
\end{equation}
and so on.
If we work in the space $(\CC^N)^{\otimes n}$, these operators
can be realized as
$e^{(i)}_{ab}:= \II^{\otimes (i-1)} \otimes
e_{ab} \otimes  \II^{\otimes (n-i)}$.
These are of course commutative $[e^{(i)}_{ab},e^{(j)}_{cd}]=0$ for
any $i \ne j$ and any $a,b,c,d \in \{1,\dots, N\}$.

According to \cite{Kazakov2007na}
the transfer matrix (\ref{R3}) can be represented as a sequence
of co-derivatives acting to the character:
\begin{equation}\label{D2}
{\sf T}_{\lambda}(x)=\Bigl ( 1+\frac{\eta \, {\sf D}_n}{x-x_n}\Bigr )
\ldots \Bigl ( 1+\frac{\eta \, {\sf D}_1}{x-x_1}\Bigr )
\chi_{\lambda}(g)\,.
\end{equation}

The Gaudin model can be obtained from here in the limit $\eta \to 0$
provided that $g$ is put equal to $e^{\eta h}$ after taking the group
derivatives. Here $h$ is an element of the Lie algebra $gl(N)$ called
the twist matrix for the Gaudin model.
The standard way is as follows. Take $\lambda =(1)$ (one box), then
$\chi_{(1)}(g)=\mbox{tr}\, g$. Let us expand:
$$
{\sf T}_{(1)}(x)=\left (1+ \eta \sum_{i=1}^{n}\frac{{\sf D}_i}{x-x_i}+
\eta^2 \sum_{i<j}\frac{{\sf D}_j {\sf D}_i}{(x-x_i)(x-x_j)}+\ldots
\right ) \mbox{tr}\, g.
$$
We have ${\sf D}_i \mbox{tr}\, g =g_i$, ${\sf D}_j g_i =
\PP_{ij}g_i$. Therefore, if $g=e^{\eta h}$, then
$$\mbox{tr}\, g =N + \eta \,
\mbox{tr}\, h +\frac{\eta^2}{2} \, \mbox{tr}\, h^2
+O(\eta^3), \quad
{\sf D}_i \mbox{tr}\, g = \II + \eta h_i +
O(\eta^2), \quad
{\sf D}_j {\sf D}_i \mbox{tr}\, g =\PP_{ij} + O(\eta  )
$$
and the $\eta$-expansion of the transfer matrix ${\sf T}_{(1)}(x)$ reads:
$$
{\sf T}_{(1)}(x)=N+\eta \Bigl ( \mbox{tr}\, h +\! \sum_i
\frac{1}{x-x_i}\Bigr ) +\eta^2 \left (\frac{1}{2}\, \mbox{tr}\, h^2 +
\! \sum_i \frac{h_i}{x-x_i} + \!
\sum_{i<j}\frac{\PP_{ij}}{(x-x_i)(x-x_j)}\right )+
\ldots
$$
The $O(1)$ and $O(\eta )$ terms are central, so the operators at the
order $\eta^2$ form a commutative family
\begin{equation}\label{TG}
H(x)=\frac{1}{2}\, \mbox{tr}\, h^2 +
\sum_i \frac{H_i}{x-x_i}\,,
\end{equation}
where
\begin{equation}\label{D3}
H_i=
\res_{x=x_i}H(x)=
h_i + \sum_{j\neq i}\frac{\PP_{ij}}{x_i-x_j}
\end{equation}
are the commuting
Gaudin Hamiltonians: $[H_i, H_j]=0$.
Below we assume that $h$ is diagonal,
$h=\mbox{diag}\,(k_1, \ldots , k_N)$,
then the Gaudin
Hamiltonians can be written more explicitly in the form
\begin{equation}\label{D4}
H_i=\sum_{a=1}^{N}k_a e^{(i)}_{aa} +\sum_{j\neq i}\sum_{a,b=1}^{N}
\frac{e^{(i)}_{ab}e^{(j)}_{ba}}{x_i-x_j}\,.
\end{equation}

It is easy to check that the operators
\beq\label{Ma}
M_a=\sum_{l=1}^{n}e_{aa}^{(l)}
\eeq
commute with the
Gaudin Hamiltonians: $[H_i, M_a]=0$ (for diagonal $h$).
Therefore, common eigenstates of the Hamiltonians can be classified
according to eigenvalues of the operators $M_a$. Let
$$
{\cal V}=\bigotimes_{i=1}^{n}V_i \,\, =
\bigoplus_{m_1, \ldots , m_N} \!\! \!\! {\cal V}(\{m_a \})
$$
be the decomposition of the Hilbert space of the Gaudin model
${\cal V}$ into the direct sum of eigenspaces for the operators
$M_a$ with the eigenvalues $m_a \in \ZZ_{\geq 0}$, $a=1, \ldots , N$.
Then the eigenstates of $H_i$'s are in the spaces ${\cal V}(\{m_a \})$.
Because $\sum_a e_{aa}=\II$ is the unit matrix,
$\sum_a M_a =n \II ^{\otimes n}$, and hence
\beq\label{ma}
\sum_{a=1}^{N}m_a =n.
\eeq
Note also that
$$
\sum_{i=1}^{n}H_i= \sum_{a=1}^{N}k_a M_a.
$$

\section{Commuting operators for the Gaudin model}

A general algebraic construction
of higher members of the Gaudin commutative family
was proposed in \cite{FFR,ER}.
In principle, their explicit form
can be found from the $\eta$-expansion of the spin chain 
transfer matrices
${\sf T}_{\lambda}(x)$ for general $\lambda$.
But there is a problem how to extract
non-trivial integrals of motion
from this expansion. 
For fundamental representations
(one-column diagrams) this
problem was solved by Talalaev \cite{Talalaev}.
Basically, Talalaev's idea
was to consider
special linear combinations of the transfer matrices ${\sf T}_{\lambda}(x)$
such that their $\eta \to 0$ limits start from higher degrees of $\eta$.
The coefficients in front of the leading terms as $\eta \to 0$
commute and can be regarded
as higher transfer matrices (generating 
functions of integrals of motion) of the Gaudin model.
Here we extend this procedure to transfer matrices
associated with diagrams of arbitrary shape.

Let us modify the definition of the transfer matrices (\ref{D2})
as follows:
\begin{equation}\label{D2a}
\tilde {\sf T}_{\lambda}(x)=\Bigl ( 1+\frac{\eta \, {\sf D}_n}{x-x_n}\Bigr )
\ldots \Bigl ( 1+\frac{\eta \, {\sf D}_1}{x-x_1}\Bigr )
\chi_{\lambda}(g-\II )\,, \quad g=e^{\eta h}.
\end{equation}
Here $g$ is put equal to $e^{\eta h}$ after taking all the derivatives.
To avoid misunderstanding, we stress that
$\tilde {\sf T}_{\lambda}(x)$ is {\it not}
the $T$-operator for the twist matrix $\tilde g = g-\II$ because
the co-derivatives act to $g$ rather than $\tilde g$.
Since $\chi_{\lambda}(g-\II )$ is a linear combination of characters
$\chi_{\mu}(g)$ with different $\mu$
and $g$-independent coefficients, $\tilde {\sf T}_{\lambda}(x)$ is a
linear combination of the ${\sf T}_{\mu}(x)$'s. More precisely,
using (\ref{char-det}) and the Cauchy-Binet formula, one can
prove that
\beq\label{char-exp}
\chi_{\lambda}(g-\II )= \sum_{\mu \subset \lambda}
c_{\lambda \mu} \chi_{\mu }(g),
\eeq
where
\beq\label{char-exp1}
c_{\lambda \mu}= (-1)^{|\lambda |-|\mu |}
\det_{1\leq i,j \leq N}
\left ( \begin{array}{c} \lambda_i +N-i \\ \mu_j +N-j
\end{array} \right ),
\eeq
$|\lambda |=\sum_{i}\lambda_i$
and $\displaystyle{\left ( \begin{array}{c} n\\ m
\end{array} \right )=\frac{n!}{m! (n-m)!}}$ is the
binomial coefficient
(see \cite[page 47, example 10]{Macdonald}).
The sum is taken over all Young diagrams $\mu$ that
are contained in $\lambda$ including the empty diagram
and $\lambda$ itself.
Therefore,
$$
\tilde {\sf T}_{\lambda}(x)=
\sum_{\mu \subset \lambda}
c_{\lambda \mu} {\sf T}_{\mu}(x).
$$
In particular,
$$
\tilde {\sf T}_{(1^k)}(x)=\sum_{l=0}^{k}(-1)^{k-l}
\left ( \begin{array}{c}N-l \\ N-k \end{array} \right )
{\sf T}_{(1^l)}(x)
$$
which agrees with Talalaev's prescription for one-column
diagrams.
As we shall see, the leading behavior of
$\tilde {\sf T}_{\lambda}(x)$ as $\eta \to 0$
is $O(\eta^{|\lambda |})$, so we can define the higher
Gaudin transfer matrices as
$${\sf T}^G_{\lambda}(x)=\lim_{\eta \to 0}\Bigl (\eta^{-|\lambda |}
\tilde {\sf T}_{\lambda}(x)\Bigr ).$$

To represent them in a more explicit form, we need to
modify the definition of the co-derivative by passing from the
group derivative to the Lie algebra derivative.
Let $h$ be an element of the Lie algebra $gl(N)$ and $f$ be any
function of $h$.
We define
\begin{equation}\label{d1}
{\sf d}f(h)=\frac{\p}{\p \varepsilon}\sum_{ab}e_{ab}
f(h+ \varepsilon {\sf e}_{ba})\Bigr |_{\varepsilon =0}.
\end{equation}
In fact this is the usual matrix derivative used in the theory
of matrix models.
For example:
\begin{align}
& {\sf d}\, (\mbox{tr} \, h)^k= k (\mbox{tr} \, h)^{k-1}\II
\quad
\mathrm{for} \quad k \in \ZZ,
\label{dtr}
\\
& %
{\sf d} \otimes h^k =
\begin{cases}
\displaystyle{\PP \! \sum_{i=0}^{k-1}h^i\otimes h^{k-i-1}}
& \mathrm{for} \quad k \in \ZZ_{\ge 0}
\\[20pt]
\displaystyle{-\PP \! \sum_{i=k}^{-1}h^i\otimes h^{k-i-1}}
& \mathrm{for} \quad k \in \ZZ_{\le -1},
\end{cases}
\label{dh}
\\
&{\sf d}\, \mathrm{det}(h)=\mathrm{det}(h) h^{-1}.
\end{align}
Similarly to (\ref{DD}), we have
\begin{equation}\label{dd}
{\sf d}_2 {\sf d}_1 f(h)=
\frac{\p}{\p \varepsilon_2}
\frac{\p}{\p \varepsilon_1}
\sum_{a_2b_2}\sum_{a_1b_1}
e^{(2)}_{a_2b_2}e^{(1)}_{a_1b_1}
f\left (
h+\varepsilon_1 {\sf e}_{b_1a_1}+
\varepsilon_2{\sf e}_{b_2a_2}\right )
\Bigr |_{\varepsilon_1=\varepsilon_2=0}
\end{equation}
and so on.
We remark that our modified co-derivative is
commutative
($ {\sf d}_1{\sf d}_2={\sf d}_2{\sf d}_1$), although
the co-derivative for the original spin chain
is non-commutative
(${\sf D}_1{\sf D}_2 \ne {\sf D}_2{\sf D}_1$).

The following lemma is crucial for our construction.

\begin{lemma}\label{lemma-limit}
Let $f(g)$ be a homogeneous scalar function
of $g\in GL(N)$ of degree $m$, i.e., $f(zg)=z^mf(g)$ for
any $z\in \CC$. Set $g=e^{\eta h}$ with $\eta \to 0$. Then
\begin{equation}\label{lemma}
\eta^k {\sf D}_k \ldots {\sf D}_1 f(g-\II )=
\eta^m \, {\sf d}_k \ldots {\sf d}_1 f (h) + \mbox{\rm terms
of higher order as $\eta \to 0$}.
\end{equation}
\end{lemma}

\noindent
{\it Sketch of proof.} We illustrate the idea of the proof by the example
$k=2$. The general case is proved in the same
way but requires too bulky formulas. Using (\ref{DD}), we can write
$$
\begin{array}{ll}
{\sf D}_2 {\sf D}_1 f(g-\II )=&
\displaystyle{\frac{\p}{\p \varepsilon_2}
\frac{\p}{\p \varepsilon_1}
\sum_{a_2b_2}\sum_{a_1b_1}
e^{(2)}_{a_2b_2}e^{(1)}_{a_1b_1}}
\\ &\\
&\times \,\,\displaystyle{
f\bigl (g-\II +\varepsilon_1{\sf e}_{b_1a_1}g+
\varepsilon_2{\sf e}_{b_2a_2}g+ (\varepsilon_1{\sf e}_{b_1a_1})
(\varepsilon_2{\sf e}_{b_2a_2})g\bigr )
\Bigr |_{\varepsilon_1=\varepsilon_2=0}}.
\end{array}
$$
After the rescaling $\varepsilon_{i} \to \eta\varepsilon_i$
the leading $\eta \to 0$ term can be written in the form
$$
{\sf D}_2 {\sf D}_1 f(g-\II )=
\frac{1}{\eta^2}\frac{\p}{\p \varepsilon_2}
\frac{\p}{\p \varepsilon_1}\sum_{a_2b_2}\sum_{a_1b_1}
e^{(2)}_{a_2b_2}e^{(1)}_{a_1b_1}
f\bigl (\eta (h+\varepsilon_2{\sf e}_{b_2a_2}+
\varepsilon_1{\sf e}_{b_1a_1})\bigr )
\Bigr |_{\varepsilon_2=\varepsilon_1 =0}+\, \ldots
$$
The homogeneity of the function $f$ and definition
(\ref{dd}) then imply that
$$
{\sf D}_2 {\sf D}_1 f(g-\II )=\eta^{m-2}\,
{\sf d}_2  {\sf d}_1 f (h) + \, \ldots
$$
Q.E.D.

The character $\chi_{\lambda}(g)$ is a homogeneous function of
degree $|\lambda |$. Therefore, by virtue of the lemma,
the family of commuting operators for the (twisted) Gaudin model
can be constructed as follows:
\begin{equation}\label{d2}
{\sf T}^G_{\lambda}(x)=\Bigl ( 1+\frac{{\sf d}_n}{x-x_n}\Bigr )
\ldots \Bigl ( 1+\frac{ {\sf d}_1}{x-x_1}\Bigr )
\chi_{\lambda}(h)
\end{equation}
or, in the polynomial normalization,
\begin{equation}\label{g1}
T^G_{\lambda}(x)=(x-x_n +{\sf d}_n)\, \ldots \,
(x-x_1 +{\sf d}_1)\chi_{\lambda}(h).
\end{equation}
For example:
\begin{equation}\label{g2}
\begin{array}{l}
{\sf T}^G_{\emptyset}(x)=1,
\\ \\
\displaystyle{
{\sf T}^G_{(1)}(x)=\mbox{tr}\, h +\sum_i \frac{1}{x-x_i}},
\\ \\
\displaystyle{
{\sf T}^G_{(1^2)}(x)=\frac{1}{2}\, (\mbox{tr}\, h)^2 +\mbox{tr}\, h
\sum_i \frac{1}{x-x_i}+\sum_{i<j}\frac{1}{(x-x_i)(x-x_j)}-
H(x)},
\\ \\
\displaystyle{
{\sf T}^G_{(2)}(x)=\frac{1}{2}\, (\mbox{tr}\, h)^2 +\mbox{tr}\, h
\sum_i \frac{1}{x-x_i}+\sum_{i<j}\frac{1}{(x-x_i)(x-x_j)}+
H(x)},
\end{array}
\end{equation}
where $H(x)$ is given by (\ref{TG}).
Note that in this approach $H(x)$ emerges from
${\sf T}_{(1^2)}(x)$ or ${\sf T}_{(2)}(x)$ rather than
${\sf T}_{(1)}(x)$.

\section{The master \texorpdfstring{\boldmath $T$}{T}-operator and the KP hierarchy}

\subsection[The master \texorpdfstring{$T$}{T}-operator]{The master {\boldmath $T$}-operator}

The master $T$-operator for the Gaudin model can be defined as
\begin{equation}\label{master1}
T^G(x, {\bf t})=(x-x_n +{\sf d}_n)\, \ldots \,
(x-x_1 +{\sf d}_1)\exp \Bigl (\sum_{k\geq 1}t_k \, \mbox{tr}\, h^k\Bigr ),
\end{equation}
where ${\bf t}=\{t_1, t_2, \ldots \}$ is an infinite set of ``time
parameters''. These operators commute for different values of the parameters:
$[T^G(x,{\bf t}), \, T^G(x',{\bf t'})]=0$.
Note that because $e^{-x \, \mbox{\scriptsize{tr}}\, h}
{\sf d}e^{x \, \mbox{\scriptsize{tr}}\, h}=x$,
the role of the variable $t_1$
is to shift $x\to x+t_1$. Namely, we can write
\begin{align*}
T^G(x, {\bf t})=e^{-x \, \mbox{\scriptsize{tr}}\, h}
({\sf d}_n -x_n)\, \ldots \,
({\sf d}_1 -x_1)\exp \Bigl (
(x+t_1)\mbox{tr}\, h +\sum_{k\geq 2}t_k \, \mbox{tr}\, h^k\Bigr ).
\end{align*}
Hence it is clear that $e^{x \, \mbox{\scriptsize{tr}}\, h}T^G(x, {\bf t})$
depends on $x,t_1$ only through their sum $x+t_1$.
In particular, we will use the relation
\begin{align}
\partial_{x}T^G(x, {\bf t})=
\partial_{t_{1}}T^G(x, {\bf t})- (\mbox{tr}\, h)
T^G(x, {\bf t}).
\label{master-diff}
\end{align}

The expansion in the Schur functions is
\begin{equation}\label{master2}
T^G(x, {\bf t})=\sum_{\lambda}T^G_{\lambda}(x)s_{\lambda}({\bf t}).
\end{equation}
The $T$-operators $T^G_{\lambda}(x)$ can be restored from the master
$T$-operator according to the formula
\begin{equation}\label{master3}
T^G_{\lambda}(x)=s_{\lambda}(\tilde \p )T^G(x, {\bf t})\Bigr |_{{\bf t}=0},
\end{equation}
where $\tilde \p =\{ \p_{t_1}, \frac{1}{2}\, \p_{t_2},
\frac{1}{3}\, \p_{t_3}, \ldots \}$.
In particular,
\begin{equation}\label{master3a}
T^G_{(1)}(x)=\p_{t_1}T^G(x, {\bf t})\Bigr |_{{\bf t}=0}, \quad
T^G_{(1^2)}(x)=\frac{1}{2}\,
(\p_{t_1}^2 -\p_{t_2})T^G(x, {\bf t})\Bigr |_{{\bf t}=0}.
\end{equation}

With a given $z\in \CC$, the following special
shift of the time variables
$$
{\bf t}\pm [z^{-1}]:= \Bigl \{t_1 \pm z^{-1}, \, t_2 \pm \frac{1}{2}\,
z^{-2}, \, t_3 \pm \frac{1}{3}\,
z^{-3}, \, \ldots \Bigr \}
$$
is often used. As we shall see below,
$T^G(x, {\bf t}\pm [z^{-1}])$ regarded as functions of
$z$ with fixed ${\bf t}$ contain an important information.
Here we only note that equation (\ref{master3}) implies that
$T^G(x, 0\pm [z^{-1}])$ is the generating series for
$T$-operators corresponding to the one-row and one-column diagrams:
\begin{equation}\label{master4}
T^G(x,  [z^{-1}])=\sum_{s\geq 0}z^{-s}T^G_{(s)}(x), \quad \quad
T^G(x,  -[z^{-1}])=\sum_{a=0}^{N}(-z)^{-a}T^G_{(1^a)}(x).
\end{equation}

\subsection[The KP hierarchy for the master \texorpdfstring{$T$}{T}-operator]{The KP hierarchy for the master {\boldmath $T$}-operator}

\subsubsection{The bilinear identity and Hirota
equations}
\label{sec-limitproof}

We are going to prove that the master $T$-operator (\ref{master1})
satisfies the bilinear identity for the KP hierarchy
\cite{DJKM83,JM83} which states that
\begin{equation}\label{hir1}
\oint_{\infty} e^{\xi ({\bf t}-{\bf t'}, z)}
T^G\left (x, {\bf t}-[z^{-1}]\right )
T^G\left (x, {\bf t'}+[z^{-1}]\right )dz =0 \quad
\mbox{for all ${\bf t}, {\bf t}'$.}
\end{equation}
Here
$\displaystyle{\xi ({\bf t},z):=\sum_{k\geq 1}t_k z^k}$ and
the integration contour is chosen in such a way that all singularities
coming from the $T^G$'s are inside it while those coming from 
$e^{\xi ({\bf t}-{\bf t'}, z)}$ are outside it. 

\noindent
{\bf Remark.}
The bilinear identity is invariant under the change of signs
of all times: if 
$T^G(x,{\bf t})$ is a solution, then $T^G(x,-{\bf t})$
is a solution, too.

Setting ${\bf t}'={\bf t}+[z_0^{-1}]
-[z_1^{-1}]-[z_2^{-1}]-[z_3^{-1}]$ and
taking the residues, one obtains from it
the 3-term
Hirota equation \cite{JM83,Sato,Hirota81,Miwa82} (the Fay identity)
\begin{equation}\label{hir20}
\begin{array}{l}
(z_0\! -\! z_1)(z_2\! -\! z_3)\, T^G \left (x, {\bf t}+\!
[z_0^{-1}]\! +\! [z_{1}^{-1}]\right )
T^G \left (x, {\bf t}+[z_{2}^{-1}]\! +\! [z_{3}^{-1}]\right )
\\ \\
\quad  + \, (z_0\! -\! z_2)(z_3\! -\! z_1)\,
T^G \left (x, {\bf t}+[z_0^{-1}]\!  +\! [z_{2}^{-1}]\right )
T^G \left (x, {\bf t}+[z_{1}^{-1}]\! +\! [z_{3}^{-1}]\right )
\\ \\
\quad  \quad
+ \, (z_0\! -\! z_3)(z_1\! -\! z_2)\, T^G
\left (x, {\bf t}+[z_0^{-1}]\! +\! [z_{3}^{-1}]\right )
T^G \left (x, {\bf t}+[z_{1}^{-1}]\! +\! [z_{2}^{-1}]\right )\, =\, 0.
\end{array}
\end{equation}
Tending $z_0\to \infty$, one obtains a simpler looking
3-term equation
\begin{equation}\label{hir2}
\begin{array}{l}
(z_2-z_3)\, T^G \left (x, {\bf t}+[z_{1}^{-1}]\right )
T^G \left (x, {\bf t}+[z_{2}^{-1}]+[z_{3}^{-1}]\right )
\\ \\
\quad  + \, (z_3-z_1)\, T^G \left (x, {\bf t}+[z_{2}^{-1}]\right )
T^G \left (x, {\bf t}+[z_{1}^{-1}]+[z_{3}^{-1}]\right )
\\ \\
\quad  \quad
+ \, (z_1-z_2)\, T^G \left (x, {\bf t}+[z_{3}^{-1}]\right )
T^G \left (x, {\bf t}+[z_{1}^{-1}]+[z_{2}^{-1}]\right )\, =\, 0.
\end{array}
\end{equation}
It appears to be equivalent to its $z_3\to \infty$ limit (the differential
Fay identity)\footnote{This can be shown in the following way.
Let us divide \eqref{hir2} by $z_{3}$ and take the limit
$z_3\to \infty$. Taking note on the fact that
the shift of ${\bf t}$ by $[z_{3}^{-1}]$ produces
$\mathrm{det}(1-hz_{3}^{-1})^{-1}=1 +
(\mathrm{tr}\, h)z_{3}^{-1}+O(z_{3}^{-2})$ in \eqref{master1},
we find that the term of order $O(z_{3}^{0})$ vanishes and
that of $O(z_{3}^{-1})$ gives (\ref{hir3}) with $\p_{t_1}$
instead of $\p_x$.
Applying the relation \eqref{master-diff}, we obtain \eqref{hir3}.
Vice versa, given \eqref{hir3}, we divide it by
$T^G(x,{\bf t}\! +\! [z_1^{-1}])T^G(x,{\bf t}\! +\!
[z_2^{-1}])$ and sum such equations for the pairs
$(z_1, z_2)$, $(z_2, z_3)$ and $(z_3, z_1)$. The result is
equation (\ref{hir2}).}
\beq\label{hir3}
\begin{array}{c}
T^G(x,{\bf t}+[z_2^{-1}])\p_x T^G(x,{\bf t}+[z_1^{-1}])-
T^G(x,{\bf t}+[z_1^{-1}])\p_x T^G(x,{\bf t}+[z_2^{-1}])
\\ \\
+\,\,\, (z_1\! -\! z_2)\Bigr [
T^G({\bf t}) T^G(x,{\bf t}\! +\! [z_1^{-1}]\! +\! [z_2^{-1}])-
T^G(x,{\bf t}\! +\! [z_1^{-1}])T^G(x,{\bf t}\! +\!
[z_2^{-1}])\Bigr ]\, = \, 0.
\end{array}
\eeq
In fact it was proved in
\cite{TakTak,Shigyo} that all
the Hirota equations of the form (\ref{hir20}), (\ref{hir2}),
(\ref{hir3}) are equivalent to each other and to the bilinear
identity (\ref{hir1}).

This means that each eigenvalue of the master $T$-operator 
is a tau-function of the KP hierarchy.
Equation (\ref{master2}) is the expansion of the tau-function
in Schur polynomials \cite{Sato,Orlov-Shiota,EH}.

\subsubsection{The quantum Giambelli formula}

All the bilinear equations for $T^G(x, {\bf t})$ follow from the
quantum Giambelli formula for
the Gaudin model:
\begin{equation}\label{det1a}
{\sf T}^{G}_{\lambda }(x)=
\det_{i,j=1, \ldots ,
d(\lambda )}{\sf T}^{G}_{\lambda_i -i, \lambda_{j}'-j}(x),
\end{equation}
where the notation
${\sf T}^{G}_{l,k}(x):={\sf T}^{G}_{(l+1, 1^k)}(x)$ is used and
$d(\lambda )$ is the number of boxes in the main diagonal of
the Young diagram $\lambda$.
Note that the quantum Giambelli formula for
the Gaudin model has the same form as the
one for the original spin chain\footnote{The quantum Giambelli 
formula for
$U_{q}(B_{n}^{(1)})$-invariant vertex models was proposed in
\cite{KOS96}.}
before the $\eta \to 0$ limit.
In the polynomial normalization we have
\begin{equation}\label{det1b}
T^{G}_{\lambda }(x)(T^G_{\emptyset}(x))^{d(\lambda )-1}=
\det_{i,j=1, \ldots ,
d(\lambda )}T^{G}_{\lambda_i -i, \lambda_{j}'-j}(x).
\end{equation}
If \eqref{det1b} holds, then
the Jacobi identity for this determinant produces
the 3-term bilinear identities (the Pl\"ucker relations) 
for the coefficients of the
Schur function expansion \eqref{master2}. This implies 
the KP hierarchy for $T^G(x, {\bf t})$
(see \cite[example 2, page 959]{JM83}). A more direct proof
is given in the appendix.

Set
\beq\label{Q}
Q(z, \zeta )=
\frac{1}{(z -\zeta)}\,
{\sf d}^{\otimes n}\! \left ( \frac{w(z)}{w(\zeta )}\right ),
\eeq
where
$$
w(z)=\bigl (\det (\II -zh)\bigr )^{-1}.
$$
The proof of the quantum Giambelli formula is based on the
following lemma:
\begin{lemma}
\label{basicid}
The operator $Q(z, \zeta )$ obeys the ``exchange relation'':
\beq\label{basic}
Q(z_1, \zeta_1)Q(z_2, \zeta_2)=
Q(z_2, \zeta_1)Q(z_1, \zeta_2).
\eeq
\end{lemma}
This is our basic identity. It is proved in  
the appendix. We give two different proofs: one through 
the limit from spin chain (Appendix \ref{sec:proof-means-limit})
and a direct proof (Appendix \ref{sec:alternative-proof}).
Note that (\ref{basic}) 
is equivalent to the very special case of the Fay identity
(\ref{hir20}) at ${\bf t}=-[z_0^{-1}]-[z_3^{-1}]$ and
$x=x_1=x_2 =\ldots =x_n$ (in this case the third term vanishes).

We will use the notation ${\sf T}^{G, n}_{\lambda }(x)$
for the $T$-operator acting on $n$ sites.

\begin{cor}
\label{rank}
The semi-infinite matrix $(M_{\alpha \beta})
:=({\sf d}_{n+1} {\sf T}^{G,n}_{\alpha ,\beta}(x))_{\alpha, \beta \geq 0}$
has rank 1.
\end{cor}

\noindent
{\it Proof of the corollary.}
We will write ${\sf T}^{G,n}_{\alpha ,\beta}(x):=
{\sf T}^{G,n}_{\alpha ,\beta}$ for brevity.
It is enough to prove that
\beq\label{B1}
\left | \begin{array}{ll}
{\sf d}_{n+1}  {\sf T}^{G,n}_{\alpha_1 \beta_1} &
{\sf d}_{n+1} {\sf T}^{G,n}_{\alpha_1 \beta_2}
\\
{\sf d}_{n+1} {\sf T}^{G,n}_{\alpha_2 \beta_1} &
{\sf d}_{n+1} {\sf T}^{G,n}_{\alpha_2 \beta_2}
\end{array} \right |=0 \quad \mbox{for any $\alpha_i, \beta_k \geq 0$.}
\eeq
At $n=0$ ${\sf T}^{G,0}_{\alpha \beta}$ is just the hook 
character $\chi_{\alpha, \beta}:=\chi_{(\alpha +1, 1^{\beta})}(h)$
and the assertion can be easily proved 
by passing to the generating function of hook characters \cite{Macdonald}
$$
E(z,\zeta )=\sum_{\alpha , \beta \geq 0}\chi_{\alpha, \beta}
z^{\alpha}(-\zeta )^{\beta}=\frac{1}{z-\zeta }\left (\frac{w(z)}{w(\zeta )}
-1\right ).
$$
Let us multiply the determinant by
$z_1^{\alpha_1}z_2^{\alpha_2}(-\zeta_1)^{\beta_1}(-\zeta_2)^{\beta_2}$ and
sum over all $\alpha_i, \beta_k$. We then see that
$$
\left | \begin{array}{ll}
{\sf d} E(z_1, \zeta_1)&
{\sf d} E(z_1, \zeta_2)
\\
{\sf d} E(z_2, \zeta_1)&
{\sf d} E(z_2, \zeta_2)\end{array}\right |=0.
$$
Indeed,
$$
{\sf d}E(z,\zeta)=\sum_{k\geq 1}\frac{z^k -\zeta ^k}{z-\zeta}\, h^{k-1}\,
\frac{w(z)}{w(\zeta)}=
\frac{w(z)/w(\zeta)}{(1-zh)(1-\zeta h)}\, ,
$$
hence the statement follows. More generally, the assertion
of Lemma \ref{basicid} means that\footnote{Due to 
commutativity of the modified co-derivatives 
${\sf d}^{\otimes k}E(z, \zeta)=
{\sf d}_{1}\dots {\sf d}_{k} E(z, \zeta)$
is equivalent to $ {\sf d}_{k}\dots {\sf d}_{1} E(z, \zeta)$.}
\beq\label{EEEE}
\left | \begin{array}{ll}
{\sf d}^{\otimes k}E(z_1, \zeta_1)&{\sf d}^{\otimes k}E(z_1, \zeta_2)
\\
{\sf d}^{\otimes k}E(z_2, \zeta_1)&{\sf d}^{\otimes k}
E(z_2, \zeta_2)\end{array}\right |=0 \quad
\mbox{for any $k\geq 1$}.
\eeq
Now we claim that vanishing of the determinant
$$
D_n=\left | \begin{array}{ll}
{\sf d}_{n+1} {\sf T}^{G,n}_{\alpha_1 \beta_1} &
{\sf d}_{n+1} {\sf T}^{G,n}_{\alpha_1 \beta_2}
\\
{\sf d}_{n+1} {\sf T}^{G,n}_{\alpha_2 \beta_1} &
{\sf d}_{n+1} {\sf T}^{G,n}_{\alpha_2 \beta_2}
\end{array} \right |
$$
for any $n\geq 0$ is equivalent
to vanishing of
$$
D_{n-1}^{(1)}=
\left | \begin{array}{ll}
{\sf d}_{n+1} {\sf d}_{n} {\sf T}^{G,n-1}_{\alpha_1 \beta_1} &
{\sf d}_{n+1} {\sf d}_{n} {\sf T}^{G,n-1}_{\alpha_1 \beta_2}
\\
{\sf d}_{n+1} {\sf d}_{n} {\sf T}^{G,n-1}_{\alpha_2 \beta_1} &
{\sf d}_{n+1} {\sf d}_{n} {\sf T}^{G,n-1}_{\alpha_2 \beta_2}
\end{array} \right |
$$
for any $n\geq 1$.
This is clear from the identity
$$\displaystyle{D_n =
\Bigl (1+\frac{{\sf d}_{n}}{x-x_n}\Bigr )
\PP_{n,n+1} D_{n-1} \PP_{n,n+1}
+\frac{1}{(x-x_{n})^{2}}D_{n-1}^{(1)}}$$
and the ``initial condition'' $D_0=0$ established above.
In a similar way, one can show that the assertion is equivalent
to vanishing of
$
\left | \begin{array}{ll}
{\sf d}^{\otimes n}\chi_{\alpha_1 \beta_1} &
{\sf d}^{\otimes n}\chi_{\alpha_1 \beta_2}
\\
{\sf d}^{\otimes n}\chi_{\alpha_2 \beta_1} &
{\sf d}^{\otimes n}\chi_{\alpha_2 \beta_2}
\end{array} \right |
$
for any $n\geq 0$. But this follows from (\ref{EEEE}) and the
corollary is thus proved.

After these preliminaries,
the proof of the
quantum Giambelli formula is easy. Suppose that it holds for some
$n\geq 0$ (for example, it holds for $n=0$, in which case it is the
usual Giambelli formula for characters; see, e.g., \cite{Macdonald}).
Let us apply
$\displaystyle{\Bigl (1+\frac{{\sf d}_{n+1}}{x-x_{n+1}}
\Bigr )}$ to the both sides. In the l.h.s. we get
${\sf T}^{G, n+1}_{\lambda }(x)$ --
the $T$-operator for the model on $n+1$ sites.
In the r.h.s. we get
$$
\det_{1 \le i,j \le d(\lambda)}
{\sf T}^{G, n}_{\lambda_i -i, \lambda_{j}'-j}(x) +
\frac{{\sf d}_{n+1}
{\displaystyle \det_{1 \le i,j \le d(\lambda)} }
{\sf T}^{G, n}_{\lambda_i -i, \lambda_{j}'-j}(x)}{x-x_{n+1}}.
$$
Using the rule of differentiating determinants and the fact that
the matrix
$$\bigl ({\sf d}_{n+1}\! {\sf T}^{G, n}_{\lambda_i -i,
\lambda_{j}'-j}(x)\bigr )_{1 \le i,j \le d(\lambda)},$$ being a
submatrix of the
$({\sf d}_{n+1} T^{G,n}_{\alpha ,\beta}(x))_{\alpha, \beta \geq 0}$,
has rank 1, we see that
the r.h.s. is equal to
$$
\det_{1 \le i,j \le d(\lambda)}
 \left ( {\sf T}^{G, n}_{\lambda_i -i, \lambda_{j}'-j}(x) +
\frac{{\sf d}_{n+1}
{\sf T}^{G, n}_{\lambda_i -i, \lambda_{j}'-j}(x)}{x-x_{n+1}}\right )=
\det_{1 \le i,j \le d(\lambda)}
 {\sf T}^{G, n+1}_{\lambda_i -i, \lambda_{j}'-j}(x).
$$
This proves the quantum Giambelli formula for the model on
$n+1$ sites.

\subsubsection[The master \texorpdfstring{$T$}{T}-operator and matrix derivatives]{The master {\boldmath $T$}-operator and matrix derivatives}
\label{sec:matrix}

Another proof of the bilinear identity (\ref{hir1}) 
can be given using the technique of 
matrix derivatives and matrix integrals. 

Let us first consider the special case 
of the master $T$-operator for the 
Gaudin model (\ref{master1}) at $x=x_1=x_2=\ldots=x_n$:
\beq\label{mtrest}
T^{G,n}({\bf t}) :={\sf d}_n\, \ldots \, 
{\sf d}_1\exp \Bigl (\sum_{k\geq 1}t_k \, \mbox{tr}\, h^k\Bigr ).
\eeq
The operator ${\sf d}$ is just the matrix derivative with 
respect to the transposed matrix $h^t$, which is well known in the 
theory of matrix models (for more details see, e.g., \cite{Mor}):
$$
{\sf d}=\frac{\p}{\p h^t}\,.
$$
One can introduce the generating function for 
the master $T$-operators which depends
on an auxiliary external matrix $A$:
\beq
T^{G}({\bf t},A)=\exp\left(\tr A\frac{\p}{\p h^{t}}\right) \exp \left (\sum_{k\geq 1}t_k \, \mbox{tr}\, h^k\right )= \exp \left (\sum_{k\geq 1}t_k \, \mbox{tr}\, \left(h+A\right)^k\right ).
\eeq
Clearly, this function generates the master $T$-operators for any $n$:
\beq
T^{G,n}({\bf t})^{i_1,\ldots,i_n}_{j_1,\ldots,j_n}=\left.\frac{\p}{\p {A}^{j_1}_{i_1}}\ldots \frac{\p}{\p A^{j_n}_{i_n}}
T^{G}({\bf t},A)\right|_{A=0}.
\eeq
Here $A=\sum_{i,j}A^{i}_{j}e_{ij}$ and 
$T^{G,n}({\bf t})=\sum_{\{i_{k},j_{k}\}_{k=1}^{n}} 
T^{G,n}({\bf t})^{i_1,\ldots,i_n}_{j_1,\ldots,j_n} 
e^{(1)}_{i_1 j_1} \cdots e^{(n)}_{i_n j_n} $. 

In a similar way, 
one can introduce a generating function for products 
of the master $T$-operators 
\beq\label{bic}
\Phi^{G,n}({\bf t}, {\bf t'}):=T^{G,n}({\bf t})T^{G,n}({\bf t'}).
\eeq
It has the form
\begin{equation}\label{phi}
\begin{array}{c}
\displaystyle{
\Phi^G({\bf t}, {\bf t'};A)=\left.
\exp\left(\tr A\frac{\p}{\p h^{t}}\frac{\p}{\p g^{t}}\right)\exp \left (\sum_{k\geq 1}t_k \, \mbox{tr}\, h^k\right )\exp \left (\sum_{k\geq 1}t'_k \, \mbox{tr}\, g^k\right )\right|_{g=h}}\\ \\
\displaystyle{=1+\left(t_1+t_1'\right)\tr h +\frac{\left(t_1+t_1'\right)^2}{2!}(\tr h)^2+(t_2+t_2')\tr h^2 + t_1t_1\tr A }\\ \\ 
\displaystyle{+\frac{\left(t_1+t_1'\right)^3}{3!}
(\tr h)^3+\left(t_1+t_1'\right)(t_2+t_2')\tr h 
\,\tr h^2+(t_3+t_3')\tr h^3} \\ \\ 
\displaystyle{+t_1t_1'(t_1+t_1')\tr A \, \tr h+2(t_1 t_2'+ t_1't_2)\tr (Ah)+\ldots} 
\end{array}
\end{equation}
This expression
can be regarded as a formal series in $A$ with coefficients 
which are functions of ${\bf t}$ and ${\bf t'}$.
Equivalently, it can be regarded as a formal series in 
${\bf t}$ and ${\bf t'}$ with coefficients depending on $A$.

We have, for example:
\begin{equation}
\begin{array}{c}
\displaystyle{
\frac{\p}{\p {A}^{j}_{i}}\Phi^{G}({\bf t},{\bf t}';A)\big|_{A=0}=\sum_{k=1}^N\left.\left(\frac{\p}{\p h^{t}}\right)^i_k
\left(\frac{\p}{\p g^{t}}\right)^k_j\exp \left (\sum_{k\geq 1}t_k \, 
\mbox{tr}\, h^k +t'_k \, \mbox{tr}\, g^k\right )\right|_{g=h}} \\ \\ 
\displaystyle{= \sum_{k=1}^NT^{G,1}({\bf t})^i_k  T^{G,1}({\bf t}')^k_j =\Phi^{G,1}({\bf t}, {\bf t'})^{i}_{j}}
\end{array}
\end{equation}
and
\begin{equation}
\begin{array}{c}
\displaystyle{
\frac{\p}{\p {A}^{j_1}_{i_1}}\frac{\p}{\p 
{A}^{j_1}_{i_1}}\Phi^{G}({\bf t},{\bf t}';A)\big|_{A=0}}\\ \\ 
\displaystyle{=\sum_{k_1,k_2=1}^N\left.\left(\frac{\p}{\p h^{t}}\right)^{i_1}_{k_1}\left(\frac{\p}{\p g^{t}}\right)^{k_1}_{j_1}
\left(\frac{\p}{\p h^{t}}\right)^{i_2}_{k_2}\left(\frac{\p}{\p g^{t}}\right)^{k_2}_{j_2}\exp \left (\sum_{k\geq 1}t_k \, \mbox{tr}\, 
h^k +t'_k \, \mbox{tr}\, g^k\right )\right|_{g=h}} \\ \\ 
\displaystyle{= \sum_{k_1,k_2=1}^NT^{G,2}({\bf t})^{i_1,i_2}_{k_1,k_2}  
T^{G,2}({\bf t}')^{k_1,k_2}_{j_1,j_2} =\Phi^{G,2}({\bf t}, {\bf t'})^{i_1,i_2}_{j_1,j_2}}
\end{array}
\end{equation}
In general, the product (\ref{bic}) is just the $n$'th matrix 
derivative of this function with respect to the (transposed) matrix $A$:
\beq
\Phi^{G,n}({\bf t}, {\bf t'})^{i_1,\ldots,i_n}_{j_1,\ldots,j_n}=\frac{\p}{\p {A}^{j_1}_{i_1}}\ldots \frac{\p}{\p A^{j_n}_{i_n}}\Phi^{G}({\bf t},{\bf t};A)\big|_{A=0}.
\eeq
Therefore, the bilinear identity (\ref{hir1}) for (\ref{mtrest}) 
and any $n$ is equivalent to the following scalar identity:
\beq\label{Hir}
\oint_{\infty}e^{\xi({\bf t-t'},z)}\Phi^G({\bf t}-\left[z^{-1}\right], {\bf t'}+\left[z^{-1}\right];A)\, dz =0.
\eeq
The proof of this identity, as well as its generalization to arbitrary $x$ and $x_i$'s, is given in Appendix \ref{Hirpr}.

\subsection[Functional relations for higher Gaudin
\texorpdfstring{$T$}{T}-operators]{Functional relations for higher Gaudin
{\boldmath $T$}-operators}

Here we present analogs of the
Cherednik-Bazhanov-Reshetikhin (CBR) determinant formulas \cite{Chered,BR90}.

The CBR determinant formulas for the original spin chain are:
\beq\label{CBR-spin}
{\sf T}_{\lambda}(x)=\det_{1\leq i,j\leq \ell (\lambda )}
{\sf T}_{(\lambda_i-i+j)}\bigl (x\! -\! (j\! -\! 1)\eta \bigr ),
\eeq
\beq\label{CBR-spin1}
{\sf T}_{\lambda}(x)=\det_{1\leq i,j\leq \lambda_{1}}
{\sf T}_{(1^{\lambda^{\prime}_i-i+j})}\bigl (x\! +\! (j\! -\! 1)\eta \bigr ).
\eeq
They are spectral parameter dependent 
analogs of the Jacobi-Trudi formula for 
characters of the group element $g$ (the twist matrix). 
In fact, \eqref{CBR-spin} and \eqref{CBR-spin1} reduce 
to (\ref{JT-det}) and
\eqref{JT-det2} in the limit $|x| \to \infty $.

We have found
the following analogs of the CBR formulas 
for the Gaudin transfer matrices: 
\beq\label{CBR}
{\sf T}^G_{\lambda}(x)=\det_{1\leq i,j\leq \ell (\lambda )}
\left ( \sum_{k=0}^{j-1}(-1)^k
\Bigl (\begin{array}{c}j\! -\! 1\\ k\end{array} \Bigr )\,
\p_x^{k} {\sf T}^G_{(\lambda_i-i+j-k)}(x)\right ),
\eeq
\beq
{\sf T}^G_{\lambda}(x)=\det_{1\leq i,j\leq  \lambda_{1}}
\left ( \sum_{k=0}^{j-1}(-1)^k
\Bigl (\begin{array}{c}j\! -\! 1\\ k\end{array} \Bigr )\,
\p_x^{k} {\sf T}^G_{(1^{\lambda^{\prime }_i-i+j-k})}(x)\right ).
 \label{CBR-dual}
\eeq
In the same ways as \eqref{CBR-spin}, 
equation (\ref{CBR}) is also 
an  analogue of the Jacobi-Trudi formula (\ref{JT-det})  and 
reduces to it
in the limit $|x| \to \infty $. 
Equation \eqref{CBR-dual} is 
a ``dual'' determinant formula (an analogue of \eqref{JT-det2}).

The determinant formulas (\ref{CBR}) and \eqref{CBR-dual}
follow from the differential Fay identity\footnote{We also remark that the quantum Giambelli formula \eqref{det1a}
follows from  \eqref{CBR}.}. A sketch of proof is given in 
Appendix \ref{appen-higher}. Here we present 
an auxiliary determinant formula obtained by
iterations of the differential Fay identity.

For any positive integer $N$ and any subset
$ \{ i_{1},i_{2},\dots, i_{m} \} \subset \{1,2,\dots, N \}$
where all  $i_k$ are different,
we introduce the operator
\begin{align}
T^{G, \{ i_{1},i_{2},\dots, i_{m} \}  }(x, {\bf t}) =
 T^{G}(x, {\bf t} +
\sum_{k=1}^{m} [z_{i_{k}}^{-1}])
= \exp
\left(\sum_{k=1}^{m}
\xi(\tilde{\partial},z_{i_{k}}^{-1})
\right) T^{G}(x, {\bf t}).
 \label{mastershift}
\end{align}
Solving \eqref{hir3} for \eqref{mastershift} recursively, 
we obtain the determinant formula
\begin{align}
T^{G, \{ i_{1},i_{2},\dots, i_{m} \}  }(x, {\bf t}) &=
\frac{
\mathrm{det}_{1 \le k,j \le m}
\left(
\sum_{l=0}^{j-1}(-1)^l
\Bigl (\begin{array}{c}j\! -\! 1\\ l\end{array} \Bigr )
\,
z_{i_{k}}^{j-m-l}
\p_x^{l}
T^{G, \{ i_{k} \}  }(x, {\bf t}) \right)
}
{
\mathrm{det}_{1 \le k,j \le m}
\left( z_{i_{k}}^{j-m}
\right)
\bigl (T^{G}(x, {\bf t})\bigr )^{m-1}
}.
\label{masterdet1}
\end{align}
Note that \eqref{hir2} is a Pl\"{u}cker identity and
\eqref{hir3} is the Jacobi identity for minors of 
the matrix $\left( z_{i_{k}}^{j-m} (1- z_{i_{k}}^{-1} \p_x )^{j-1}
T^{G, \{ i_{k} \}  }(x, {\bf t}) \right)_{1 \le k,j \le m}$  
entering \eqref{masterdet1}. 
As is shown in Appendix \ref{appen-higher},
\eqref{masterdet1} is 
a generating function for \eqref{CBR}.  

\subsection{The Baker-Akhiezer functions}

According to the general scheme, the Baker-Akhiezer (BA)
function and its adjoint corresponding to the tau-function (\ref{master1})
are given by the formulas
\beq\label{BA1}
\psi (x, {\bf t};z)=e^{xz +\xi ({\bf t}, z)}
(T^G(x, {\bf t}))^{-1}\, T^G\bigl (x, {\bf t}-[z^{-1}]\bigr ),
\eeq
\beq\label{BA2}
\psi^{*} (x, {\bf t};z)=e^{-xz -\xi ({\bf t}, z)}
(T^G(x, {\bf t}))^{-1}\, T^G\bigl (x, {\bf t}+[z^{-1}]\bigr ).
\eeq
For brevity, we will refer to both $\psi$ and $\psi^*$ as
BA functions. In terms of the BA functions,
the bilinear identity (\ref{hir1}) can be written as
\beq\label{hir1psi}
\oint_{\infty}\psi (x, {\bf t};z)
\psi^{*} (x, {\bf t'};z) dz =0.
\eeq

Using the definition (\ref{master1}), we have:
\beq\label{BA3}
\psi (x, {\bf t};z)=z^{-N}e^{xz +\xi ({\bf t}, z)}
(T^G(x, {\bf t}))^{-1} \!
{\prod_{i=1}^{n}}
(x\! -\! x_i \! +\! {\sf d}_i)
\left [ \det \left (z \! -\! h\right )
\exp \Bigl (\sum_{k\geq 1}t_k \, \mbox{tr}\, h^k\Bigr )\right ],
\eeq
\beq\label{BA4}
\psi ^*(x, {\bf t};z)=z^N e^{-xz -\xi ({\bf t}, z)}
(T^G(x, {\bf t}))^{-1} \!
{\prod_{i=1}^{n}}
(x\! -\! x_i \! +\! {\sf d}_i)
\left [ \frac{1}{\det \left (z\! -\! h\right )}
\exp \Bigl (\sum_{k\geq 1}t_k \, \mbox{tr}\, h^k\Bigr )\right ],
\eeq
where we write simply $\det (z-h)$ instead of
$\det (\II z -h)$. From these formulas we see that
$e^{-xz -\xi ({\bf t}, z)}\psi (x, {\bf t};z)$ is a polynomial
in $z^{-1}$ of degree $N$ while
$e^{xz +\xi ({\bf t}, z)}\psi ^* (x, {\bf t};z)$ is a rational
function of $z$ with poles at the points $z=k_i$ (eigenvalues
of the matrix $h$) of at least first order (because of the
$\det (z-h)$ in the denominator). Moreover, since each co-derivative
raises the order of the poles, these poles may be actually of a higher
order, up to $n+1$. Also, as is seen from the
second formula, this function has the $N$-th order zero
at $z=0$. (We assume that $k_i\neq 0$.)

Regarded as functions of $x$, both $e^{-xz}\psi (x, {\bf t};z)$ and
$e^{xz}\psi ^* (x, {\bf t};z)$ are rational functions of $x$ with
$n$ zeros and $n$ poles which are simple in general position.
From (\ref{master1}) and (\ref{BA3}), (\ref{BA4}) it follows that
\beq\label{BA5}
\lim_{x\to \infty}\Bigl (e^{-xz -\xi ({\bf t}, z)}\psi (x, {\bf t};z)
\Bigr )=z^{-N}\det (z-h),
\eeq
\beq\label{BA6}
\lim_{x\to \infty}\Bigl (e^{xz +\xi ({\bf t}, z)}\psi ^* (x, {\bf t};z)
\Bigr )=z^{N}(\det (z-h))^{-1}.
\eeq

The BA functions for any solution of the KP hierarchy satisfy the following differential
equations:
\beq\label{BA7}
\p_{t_2}\psi (x, {\bf t};z)=\p_x^2 \psi (x, {\bf t};z) +
2u(x, {\bf t}) \psi (x, {\bf t};z),
\eeq
\beq\label{BA8}
-\p_{t_2}\psi ^*(x, {\bf t};z)=\p_x^2 \psi ^* (x, {\bf t};z) +
2u(x, {\bf t}) \psi ^* (x, {\bf t};z),
\eeq
where
\beq\label{BA9}
u(x, {\bf t})=\p_x^2 \log T^G(x, {\bf t}).
\eeq

We also note the formulas for the {\it stationary} BA
functions $\psi (x,z):=\psi (x,0;z)$, $\psi^* (x,z):=\psi^* (x,0;z)$
which directly follow from (\ref{BA3}), (\ref{BA4}):
\beq\label{BA3st}
\psi (x, z)=z^{-N}e^{xz}
{\prod_{i=1}^{n}}
\Bigl (1+\frac{{\sf d}_i}{x-x_i}\Bigr )
\det (z-h),
\eeq
\beq\label{BA4st}
\psi ^*(x, z)=z^N e^{-xz}
{\prod_{i=1}^{n}}
\Bigl (1+\frac{{\sf d}_i}{x-x_i}\Bigr )
\frac{1}{\det (z-h)}.
\eeq

Below we will also need the relation
\beq\label{BA10}
\p_{t_m}\p_{t_1}\log T^G (x, {\bf t})=
\mbox{res}_{\infty}\bigl (\psi (x, {\bf t};z\bigr )
\psi^* (x, {\bf t};z)z^m \, dz \bigr ).
\eeq
(Here $\mbox{res}_{\infty}(\ldots )\equiv \frac{1}{2\pi i}
\oint_{\infty}(\ldots )$ and $ \frac{1}{2\pi i}
\oint_{\infty}z^{-1}dz =1$.)
This relation follows from the bilinear identity (\ref{hir1})
by applying $\p_{t_m}$ and putting $t'_k=t_k$ afterwards.

\section{Zeros of the master \texorpdfstring{\boldmath $T$}{T}-operator as the Calogero-Moser
particles}

The eigenvalues of the master $T$-operator are polynomials in
the spectral parameter $x$:
\begin{equation}\label{CM1}
T^G(x, {\bf t})=e^{t_1 \, {\scriptsize{\mbox{tr}}\, h}
+t_2 \, {\scriptsize{\mbox{tr}}\, h^2 +\ldots}}
\prod_{k=1}^{n}(x+t_1 -x_k(t_2, t_3, \ldots )).
\end{equation}
The roots of each eigenvalue have their own
dynamics in the times $t_k$. This dynamics is known \cite{Kricheverrat1,*Kricheverrat2}
to be given by the rational CM model \cite{Calogero,*Moser}.
The inhomogeneity parameters of the Gaudin model play the role of
coordinates of the CM particles at $t_i=0$:  $x_j=x_j(0)$.
In particular, we have
$ T^G(x,0)=T^G_{\emptyset}(x)=\prod_{k=1}^{n}(x-x_k)$.

Using (\ref{master3a}), we easily obtain the formula
for ${\sf T}^G_{(1)}(x)= T^G_{(1)}(x)/T^G_{\emptyset}(x)$
given in (\ref{g2}). For ${\sf T}^G_{(1^2)}(x)$ the second
equation in (\ref{master3a}) yields
$$
{\sf T}^G_{(1^2)}(x)=\chi_{(1^2)}(h)+
\mbox{tr}\, h
\sum_i \frac{1}{x-x_i}+\sum_{i<j}\frac{1}{(x-x_i)(x-x_j)}
+\frac{1}{2}\sum_i \frac{\dot x_i}{x-x_i},
$$
where $\dot x_i =\p_{t_2}x_i(t_2)\Bigr |_{t_2=0}$.
Comparing with the third equation in (\ref{g2}), we conclude that
the initial velocities are proportional to eigenvalues of the
Gaudin Hamiltonians:
\begin{equation}\label{CM2}
\dot x_i =-2 H_i.
\end{equation}
This unexpected connection between the quantum Gaudin model 
and the classical CM model was observed in 
\cite{MTV1,*MTV2} using a different reasoning.

\subsection{Dynamics of poles}

\subsubsection{The Lax pair for the CM model}

Following Krichever's work
\cite{Kricheverrat1,*Kricheverrat2}, let us derive equations of motion
for the $t_2$-dynamics of the $x_i$'s. It is convenient to denote
$t_2=t$ and put all other times to zero since they are irrelevant for
this derivation.
From (\ref{BA9}) we see that
\beq\label{CM3}
u=-\sum_{i=1}^{n}\frac{1}{(x-x_i)^2}.
\eeq
The method of \cite{Kricheverrat1,*Kricheverrat2} is to perform the pole expansion
of the linear problem (\ref{BA7}) for the BA function $\psi$. From
(\ref{BA3}) we have
\beq\label{CM4}
\psi = e^{xz+t z^2}\left (c_0 (z) +
\sum_{i=1}^{n}\frac{c_i(z,t)}{x-x_i (t)}\right ),
\eeq
where $c_0(z)=\det (\II -z^{-1}h)$ (see (\ref{BA5})). Plugging this into (\ref{BA7}), we obtain
$$
e^{-xz-t z^2}\bigl (\p_t -\p_x^2\bigr )\left [ e^{xz+t z^2}\Bigl (c_0  +
\sum_{i=1}^{n}\frac{c_i}{x-x_i}\Bigr )\right ]
+2 \Bigl (\sum_{i=1}^{n}\frac{1}{(x-x_i)^2}\Bigr )\Bigl (c_0  +
\sum_{i=1}^{n}\frac{c_i}{x-x_i}\Bigr )=0.
$$
The l.h.s. is a rational function of $x$ with first and second
order poles at $x=x_i$ (possible
poles of the third order cancel automatically) vanishing at infinity.
Therefore, to solve the linear problem it is enough to cancel
all the poles. Equating the coefficients in front of each pole
to zero, we get the following system of equations for
$i=1, \ldots , n$:
$$
\left \{
\begin{array}{l}\displaystyle{
(\dot x_i +2z)c_i +2\sum_{k\neq i}\frac{c_k}{x_i-x_k}=-2c_0
\quad \mbox{(from 2-nd order poles)}},
\\ \\
\displaystyle{\dot c_i  +2c_i \sum_{k\neq i}\frac{1}{(x_i-x_k)^2}
-2\sum_{k\neq i}\frac{c_k}{(x_i-x_k)^2}=0
\quad \mbox{(from 1-st order poles).}}
\end{array}
\right.
$$
These equations can be rewritten in the matrix form:
\beq\label{CM5}
\left \{
\begin{array}{l}
(z\II -Y){\sf c}=-c_0 (z){\sf 1}
\\ \\
{\sf \dot c}=T{\sf c},\end{array}
\right.
\eeq
where ${\sf c}=(c_1, c_2, \ldots , c_n)^{\sf t}$,
${\sf 1} =(1, 1, \ldots , 1)^{\sf t}$ are $n$-component
vectors and the $n \! \times \!n$ matrices $Y=Y(t)$, $T=T(t)$ are
given by
\beq\label{CM6}
Y_{ik}= -p_i \delta_{ik}-
\frac{1-\delta_{ik}}{x_i-x_k}\,, \quad \quad
p_i:=\frac{1}{2}\, \dot x_i,
\eeq
\beq\label{CM7}
T_{ik}=-\delta_{ik}\sum_{j\neq i}\frac{2}{(x_i-x_j)^2}+
\frac{2(1-\delta_{ik})}{(x_i-x_k)^2}.
\eeq
The compatibility condition of the system (\ref{CM5}) is
\beq\label{CM8}
\dot Y=[T, \, Y].
\eeq
It is the Lax representation for the equations of motion of the
CM model:
\beq\label{CM9}
\ddot x_i = -8\sum_{j\neq i} \frac{1}{(x_i-x_j)^3}.
\eeq
The matrices $Y,T$ form the Lax pair for the CM model. 

In a similar way, plugging the adjoint BA function
\beq\label{CM4a}
\psi ^*= e^{-xz-t z^2}\left (c_0^{-1} (z) +
\sum_{i=1}^{n}\frac{c_i^*(z,t)}{x-x_i (t)}\right )
\eeq
to the adjoint linear problem (\ref{BA8}), we get
\beq\label{CM5a}
\left \{
\begin{array}{l}
(z\II -Y^{\sf t}){\sf c}^*=c_0^{-1} (z){\sf 1}
\\ \\
{\sf \dot c}^*=-T{\sf c}^*, \end{array}
\right.
\eeq
where ${\sf c}^*=(c_1^*, c_2^*, \ldots , c_n^*)^{\sf t}$.

Using (\ref{CM5}), (\ref{CM5a}) and recalling that
$c_0(z)=z^{-N}\det (z -h)$, we find the solutions for the vectors
${\sf c}$, ${\sf c}^*$:
\beq\label{cc}
\begin{array}{l}
{\sf c}(z,t)=-z^{-N}\det (z -h)(z -Y(t))^{-1}{\sf 1}\,,
\\ \\
{\sf c}^*(z,t)=z^{N}(\det (z -h))^{-1}
(z -Y^{\sf t}(t))^{-1}{\sf 1}.
\end{array}
\eeq
Set $X(t)=\mbox{diag} (x_1(t), \ldots , x_n(t))$.
For the functions $\psi$, $\psi^*$ themselves we then have:
\beq\label{cc1}
\begin{array}{l}\displaystyle{
\psi =c_0(z)\, e^{xz+tz^2}\Bigl (
1-{\sf 1}^{\sf t}(x -X(t))^{-1}(z -Y(t))^{-1}{\sf 1}\Bigr )},
\\ \\
\displaystyle{\psi^*=c_0^{-1}(z)e^{-xz-tz^2}\Bigl (
1+{\sf 1}^{\sf t}(z -Y(t))^{-1}(x -X(t))^{-1}{\sf 1}\Bigr )}.
\end{array}
\eeq

\subsubsection[Properties of the matrices \texorpdfstring{$X(t), Y(t), T(t)$}{X(t), Y(t), T(t)}]{Properties of the matrices {\boldmath $X(t), Y(t), T(t)$}}

Here we list some properties of the matrices $X, Y, T$
to be used in what follows.

As is well known (and easy to check),
the matrices $X$, $Y$ satisfy the commutation
relation
\beq\label{comm}
[X, \, Y]=\II -{\sf 1}\otimes {\sf 1}^{\sf t}
\eeq
(here ${\sf 1}\otimes {\sf 1}^{\sf t}$ is the $n\! \times \! n$
matrix of rank $1$ with all entries equal to $1$).

\begin{lemma} For any $k\geq 0$ it holds \end{lemma}
\beq\label{trYk}
{\sf 1}^{\sf t} Y^k {\sf 1} = \mbox{tr}\, Y^k.
\eeq

\noindent
Indeed, we have: ${\sf 1}^{\sf t} Y^k {\sf 1}=
\mbox{tr}\, \bigl ({\sf 1}\otimes {\sf 1}^{\sf t}\cdot Y^k\bigr )=
\mbox{tr}\, \bigl ((\II -[X, \, Y])Y^k\bigr )=
\mbox{tr}\, Y^k -\mbox{tr}\, [X, \, Y^{k+1}]$ but the last trace
is $0$ as trace of a commutator.

Let $E_{ik}$ be the basis matrices of $gl(n)$ having
just one non-zero element (equal to $1$) at the place $ik$:
$(E_{ik})_{i'k'}=\delta_{ii'}\delta_{kk'}$. The following
relations are easy to verify directly:
\beq\label{ET}
\frac{\p Y}{\p p_i}=-E_{ii}, \quad \quad
\frac{\p Y}{\p x_i}=\frac{1}{2}
\bigl [E_{ii}, T\bigr ].
\eeq

\subsubsection{Integrals of motion}

The matrix $Y$ is the Lax matrix for the CM model. 
As is seen from (\ref{CM8}), the time evolution preserves
its spectrum, i.e., the coefficients 
${\cal J}_k$ of the characteristic polynomial
\beq\label{char-pol}
\det (z\II -Y(t))=\sum_{k=0}^{n}{\cal J}_k z^{n-k}
\eeq
are integrals of motion. The highest integral, ${\cal J}_n$, 
was found explicitly in \cite{Sawada-Kotera}, where a recurrence 
procedure for finding all other integrals of motion was also 
suggested. 
In fact this procedure is equivalent to the following
explicit expression for the characteristic polynomial:
\beq\label{explicit}
\det \bigl (z\II -Y(t)\bigr )=\exp \Bigl ( \sum_{i<j}
\frac{\p_{p_i}\p_{p_j}}{(x_i-x_j)^2}\Bigr ) 
\prod_{l=1}^{n}(z+p_l).
\eeq
Note that this expression
is well-defined because the sum obtained after expansion 
of the exponential function in the r.h.s. contains a finite 
number of non-zero terms.

\subsection{Eigenvalues of the Lax matrix}

The singularities of the ${\sf c}(z,t)$, ${\sf c}^*(z,t)$ as functions
of $z$ are the same as singularities of the functions
$\psi$, $\psi^*$ in the finite part of the complex plane.
From (\ref{BA3}) we see that ${\sf c}(z,t)$ has the pole of order
$N$ at $z=0$ and no other poles. At the same time the first equation in
(\ref{cc}) states that there are possible poles at eigenvalues of
the matrix $Y(t)$ (which do not depend on time).
Therefore, they must be canceled by zeros
of $\det (z -h)$ which are at $z=k_i$ and are assumed to be simple.
If all eigenvalues of $Y$ are distinct, such cancellation
is possible only if $n\leq N$. However, the most interesting
setting for the quantum spin chains and
for the Gaudin model in particular is quite opposite: $n>N$
or even $n\gg N$
(large chain length at a fixed rank of the symmetry algebra).
Therefore, we conclude that in this case
\begin{itemize}
\item {\it The Lax matrix $Y$ has
multiple eigenvalues $k_i$ with multiplicities $m_i\geq 1$ such that
$m_1 +\ldots +m_N =n$}.
\end{itemize}

\noindent
On the first glance, a multiple eigenvalue $k_i$
might lead to an unwanted pole of $\psi$ at $z=k_i$
coming from the higher order pole of
the matrix $(z - Y)^{-1}$ which now
cannot be compensated by the {\it simple} zero of $\det (z -h)$.
In fact the higher order poles do not appear in the vector
$(z - Y)^{-1}{\sf 1}$ because ${\sf 1}$
is a special vector for the matrix $Y$ which
can be decomposed into $N$
Jordan blocks of sizes $m_i\! \times \! m_i$.
However, they do appear in the vector
$(z - Y^{\sf t})^{-1}{\sf 1}$ and the function $\psi^*$ has
multiple poles at $z=k_i$ (with multiplicities $m_i+1$).

Another way to see that eigenvalues of the Lax matrix $Y$
are the same as eigenvalues of the twist matrix $h$ (with appropriate
multiplicities) is to compare expansions of (\ref{BA3st}) and
(\ref{cc1}) at large $x$. From (\ref{BA3st}) we have:
$$
\psi (x,z)=\det (1-z^{-1}h) e^{xz} \left (
1-\frac{1}{x}\sum_i \sum_a \frac{e_{aa}^{(i)}}{z-k_a} +
O(x^{-2})\right ).
$$
The expansion of (\ref{cc1}) at $t=0$ gives (the lemma
from the previous subsection is to be used):
$$
\psi (x,z)=\det (1-z^{-1}h) e^{xz} \left (
1-\frac{1}{x}\, \mbox{tr} \,\frac{1}{z-Y_0} \, +O(x^{-2})\right ),
$$
where we set $Y_0:=Y(0)$.
Therefore, we conclude that
$$
\mbox{tr} \,\frac{1}{z-Y_0} =
\sum_i \sum_a \frac{e_{aa}^{(i)}}{z-k_a}
$$
and, since $\mbox{tr}\, (z-Y_0)^{-1}=\p_z \log \det (z-Y_0)$, we have
\beq\label{cc2}
\det (z-Y_0)=\prod_{a=1}^{N}(z-k_a)^{\sum_{i=1}^{n}e_{aa}^{(i)}}
=\prod_{a=1}^{N}(z-k_a)^{M_a},
\eeq
where $M_a$ is the operator (\ref{Ma}). Hence we see that
the $M_a$ is the ``operator multiplicity'' of the
eigenvalue $k_a$. In the sector ${\cal V}(\{m_a\})$ the
multiplicity becomes equal to $m_a$.

\subsection{Equations of motion in the hamiltonian form}

The hamiltonian form of equations of motion is
\beq\label{CM10}
\left (\begin{array}{l}\dot x_i \\ \dot p_i \end{array} \right )=
\left (\begin{array}{r} \p_{p_i} {\cal H}_2 \\
- \p_{x_i} {\cal H}_2 \end{array} \right )
\eeq
with the Hamiltonian
\beq\label{CM11}
{\cal H}_2= \mbox{tr}\, Y^2 = \sum_i p_i^2 -
\sum_{i<j}\frac{2}{(x_i-x_j)^2}\,.
\eeq
This result was generalized to the whole hierarchy in \cite{Shiota}:
\beq\label{CM12}
\left (\begin{array}{l}\p_{t_k}x_i \\ \p_{t_k} p_i \end{array} \right )=
\left (\begin{array}{r} \p_{p_i} {\cal H}_k \\
- \p_{x_i} {\cal H}_k \end{array} \right ),
\quad \quad {\cal H}_k =\mbox{tr}\, Y^k.
\eeq
The ${\cal H}_k$'s are higher integrals of motion (Hamiltonians) 
for the CM
model. They are known to be in involution 
\cite{Sawada-Kotera,Woj,OP}. This agrees with commutativity 
of the KP flows. The integrals ${\cal H}_k$ are connected with the
integrals ${\cal J}_k$ introduces in (\ref{char-pol}) by Newton's
formula \cite{Macdonald}
\beq\label{Newton}
\sum_{k=0}^{n}{\cal J}_{n-k}{\cal H}_k =0
\eeq
(here we set ${\cal H}_0=\mbox{tr}\, Y^0= n$).

For completeness, we give a short derivation of (\ref{CM12})
which is a version of the argument from \cite{Shiota}.
The main technical tool is equation (\ref{BA10}) which states that
$$
\sum_i \frac{\p_{t_m}x_i}{(x-x_i)^2}=
\mbox{res}_{\infty} \left [
\Bigl (c_0 +\sum_i \frac{c_i}{x-x_i}\Bigr )
\Bigl (c_0^{-1} +\sum_i \frac{c_i^*}{x-x_i}\Bigr ) z^m dz\right ].
$$
Matching coefficients in front of the poles at $x_i$, we get
\beq\label{CM14}
\p_{t_m}x_i=
\mbox{res}_{\infty} \bigl (c_i({\bf t}, z)c_i^*({\bf t}, z)\,
z^m dz \bigr ).
\eeq
Using (\ref{cc}), we have:
$$
c_i^* c_i = - \, \Bigl ({\sf 1}^{\sf t} \frac{1}{z-Y}\Bigr )_i
\Bigl (\frac{1}{z-Y}{\sf 1}\Bigr )_i =
-\, {\sf 1}^{\sf t}\frac{1}{z-Y} E_{ii}
\frac{1}{z-Y}{\sf 1}.
$$
Equations
(\ref{ET}) and (\ref{trYk}) allow us to rewrite the result
as follows:
$$
c_i^* c_i =\frac{\p}{\p p_i}\left ({\sf 1}^{\sf t}
\frac{1}{z-Y}{\sf 1}\right )
=\frac{\p}{\p p_i} \, \mbox{tr}\, \frac{1}{z-Y}\,,
$$
hence $\p_{t_m}x_i=\p_{p_i} \mbox{tr}\, Y^m$.
This proves the first equality in (\ref{CM12}).
Next, applying $\p_{t_2}$ to (\ref{CM14}), we get:
$$
-\p_{t_m}p_i = -\frac{1}{2}\, \p_{t_2} (\p_{t_m}x_i)=
-\frac{1}{2}\,
\mbox{res}_{\infty} \bigl (
(c_i \dot c_i^* + \dot c_i c_i^*)z^m dz\bigr )
$$
$$
=\, \frac{1}{2}\,
\mbox{res}_{\infty} \bigl (
(c_i (T {\sf c}^*)_i -c_i^* (T {\sf c})_i)z^m dz\bigr )=
-\frac{1}{2}\, \mbox{res}_{\infty} \Bigl (
{\sf c^*}^{\sf t} [E_{ii}, T] {\sf c} \, z^m dz\Bigr ).
$$
We continue the chain of equalities
using the second formula in (\ref{ET}):
$$
\p_{t_m}p_i = \mbox{res}_{\infty} \Bigl (
{\sf c^*}^{\sf t} \, \frac{\p Y}{\p x_i} \, {\sf c} \, z^m dz\Bigr )
=\mbox{res}_{\infty} \Bigl (
{\sf 1}^{\sf t}\frac{1}{z-Y}  \, \frac{\p Y}{\p x_i} \,
\frac{1}{z-Y}{\sf 1}\, z^m dz \Bigr )
$$
$$
=\, -\, \frac{\p}{\p x_i} \, \mbox{res}_{\infty} \Bigl (
{\sf 1}^{\sf t}\frac{1}{z-Y}  \, {\sf 1}\, z^m dz \Bigr )=
\, -\, \frac{\p}{\p x_i} \, \mbox{tr} \, Y^m.
$$
This is the second equality in (\ref{CM12}).

\subsection[Determinant formula for the master \texorpdfstring{$T$}{T}-operator]{Determinant formula for the master {\boldmath $T$}-operator}

The results of \cite{Shiota} imply an explicit
determinant representation of the tau-function.
It is easy to adopt it for the master $T$-operator $T^G(x, {\bf t})$
(\ref{CM1}). Let $X_0 =X(0)$ be the diagonal matrix
$X_0=\mbox{diag}(x_1, x_2, \ldots , x_n)$,
where $x_i=x_i(0)$ and $Y_0$ be the
Lax matrix (\ref{CM6}) at ${\bf t}=0$, with the diagonal elements
being the Gaudin Hamiltonians $H_i =-p_i (0)$ (see (\ref{CM2})).
Then
\beq\label{CM13}
T^G(x, {\bf t})=e^{\sum_{k\geq 1}t_k{\scriptsize{\mbox{tr}}\, h^k}}
\det \left ( x\II -X_0 +\sum_{k\geq 1}kt_k Y_0^{k-1}\right ).
\eeq
Plugging this into (\ref{BA1}), (\ref{BA2}) we find formulas for
the stationary BA functions:
\beq\label{BA1a}
\psi (x,z)=z^{-N}\det (z - h) e^{xz}\,
\frac{\det \left (x\! -\! X_0\! -\! (z \! -\!
Y_0)^{-1}\right )}{\det (x-X_0)},
\eeq
\beq\label{BA2a}
\psi^{*} (x,z)=z^N e^{-xz}\,
\frac{\det \left (x\! -\! X_0\! +\! (z \! -\!
Y_0)^{-1}\right )}{\det (z \! -\! h) \det (x-X_0)}.
\eeq

Let us show that these formulas are equivalent to the
stationary versions of (\ref{cc1}).
Using commutation relation (\ref{comm}), we have:
$$
\det \Bigl ( (x-X_0)(z-Y_0) - \II \Bigr )=
\det \Bigl ( (z-Y_0)(x-X_0) - {\sf 1}\otimes {\sf 1}^{\sf t}\Bigr )
$$
$$
=\,\, \det (x-X_0) \, \det (z-Y_0)\, \Bigl (
1 - {\sf 1}^{\sf t}(x-X_0)^{-1}(z-Y_0)^{-1}{\sf 1}\Bigr )
$$
and
$$
\det \Bigl ( (z-Y_0)(x-X_0) + \II \Bigr )=
\det \Bigl ( (x-X_0)(z-Y_0)+ {\sf 1}\otimes {\sf 1}^{\sf t}\Bigr )
$$
$$
=\,\, \det (x-X_0) \, \det (z-Y_0)\, \Bigl (
1 + {\sf 1}^{\sf t}(z-Y_0)^{-1}(x-X_0)^{-1}{\sf 1}\Bigr ),
$$
which shows that (\ref{cc1}) and (\ref{BA1a}), (\ref{BA2a}) are
indeed equivalent at ${\bf t}=0$.

\subsection{Spectrum of the
Gaudin Hamiltonians from the classical CM model}

It follows from the above arguments that eigenvalues of the
Gaudin Hamiltonians $H_i$, $i=1, \ldots , n$ (\ref{D4}),
can be found in the framework of the classical
CM system with $n$ particles as follows. Consider the matrix
\beq\label{S1}
Y_0= \left ( \begin{array}{ccccc}
H_1 & \displaystyle{\frac{1}{x_2-x_1}} &\displaystyle{\frac{1}{x_3-x_1}} &
\ldots & \displaystyle{\frac{1}{x_n-x_1}}
\\ &&&& \\
 \displaystyle{\frac{1}{x_1-x_2}} & H_2 & \displaystyle{\frac{1}{x_3-x_2}} &
 \ldots & \displaystyle{\frac{1}{x_n-x_2}}
 \\ &&&& \\ \vdots & \vdots & \vdots & \ddots & \vdots
 \\ &&&& \\
 \displaystyle{\frac{1}{x_1-x_n}} & \displaystyle{\frac{1}{x_2-x_n}}&
 \displaystyle{\frac{1}{x_3-x_n}} & \ldots & H_n
\end{array}\right ).
\eeq
The spectrum of $H_i$'s in the space ${\cal V}(\{ m_a\})$ is determined
by the conditions
\beq\label{S2}
\mbox{tr}\, Y_0^j = \sum_{a=1}^{N} m_a k_a^j \quad
\mbox{for all} \quad j\geq 1,
\eeq
i.e., given the initial coordinates $x_i$ and the action variables
${\cal H}_j =\mbox{tr}\, Y_0^j$ one has to find possible values
of the initial
momenta $p_i =-H_i$. 
Taking into account equations (\ref{char-pol}) and (\ref{explicit}),
we can represent the equations for $H_i$ in the form of the equality
\beq\label{S3}
\exp \Bigl (\sum\limits_{i<j}x_{ij}^{-2}
\p_{H_i}\p_{H_j}\Bigl )\prod_{l=1}^{n}(z-H_l)=
\prod_{a=1}^{N}(z-k_a)^{m_a}\,, \quad \quad x_{ij}\equiv x_i-x_j
\eeq
which has to be satisfied identically in $z$. This identity
is equivalent to $n$ algebraic equations for $n$ quantities
$H_1, \ldots , H_n$.

In other words,
the eigenstates of the Gaudin Hamiltonians correspond to 
the intersection points of two Lagrangian submanifolds:
one obtained by fixing the $x_i$'s and the other obtained by 
fixing the ${\cal H}_i$'s, with values of the latter being 
determined by eigenvalues of the twist matrix.
This purely classical prescription appears to be
equivalent to the Bethe ansatz solution and 
solves the spectral problem for the quantum Gaudin Hamiltonians.

{\bf Example.} Consider the vector ${\sf v}_a \in \CC^N$ with
components $({\sf v}_a)_b =\delta_{ab}$. Since
$\PP_{ij} ({\sf v}_a)^{\otimes n} = ({\sf v}_a)^{\otimes n}$, the vector
$({\sf v}_a)^{\otimes n}$ is an eigenstate for the Gaudin
Hamiltonians $H_i$ with the eigenvalues
$\displaystyle{k_a +\sum_{j\neq i}\frac{1}{x_i-x_j}}$.
It is also an eigenvector for the operators $M_b$ with
eigenvalues $m_b=n\delta_{ab}$. The matrix (\ref{S1}) in this case
is the $n\! \times \! n$
Jordan block with the only eigenvector ${\sf 1}$ with eigenvalue
$k_a$ and $\mbox{tr}\, Y_0^j =nk_a^j$.

\section{Concluding remarks}

We have shown that the most general 
generating function of commuting $T$-operators in the Gaudin model
(the master $T$-operator) satisfies bilinear equations of the classical 
KP hierarchy. This implies that each eigenvalue 
of the master $T$-operator is a classical 
tau-function. By construction, these tau-functions appear to be 
polynomials in the spectral parameter $x$ of the Gaudin model
which can be identified with the KP time $t_1$. The dynamics
of zeros of polynomial tau-functions leads to a close connection
with integrable many-body problems. 

This result immediately leads to the important conclusion
that the eigenstates of the Gaudin Hamiltonians are naturally
labeled by intersection points of two Lagrangian submanifolds
in the phase space of the classical Calogero-Moser system
of particles.
This is a degenerate case of a more general correspondence between
quantum spin chains (of the XXX and XXZ types) and the classical
Ruijsenaars-Schneider model outlined earlier in \cite{AKLTZ11,zabrodin2013master,*Zabrodin2012gx}.
Presumably, this ``quantum-classical'' 
correspondence extends to models with
elliptic $R$-matrices and to supersymmetric integrable models.

Recently, the link between quantum spin chains
and integrable many-body problems of classical mechanics has been
discussed \cite{NRS11,GK13} in the context of supersymmetric gauge theories.
Physical consequences of this ``quantum-classical correspondence'' 
are yet to be recognized and articulated while its mathematical roots 
lie deeply in quantum theory of Hitchin's 
integrable systems \cite{BD,*BD2,NRS11}.

In the main text the Planck constant in the Gaudin model
was set to $1$. Here we would like to remark that if one
introduces the Planck constant $\hbar$ in the Gaudin model, then
the master $T$-operator satisfies the $\hbar$-version of the
KP hierarchy \cite{TakTak} which is obtained by the transformation
$t_{k} \to \hbar^{-1} t_{k}$ for $ k \ge 1$ 
and $x \to \hbar^{-1} x$. The 
coupling constant of the
CM model (the coefficient $2$ in the numerator of 
the second term in the 
r.h.s. of \eqref{CM11}) becomes $2\hbar^2$.

Because of the space limitation, in this paper
we have not addressed the construction 
of Baxter's $Q$-operators for the Gaudin model. 
In fact the master $T$-operator unifies them in one
commuting family with the higher $T$-operators.
Namely,
taking residues of $T^G(x, {\bf t}+[z])$ at the poles
(at the points $k_i$ which are eigenvalues of the 
twist matrix)
one can introduce a family of $2^{N}$ $Q$-operators. 
The bilinear identities among them define 
the B\"{a}cklund 
transformations from which the Bethe equations follow. 
For the original spin chains, the B\"{a}cklund 
transformations were discussed in
 \cite{KLWZ97,Z971,*Z972,KSZ08,*Z2008,Tsuboi09} on 
the level of the eigenvalues 
of the transfer matrices, 
and in \cite{KLT10,AKLTZ11,zabrodin2013master,*Zabrodin2012gx} on the level of operators. 
We plan to extend this approach to the Gaudin model 
in a separate publication.

\section*{Acknowledgments}
\addcontentsline{toc}{section}{Acknowledgments}
We thank A.Gorsky, V.Kazakov, S.Khoroshkin, I.Krichever,
M.Olshanetsky, T.Ta\-ke\-be and A.Zo\-tov for discussions.
The work of A.A. is supported in part by RFBR grants 13-02-00457, and by ERC Starting Independent Researcher Grant StG No. 204757-TQFT. The work of S.L. is supported by the ERC Advanced grant No.290456. The work of Z.T. is supported by the Australian Research Council.
The work of A.Z. is supported in part by RFBR grant 14-02-00627, and by Ministry of
Science and Education of Russian Federation under contract 8207. The work of A.A. and A.Z. is supported in part by grant NSh-1500.2014.2 of president of Russian Federation for support of leading scientific schools.

\section*{Note added}
After the arXiv version of the current paper 
[arXiv:1306.1111] appeared, 
a paper [A. Gorsky, A. Zabrodin, A. Zotov, 
{\it Spectrum of Quantum Transfer Matrices via Classical 
Many-Body Systems}, arXiv:1310.6958] appeared on the 
arXiv. 
They clarified the relationship between inhomogeneous quantum spin chains and classical integrable many-body systems. 

\newpage

\section*{Appendix: Proofs and details}
\addcontentsline{toc}{section}{Appendix: Proofs and details}
\def\theequation{\thesubsection\arabic{equation}}
\def\theHequation{\theequation}
\setcounter{equation}{0}

In this appendix, we present some details 
omitted in the main text. 
Some of the facts related to the KP hierarchy 
are consequences of the general theory known in the
literature (see, e.g., \cite{DJKM83,JM83,Sato,TakTak,EH}). 
However, to make the text self-contained, we 
give here some details of the calculations in our notations. 

\def\thesubsection{A}
\subsection{KP hierarchy from Giambelli relation}
\label{appen-KP}
\setcounter{equation}{0}

Here we show that if the coefficients $c_{\lambda}$ of the 
expansion
$\displaystyle{
\tau ({\bf t})=\sum_{\lambda}c_{\lambda}s_{\lambda}({\bf t})}
$
satisfy the Giambelli-like relation
\beq\label{giamb1}
c_{\lambda}=(c_{\emptyset})^{1-d}
\det_{1\leq i,j\leq d(\lambda )}c_{(\lambda_i -i|
\lambda '_j -j)},
\eeq
where 
$c_{(l|k)}:=c_{(l+1, 1^k)}$ are the coefficients for the hooks, 
then $\tau ({\bf t})$ is a KP tau-function, 
i.e., it solves the bilinear identity for the KP
hierarchy. We assume $c_{\emptyset}\neq 0$.

In this section the Frobenius notation for the Young diagrams 
is convenient.
Given a Young diagram $\lambda =
(\lambda_1 , \ldots , \lambda_{\ell})$, let
$(\vec \alpha |\vec \beta )=(\alpha_1, \ldots , \alpha_{d}|
\beta_1 , \ldots , \beta_{d})$ be the Frobenius notation
for the diagram $\lambda$ \cite{Macdonald}. 
Here $\alpha_i =\lambda_i -i$,
$\beta_i =\lambda'_i -i$.  In other words,
$\alpha_i$ is the length of the part of the
$i$-th row to the right from the
main diagonal and $\beta_i$ is the length of the part of the $i$-th column
under the main diagonal (not counting the diagonal box).
Clearly, $\alpha_1 > \alpha_2 >\ldots > \alpha _d\geq 0$ and
$\beta_1 > \beta_2 >\ldots > \beta _d\geq 0$. 

Using this notation, we can represent (\ref{giamb1}) and the 
Giambelli identity for the Schur functions as $d \times  d$ determinants:
$\displaystyle{
c_{\lambda}=\det_{1\leq i,j\leq d(\lambda )}c_{(\alpha_{i} |
\beta_j )}}$,
$\,\,\,\,\displaystyle{s_{\lambda}({\bf t})=\det_{1\leq i,j\leq d(\lambda )}
s_{(\alpha_i |\beta_j )}({\bf t})}
$.
Plugging these into the expansion and separating the contribution
of the empty diagram, we get:
$$
\tau ({\bf t})=c_{\emptyset} \left (
1+\sum_{d\geq 1} c_{\emptyset}^{-d}
\!\!\! \!\!\!  \sum_{ 
{\alpha_1 >\alpha_2 >\ldots >\alpha_d \geq 0}\atop{\beta_1 >
\beta_2 >\ldots >\beta_d \geq 0}}
\det_{1\leq i,j\leq d}c_{(\alpha_i |\beta_j)}
\det_{1\leq i,j\leq d}s_{(\alpha_i |\beta_j)}({\bf t})
\right ).
$$
The sums over $\beta_j$ in each term 
can be handled with the help of the 
Cauchy-Binet formula:
$$
\tau ({\bf t})=c_{\emptyset} \left (
1+\sum_{d\geq 1} 
\sum_{ \alpha_1 >\alpha_2 >\ldots >\alpha_d \geq 0}
\det_{1\leq i,j\leq d}\Bigl ((c_{\emptyset})^{-1}\sum_{k\geq 0}
c_{(\alpha_i |k)}
s_{(\alpha_j |k)}({\bf t})\Bigr ) 
\right ).
$$
Next, it is not difficult to notice 
that the sums over $\alpha_i$ and $d$ represent the expansion 
of the determinant $\det_{ij}(\delta_{ij}+A_{ij})$ in terms of
diagonal minors of the matrix $A_{ij}$. Therefore, we have
\beq\label{taucs1}
\tau ({\bf t})=c_{\emptyset} \det_{i,j\geq 0}\Bigl (
\delta_{ij} +(c_{\emptyset})^{-1}\! \sum_{k\geq 0}c_{(i|k)}
s_{(j|k)}({\bf t})\Bigr ).
\eeq

To proceed, we need the identity \cite{Macdonald}
\beq\label{hhs}
\sum_{k\geq 0}h_{i-k}({\bf t})h_{k-j}(-{\bf t})=
\left \{ \begin{array}{l}
\delta_{ij} \quad \quad \quad \quad  \quad \quad  \,\,\,\,\,
\mbox{if $j\geq 0$}
\\ \\
(-1)^j s_{(i|-j-1)}({\bf t}) \quad \mbox{if $j<0$.}
\end{array}\right.
\eeq
It means, in particular, that the 
lower-triangular matrices (with $1$'s on the main diagonal)
$h_{i-j}({\bf t})$ and $h_{i-j}({-\bf t})$ are inverse to each other.
(Recall that 
$h_{k}({\bf t})=0$ at $k<0$.)
This is obvious from the fact that the product 
of their generating functions is $1$. Using this identity, we 
rewrite (\ref{taucs1}) as follows:
$$
\begin{array}{rl}
\tau ({\bf t})= & \displaystyle{c_{\emptyset} \det_{i,j\geq 0}\Bigl (\,
\underbrace{\sum_{l\geq 0}h_{i-l}({\bf t}) 
h_{l-j}(-{\bf t})}_{\delta_{ij}} \, +\,
(c_{\emptyset})^{-1}\!
\sum_{l\geq 0}\sum_{k\geq 0}(-1)^{k+1}h_{i-l}({\bf t})
h_{l+k+1}(-{\bf t})c_{(j|k)}\Bigr )}
\\ &\\
=&\displaystyle{c_{\emptyset} \, \underbrace{\det_{i,j\geq 0}
\bigl ( h_{i-j}({\bf t})\bigr )}_{1}\,\, \det_{i,j\geq 0}
\Bigl (h_{i-j}(-{\bf t}) + 
(c_{\emptyset})^{-1}\! \sum_{k<0}(-1)^k h_{i-k}(-{\bf t})
c_{(j|-k-1)}\Bigr ).}
\end{array}
$$
Finally, we conclude that equation (\ref{taucs1}) can be represented
in the form
\beq\label{taucs2}
\tau ({\bf t})=c_{\emptyset}\det_{i,j\geq 0}\Bigl (\,
\sum_{k\in \z }S_{kj} h_{i-k}({-\bf t})\Bigr ),
\eeq
where the $\ZZ \! \times \! \ZZ_{\geq 0}$ matrix $S$ is given by
$$
S_{kj}=\left \{ \begin{array}{l}
(-1)^k c_{(j|-k-1)}/ c_{\emptyset}, \quad k<0
\\ \\ \delta_{kj}\,,   \quad \quad \quad  \quad \quad \quad\quad k \geq 0.
\end{array}
\right.
$$
One can see that this formula is the 
general solution to the KP hierarchy given in the 
form of determinant of a semi-infinite matrix \cite{DJKM83,Sato}. 
The unusual argument of $h_{i-l}$ 
($-{\bf t}$ instead of ${\bf t}$)
does not spoil anything 
because the KP hierarchy is invariant under changing signs of all 
times (see the remark after equation (\ref{hir1})).
Therefore, (\ref{taucs1}) provides the general
solution to the KP hierarchy and thus solves the bilinear 
identity for any $c_{(j|k)}$.

\def\thesubsection{B}
\subsection[Details on functional relations for 
higher Gaudin \texorpdfstring{$T$}{T}-operators]{Details on functional relations for 
higher Gaudin {\boldmath $T$}-operators}
\label{appen-higher}
\setcounter{equation}{0}

\subsubsection*{From the determinant \eqref{masterdet1} to CBR formula \eqref{CBR}}
The equivalence of \eqref{masterdet1} and \eqref{CBR}
can be shown in the following way\footnote{See also discussions on the generation function of the
transfer matrices in \cite{KLT10,Tsuboi09}.}.
First, let us rewrite \eqref{masterdet1} as follows
\begin{multline}
\mathrm{det}_{1 \le k,j \le m}
\left( z_{i_{k}}^{j-k}
\right)
\exp
\left(\sum_{k=1}^{m}
\xi(\tilde{\partial},z_{i_{k}}^{-1})
\right)
T^{G }(x, {\bf t}) =
\\
=
\frac{
\mathrm{det}_{1 \le k,j \le m}
\left(
\sum_{l=0}^{j-1}(-1)^l
\Bigl (\begin{array}{c}j\! -\! 1\\ l\end{array} \Bigr )
\,
z_{i_{k}}^{j-k-l}
\p_x^{l}
\exp
\left(
\xi(\tilde{\partial},z_{i_{k}}^{-1})
\right)
T^{G }(x, {\bf t}) \right)
}
{
\bigl (T^{G}(x, {\bf t})\bigr )^{m-1}
}.
\label{masterdet1b}
\end{multline}
Using \eqref{Schur-p} and expanding the determinant
in both side of \eqref{masterdet1b}, we obtain
\begin{multline}
\sum_{\sigma \in S_{m}}
\mathrm{sgn}(\sigma)
\sum_{a_{1}=0}^{\infty}
\dots
\sum_{a_{m}=0}^{\infty}
\prod_{k=1}^{m}
h_{a_{k}}(\tilde{\partial})T^{G}(x, {\bf t})
z_{i_{k}}^{\sigma(k)-k-a_{k}} =
\\
=
\sum_{\sigma \in S_{m}}
\mathrm{sgn}(\sigma)
\sum_{b_{1}=0}^{\sigma(1)-1}
\dots
\sum_{b_{m}=0}^{\sigma(m)-1}
\sum_{a_{1}=0}^{\infty}
\dots
\sum_{a_{m}=0}^{\infty}
\prod_{k=1}^{m}
\Bigl\{
(-1)^{b_{k}}
\Bigl (\begin{array}{c}\sigma(k)\! -\! 1\\ b_{k} \end{array}
\Bigr )
\\
\times
\p_x^{b_{k}}
h_{a_{k}}(\tilde{\partial})T^{G}(x, {\bf t})
z_{i_{k}}^{\sigma(k)-k-b_{k}-a_{k}}
\Bigr\}
\bigl (T^{G}(x, {\bf t})\bigr )^{1-m} ,
 \label{masterexpan}
\end{multline}
where $S_{m}$ is the permutation group on $\{1,2,\dots, m \}$
and $\mathrm{sgn}(\sigma)$ is the signature of $\sigma \in S_{m}$.
Then, comparing the coefficient of
$z_{i_{1}}^{-\lambda_{1}} z_{i_{2}}^{-\lambda_{2}}\cdots z_{i_{m}}^{-\lambda_{m}}$
in the both side of \eqref{masterexpan}, we get
\begin{align}
s_{\lambda}(\tilde{\partial})
T^{G }(x, {\bf t})
=
\frac{
\mathrm{det}_{1 \le k,j \le m}
\left(
\sum_{l=0}^{j-1}(-1)^l
\Bigl (\begin{array}{c}j\! -\! 1\\ l\end{array} \Bigr )
\,
\p_x^{l}
h_{\lambda_{k}-k+j-l}(\tilde{\partial})
T^{G }(x, {\bf t}) \right)
}
{
\bigl (T^{G}(x, {\bf t})\bigr )^{m-1}
} ,
\end{align}
where we used \eqref{JT-det}.
This reduces to \eqref{CBR} at ${\bf t}=0$ 
after a renormalization\footnote{
One has to use the relation
\begin{multline*}
\sum_{k=0}^{j-1}(-1)^k
\Bigl (\begin{array}{c}j\! -\! 1\\ k\end{array} \Bigr )\,
\p_x^{k} T^G_{(\lambda_i-i+j-k)}(x)
=
\phi(x)
\sum_{k=0}^{j-1}(-1)^k
\Bigl (\begin{array}{c}j\! -\! 1\\ k\end{array} \Bigr )\,
\p_x^{k} {\sf T}^G_{(\lambda_i-i+j-k)}(x)
\\
+
\sum_{l=1}^{j-1}
(-1)^{j-l}
\Bigl (\begin{array}{c}j\! -\! 1\\ l-1 \end{array} \Bigr )\,
(\p_x^{j-l} \phi(x))
\sum_{k=0}^{l-1}(-1)^k
\Bigl (\begin{array}{c}l\! -\! 1\\ k\end{array} \Bigr )\,
\p_x^{k} {\sf T}^G_{(\lambda_i-i+l-k)}(x),
\end{multline*}
where $\phi(x)=\prod_{k=1}^{n}(x-x_{k})$.
}
and the identification $m=\ell(\lambda)$.
\subsubsection*{The dual form of the CBR \eqref{CBR-dual}}
Equation \eqref{CBR-dual} can be proved, for example,
by repeating a similar argument as above
with the formulas \eqref{JT-det2} and \eqref{Schur-ele}
for the equation obtained by the shift
${\bf t} \to {\bf t}-[z_{1}^{-1}]-[z_{2}^{-1}]$ in \eqref{hir3}.
\subsubsection*{From CBR \eqref{CBR} to the quantum Giambelli 
formula \eqref{det1a}}
Let us introduce the notation
\begin{align}
t_{i,j}(x)= \sum_{k=0}^{j-1}(-1)^k
\Bigl (\begin{array}{c}j\! -\! 1\\ k\end{array} \Bigr )\,
\p_x^{k} {\sf T}^G_{(-i+j-k)}(x).
\end{align}
Then \eqref{CBR} can be rewritten as
\begin{align}
{\sf T}^G_{\lambda}(x)=\det_{1\leq i,j\leq \ell (\lambda )}
\left ( t_{i-\lambda_{i},j}\right).
 \label{CBR2}
\end{align}
Using the same argument as in the Appendix C 
of \cite{AKLTZ11} for \eqref{CBR2}, one can prove \eqref{det1a}. 
\subsubsection*{The determinant \eqref{masterdet1} solves the
 Fay identities \eqref{hir20}-\eqref{hir3}}

We will show that the determinant \eqref{masterdet1} satisfies 
the following identities: 
\begin{multline}
(z_{i_{m-3}} - z_{i_{m-2}})(z_{i_{m-1}} - z_{i_{m}})
T^{G, I \sqcup \{ i_{m-3}, i_{m-2} \}   }(x, {\bf t})
T^{G, I \sqcup \{ i_{m-1}, i_{m} \}  }(x, {\bf t}) 
\\
-
(z_{i_{m-3}} - z_{i_{m-1}})(z_{i_{m-2}} - z_{i_{m}})
T^{G, I \sqcup \{ i_{m-3}, i_{m-1} \}   }(x, {\bf t})
T^{G, I \sqcup \{ i_{m-2}, i_{m} \}  }(x, {\bf t})
\\
+
(z_{i_{m-3}} - z_{i_{m}})(z_{i_{m-2}} - z_{i_{m-1}})
T^{G, I \sqcup \{ i_{m-3}, i_{m} \}   }(x, {\bf t})
T^{G, I \sqcup \{ i_{m-2}, i_{m-1} \}  }(x, {\bf t})
=0, 
\label{Fay-gen0}
\end{multline}
where $I=\{i_{1},i_{2},\dots, i_{m-4} \}$, 
\begin{multline}
(z_{i_{m-1}} - z_{i_{m}}) 
T^{G, I \sqcup \{ i_{m-2} \}   }(x, {\bf t})
T^{G, I \sqcup \{ i_{m-1}, i_{m} \}  }(x, {\bf t}) 
\\
+
(z_{i_{m}} - z_{i_{m-2}}) 
T^{G, I \sqcup \{ i_{m-1} \}   }(x, {\bf t})
T^{G, I \sqcup \{ i_{m-2}, i_{m} \}  }(x, {\bf t}) 
\\
+
(z_{i_{m-2}} - z_{i_{m-1}}) 
T^{G, I \sqcup \{ i_{m} \}   }(x, {\bf t})
T^{G, I \sqcup \{ i_{m-2}, i_{m-1} \}  }(x, {\bf t}) 
=0, 
\label{Fay-gen}
\end{multline}
where $I=\{i_{1},i_{2},\dots, i_{m-3} \}$, and 
\begin{multline}
(z_{i_{m-1}}^{-1} - z_{i_{m}}^{-1}) 
T^{G, I   }(x, {\bf t})
T^{G, I \sqcup \{ i_{m-1}, i_{m} \}  }(x, {\bf t}) 
=
\\
=
z_{i_{m-1}}^{-1} 
T^{G, I \sqcup \{ i_{m-1} \}  }(x, {\bf t}) 
(1 - z_{i_{m}}^{-1} \p_{x} )
T^{G, I \sqcup \{ i_{m} \}  }(x, {\bf t}) 
\\ 
-
z_{i_{m}}^{-1} 
T^{G, I \sqcup \{ i_{m} \}  }(x, {\bf t}) 
(1 - z_{i_{m-1}}^{-1} \p_{x} )
T^{G, I \sqcup \{ i_{m-1} \}  }(x, {\bf t}) , 
\label{diff-Fay-gen}
\end{multline}
where $I=\{i_{1},i_{2},\dots, i_{m-2} \}$. 
Note that 
\eqref{Fay-gen0} for $(m,i_{1},i_{2},i_{3},i_{4})=(4,0,1,2,3)$ 
corresponds to the Fay identity \eqref{hir20},
\eqref{Fay-gen} for $(m,i_{1},i_{2},i_{3})=(3,1,2,3)$ 
corresponds to  \eqref{hir2}, and 
\eqref{diff-Fay-gen} for  $(m,i_{1},i_{2})=(2,1,2)$ 
 corresponds to the differential Fay identity \eqref{hir3}. 

For any $m \times n$ matrix 
${\mathcal M} $, we will denote a minor determinant whose 
$j_{1},j_{2},\dots,j_{\alpha} $-th rows and $k_{1},k_{2},\dots,k_{\beta}$-th columns removed from 
it as 
${\mathcal M} \begin{bmatrix}
j_{1}, & j_{2}, & \dots, j_{\alpha} \\
k_{1}, & k_{2}, & \dots, k_{\beta}
\end{bmatrix}$,
where $m-\alpha =n -\beta$, $j_{1} <j_{2}\cdots < j_{\alpha} $ and 
$k_{1} < k_{2} < \dots < k_{\beta}$. 
We will use the following identities for determinants:
\begin{align}
&
{\mathcal M} \begin{bmatrix}
j_{1}, &  j_{2}\\
{} 
\end{bmatrix}
{\mathcal M} \begin{bmatrix}
j_{3}, &  j_{4}\\
{} 
\end{bmatrix}
-
{\mathcal M} \begin{bmatrix}
j_{1}, &  j_{3}\\
{} 
\end{bmatrix}
{\mathcal M} \begin{bmatrix}
j_{2}, &  j_{4}\\
{} 
\end{bmatrix}
+
{\mathcal M} \begin{bmatrix}
j_{1}, &  j_{4}\\
{} 
\end{bmatrix}
{\mathcal M} \begin{bmatrix}
j_{2}, &  j_{3}\\
{} 
\end{bmatrix}
=0,  \label{plucker1}
\\
&
{\mathcal M} \begin{bmatrix}
j_{1} \\
{} 
\end{bmatrix}
{\mathcal M} \begin{bmatrix}
j_{2}, &  j_{3} \\
k_{1} 
\end{bmatrix}
-
{\mathcal M} \begin{bmatrix}
j_{2} \\
{} 
\end{bmatrix}
{\mathcal M} \begin{bmatrix}
j_{1},&  j_{3} \\
k_{1} 
\end{bmatrix}
+
{\mathcal M} \begin{bmatrix}
j_{3} \\
{} 
\end{bmatrix}
{\mathcal M} \begin{bmatrix}
j_{1}, &  j_{2} \\
k_{1} 
\end{bmatrix}
=0, 
\label{plucker2}
\\[8pt]
& 
{\mathcal M} \begin{bmatrix}
\quad \\
\quad 
\end{bmatrix}
{\mathcal M} \begin{bmatrix}
j_{1}, &  j_{2} \\
k_{1}, &  k_{2}
\end{bmatrix}
-
{\mathcal M} \begin{bmatrix}
j_{1}  \\
k_{1} 
\end{bmatrix}
{\mathcal M} \begin{bmatrix}
 j_{2} \\
 k_{2}
\end{bmatrix}
+
{\mathcal M} \begin{bmatrix}
j_{1}  \\
k_{2} 
\end{bmatrix}
{\mathcal M} \begin{bmatrix}
 j_{2} \\
 k_{1}
\end{bmatrix}
=0. 
\label{jacobi}
\end{align}
\eqref{plucker1}-\eqref{plucker2} are specialization of 
the so-called Pl\"ucker identity 
and 
\eqref{jacobi} is the Jacobi identity.

The Jacobi identity \eqref{jacobi} for the matrix 
$$
{\mathcal M}=\left(z_{i_{k}}^{j-m}(1-z_{i_{k}}^{-1} 
\p_{x})^{j-1} T^{G, \{ i_{k} \}  }(x, {\bf t}) \right)_{1 \le k,j \le m}$$
for $j_{1}=k_{1}=m-1$ and  $j_{2}=k_{2}=m$ 
corresponds to \eqref{diff-Fay-gen}. 
In fact we obtain the relations 
\begin{align}
&
{\mathcal M} 
\begin{bmatrix}
 \quad \\
 \quad 
\end{bmatrix}
=\Delta(I \sqcup \{ i_{m-1}, i_{m}  \})
\left( T^{G }(x, {\bf t}) \right)^{m-1}  
T^{G, I \sqcup \{i_{m-1}, i_{m} \}  }(x, {\bf t}) ,
\label{mino1}
\\
&
{\mathcal M} 
\begin{bmatrix}
 m-1, & m \\
 m-1, & m 
\end{bmatrix}
=\Delta(I )
\left(  \prod_{k \in I} z_{k}^{-2} \right)
\left( T^{G }(x, {\bf t}) \right)^{m-3}  
T^{G, I  }(x, {\bf t}) ,
\\
&
{\mathcal M} \begin{bmatrix}
 m-1 \\
 m-1
\end{bmatrix}
=\Delta(I \sqcup \{i_{m} \})
\left(  \prod_{k \in I} z_{k}^{-1} \right)
\left( T^{G }(x, {\bf t}) \right)^{m-2}  \times 
\nonumber 
\\
&
\qquad 
\times 
\left(
\left(1 -z_{i_{m}}^{-1} \p_{x} \right) 
+
z_{i_{m}}^{-1}
\left(
\sum_{k \in I}z_{k} - 
(m-2)\frac{\p_{x} T^{G }(x, {\bf t})}{ T^{G}(x, {\bf t})}
\right)
\right)
T^{G, I \sqcup \{ i_{m} \}  }(x, {\bf t}) , 
\label{mino3}
\\
&
{\mathcal M} 
\begin{bmatrix}
m \\
  m 
\end{bmatrix}
=\Delta(I \sqcup \{i_{m-1} \} )
\left(  \prod_{k \in I \sqcup \{i_{m-1} \} } z_{k}^{-1} \right)
\left( T^{G }(x, {\bf t}) \right)^{m-2}  
T^{G, I \sqcup \{i_{m-1} \}  }(x, {\bf t}) ,
\\
&
{\mathcal M} 
\begin{bmatrix}
m-1 \\
  m 
\end{bmatrix}
=\Delta(I \sqcup \{i_{m} \} )
\left(  \prod_{k \in I \sqcup \{i_{m} \} } z_{k}^{-1} \right)
\left( T^{G }(x, {\bf t}) \right)^{m-2}  
T^{G, I \sqcup \{i_{m} \}  }(x, {\bf t}) ,
\\
&
{\mathcal M} \begin{bmatrix}
 m \\
 m-1
\end{bmatrix}
=\Delta(I \sqcup \{i_{m-1} \})
\left(  \prod_{k \in I} z_{k}^{-1} \right)
\left( T^{G }(x, {\bf t}) \right)^{m-2}  \times 
\nonumber 
\\
&
\qquad 
\times 
\left(
\left(1 -z_{i_{m-1}}^{-1} \p_{x} \right) 
+
z_{i_{m-1}}^{-1}
\left(
\sum_{k \in I}z_{k} - 
(m-2)\frac{\p_{x} T^{G }(x, {\bf t})}{ T^{G}(x, {\bf t})}
\right)
\right)
T^{G, I \sqcup \{ i_{m-1} \}  }(x, {\bf t}) ,
\label{mino6}
\end{align}
where $I=\{i_{1},i_{2},\dots, i_{m-2} \}$ and 
 $\Delta(\{i_{1},i_{2},\dots, i_{m}\})=
 \prod_{1 \le a <b \le m}(z_{i_{a}}^{-1} - z_{i_{b}}^{-1})$. 
Substituting the above relations to 
\eqref{jacobi}, we obtain  \eqref{diff-Fay-gen}.
Among \eqref{mino1}-\eqref{mino6}, 
equations \eqref{mino3} and \eqref{mino6} are rather non-trivial. 
Equations \eqref{mino3} and \eqref{mino6} can be proved using 
the Leibnitz rule for the derivative with respect to $x$, 
taking linear combinations of columns in determinants 
and using the following identity valid 
for any matrix $(A_{jk})_{1 \le k,j \le m}$ 
and parameters $z_{k}$: 
\begin{align}
\sum_{p=1}^{m}\det_{1 \le k,j \le m}
\left( z_{k}^{\delta_{jp}} A_{kj}\right)
= \left( \sum_{p=1}^{m} z_{p} \right) \det_{1 \le k,j \le m}
\left( A_{kj}\right).
\end{align}

In a similar way, one can show that 
the Pl\"{u}cker identity \eqref{plucker1} for the matrix 
${\mathcal M}=\left(z_{i_{k}}^{j-m+2}(1-z_{i_{k}}^{-1} \p_{x})^{j-1} 
T^{G, \{ i_{k} \}  }(x, {\bf t}) \right)_{1 \le k \le m,
1 \le j \le m-2}$ 
for $(j_{1},j_{2},j_{3},j_{4})=(m-3,m-2,m-1,m)$ 
corresponds to \eqref{Fay-gen0}, 
and 
the Pl\"{u}cker identity \eqref{plucker2} for the matrix 
$$
{\mathcal M}=\left(z_{i_{k}}^{j-m+1}(1-z_{i_{k}}^{-1} 
\p_{x})^{j-1} T^{G, \{ i_{k} \}  }(x, {\bf t}) \right)_{1 \le k \le m,
1 \le j \le m-1}$$
for $(j_{1},j_{2},j_{3},k_{1})=(m-2,m-1,m,m-1)$ 
corresponds to \eqref{Fay-gen}.

\def\thesubsection{C}
\subsection{Two proofs of Lemma \ref{basicid}}
\label{appen-lemma}
\setcounter{equation}{0}

\subsubsection{Proof by means of the limit from spin chain}
\label{sec:proof-means-limit}

The proof is based on the limit from the spin chain,
where the relation
\beq\label{proof1}
\left | \begin{array}{ll}
{\sf D}^{\otimes k}\chi_{\alpha_1, \beta_1}(g) &
{\sf D}^{\otimes k}\chi_{\alpha_1, \beta_2}(g)
\\
{\sf D}^{\otimes k}\chi_{\alpha_2, \beta_1}(g) &
{\sf D}^{\otimes k}\chi_{\alpha_2, \beta_2}(g)
\end{array}
\right |=0
\eeq
follows from the Hirota equations for the master $T$-operator
(or from the quantum Giambelli formula
for the transfer matrices) proved in \cite{AKLTZ11}.

Applying the general formula (\ref{char-exp1}) for the hook
characters $\chi_{\alpha , \beta}(g)=
\chi_{(\alpha +1, 1^{\beta})}(g)$,
we get
\begin{multline}
\chi_{\alpha , \beta}(g-\II )=
\sum_{\alpha '=0}^{\alpha}\sum_{\beta '=0}^{\beta}
(-1)^{\alpha-\alpha^{\prime}+\beta-\beta^{\prime}} \left (\begin{array}{c}
N+\alpha \\ N+\alpha ' \end{array}\right )
\left (\begin{array}{c}
N-\beta '-1 \\ N-\beta -1 \end{array}\right )
\chi_{\alpha ', \beta '}(g)
\\
+\sum_{j=0}^{\beta}(-1)^{\alpha+1-j}
\begin{pmatrix}
N+\alpha+\beta-j \\
N-1
\end{pmatrix}
\begin{pmatrix}
N \\
j
\end{pmatrix}.
\label{proof2}
\end{multline}
Alternatively, this expansion can be derived by comparison
of the generating functions.
The generating function of the hook characters is
$$
E^{(g)}(z, \zeta )=\sum_{\alpha , \beta \geq 0}
\chi_{\alpha , \beta}(g) z^{\alpha}(-\zeta )^{\beta} =
\frac{1}{z-\zeta}\left (\frac{w(z)}{w(\zeta ) }-1\right ).
$$
It is easy to obtain the following relation between
generating functions for the characters
$\chi_{\alpha , \beta}(g)$ and
$\chi_{\alpha , \beta}(g-\II )$:
$$
E^{(g-\III )}(z, \zeta )=(1+z)^{-N-1}(1+\zeta )^{N-1}
E^{(g)}(\tilde z, \tilde \zeta ) + E^{(-\III )}(z, \zeta ),
$$
where
$
\displaystyle{\tilde z = \frac{z}{1+z}}$,
$\displaystyle{\tilde \zeta = \frac{\zeta}{1+\zeta}}
$. Therefore,
$$
\sum_{\alpha , \beta \geq 0}
\chi_{\alpha , \beta}(g\! -\! \II ) z^{\alpha}(-\zeta )^{\beta} =
\!\! \!\! \sum_{\alpha ' , \beta ' \geq 0}\!\!
\chi_{\alpha ', \beta '}(g) z^{\alpha '}(1+z)^{-\alpha '-N-1}
(-\zeta )^{\beta '}(1+\zeta )^{-\beta ' +N-1} \, +
E^{(-\III )}(z, \zeta ).
$$
This implies the relation for characters (\ref{proof2}).

The important thing is that the
coefficient in front of $\chi_{\alpha ', \beta '}(g)$ in (\ref{proof2})
factorizes into a product of $\alpha '$ and $\beta '$-dependent
factors.
Using Lemma \ref{lemma-limit}
we can write
\begin{align}
{\sf d}^{\otimes k}\chi_{\alpha , \beta}(h)=
\lim_{\eta \to 0}\left [
\eta^{k-1-\alpha -\beta}{\sf D}^{\otimes k}\chi_{\alpha, \beta}(g-\II )
\right ], \quad g=e^{\eta h}.
\label{hook-limit}
\end{align}
Substituting (\ref{proof2}) into (\ref{hook-limit})
and taking note on the relation \eqref{proof1}, we see that
\beq\label{proof3}
\left | \begin{array}{ll}
{\sf d}^{\otimes k}\chi_{\alpha_1, \beta_1}(h) &
{\sf d}^{\otimes k}\chi_{\alpha_1, \beta_2}(h)
\\
{\sf d}^{\otimes k}\chi_{\alpha_2, \beta_1}(h) &
{\sf d}^{\otimes k}\chi_{\alpha_2, \beta_2}(h)
\end{array}
\right |=0 \quad \quad \mbox{for any $\alpha_i, \beta_j \in \ZZ_{\ge 0}$}
\quad  \mbox{and $k \in \ZZ_{\ge 1}$},
\eeq
which is (\ref{EEEE}).

\subsubsection{Direct proof}
\label{sec:alternative-proof}

We will now provide an alternative proof of Lemma
\ref{basicid} which  does not rely on the results proven in  \cite{Kazakov2007na,KLT10,AKLTZ11},
although it uses some ideas similar in spirit to the ones used in these
papers. The same argument will also prove the commutation of the
operators $Q(z, \zeta)$. More precisely, we will show that 
\begin{gather}
  \label{eq:23}
  \mathcal{S}_z\left( Q(z_1, \zeta_1)~%
    Q(z_2,
    \zeta_2)\right)=\mathcal{S}_{z,\zeta}\left( Q(z_1, \zeta_1)~%
    Q(z_2,
    \zeta_2)\right)=0\,,\\\textrm{where }
  \begin{cases}
    \mathcal{S}_z\left(F(z_1,z_2,\zeta_1,\zeta_2)\right)\equiv
    F(z_1,z_2,\zeta_1,\zeta_2)-F(z_2,z_1,\zeta_1,\zeta_2)\\
 \mathcal{S}_{z,\zeta}\left(F(z_1,z_2,\zeta_1,\zeta_2)\right)\equiv
    F(z_1,z_2,\zeta_1,\zeta_2)-F(z_2,z_1,\zeta_2,\zeta_1) \,.
  \end{cases}
\end{gather}
The $\mathcal{S}_z$ case is the assertion of Lemma \ref{basicid}. The
$\mathcal{S}_{z,\zeta}$ case is the commutativity 
of these operators that can also be
proven from the Yang-Baxter equation. Our argument is universal and works for both cases, thus, later on we denote both $\mathcal{S}_z$ and $\mathcal{S}_{z,\zeta}$ by the same symbol $\mathcal{S}$.

In order to prove \eqref{eq:23}, one can 
first notice that 
\begin{gather}
\label{eq:24}    {\sf d} \, w(z) = \frac {z}{1-h z} w(z),
 ~\hspace{1cm}~ {\sf d}^{\otimes 2} \, w(z)
    = \left (\II+\PP_{1,2}\right) ~%
    \left(\frac {z}{1-h z}\otimes \frac
      {z}{1-h z} \right) w(z),
\\
\label{eq:25} \textrm{and generally } ~\hspace{1cm}~ {\sf d}^{\otimes n} \, w(z)
    = \sum_{\sigma\in S_n}\PP_\sigma~%
 \left(\frac {z}{1-h
        z}\right)^{\otimes n} w(z)\,,
\end{gather}
where $S_{n}$ is the permutation group over the set 
$\{1,2,\dots, n \}$; and 
$\PP_{\sigma}$ denotes the permutation operator associated to
the permutation $\sigma$, i.e. the operator such that for any vectors $v_1,
 v_2, \ldots v_n \in \CC^N$, we have 
$\PP_{\sigma}\,\, v_1\otimes v_2\otimes\ldots\otimes v_n = v_{\sigma(1)}\otimes v_{\sigma(2)}\otimes\ldots\otimes v_{\sigma(n)}$. For instance, for the transposition $\tau_{(i,j)}$
(which permutes indices $i$ and $j$,  $\tau_{(i,j)}(i)=j$, $\tau_{(i,j)}(j)=i$, and does not act on all other indices, $\tau_{(i,j)}(k)=k$ if $k\neq i,j$), we have
$\PP_{\tau_{(i,j)}}=\PP_{i,j}$. 

The expression \eqref{eq:25} is easily proven by recurrence over the 
number of spins, by noticing that in
 ${\sf d_n} \, {\sf d}^{\otimes (n-1)} \, w(z) $, 
the derivative ${\sf d_n}$ can (due to the Leibnitz rule for
derivatives) either 
act on $w(z)$ to give $( {\sf
  d}^{\otimes (n-1)} \, w(z))\otimes \frac z {1-h z}$ which corresponds,
in equation \eqref{eq:25}, to the permutations such that
$\sigma(n)=n$,
or it can act on any factor\footnote{We remind that
$h^{(i)}$ denotes the operator $\II^{\otimes i-1}\otimes h\otimes
\II^{\otimes n-i}$, so that $\left(\frac {z}{1-h
        z}\right)^{\otimes n}=
\prod_{i=1}^n \frac  z {1-h^{(i)} z}
$.
} $\frac  z {1-h^{(i)} z}$ of $\left(\frac {z}{1-h
        z}\right)^{\otimes (n-1)}$ to produce
    $\sum_{\sigma\in S_{n-1}}\PP_{i,n}\PP_{\sigma} \left(\frac {z}{1-h
        z}\right)^{\otimes n}$, which produces, in equation
    \eqref{eq:25}, all the terms where $\sigma(n) \ne n$.

Moreover, for arbitrary numbers $\alpha_1,\alpha_2,\ldots \alpha_m$
and $z_1,z_2,\ldots z_m$ the Leibnitz rule for derivatives of products gives
\begin{gather}
  \begin{aligned}
\label{eq:26}
    {\sf d}\, \prod_{k=1}^m w(z_k)  ^{\alpha_k}  =& \sum_{k=1}^{m} \alpha_k\frac {z_k}{1-h
      z_k} \prod_{k=1}^m w(z_k) ^{\alpha_k} ,\\
    {\sf d}^{\otimes 2} \, \prod_{k=1}^m w(z_k)  ^{\alpha_k}  =&\left(
    \left(\sum_{k=1}^{m} \alpha_k \frac {z_k}{1-h z_k}\right)^{\otimes 2} + \PP_{1,2}
    \sum_{k=1}^{m}\alpha_k \left(\frac {z_k}{1-h z_k}\right)^{\otimes 2}\right) \,\,\,\prod_{k=1}^m
    w(z_k)^{\alpha_k}.
  \end{aligned}
\end{gather}
Noticing that for any operators $A$, $B$, the tensor product $A\otimes B$ can be written as $(A\otimes \II)~%
 (\II\otimes B)$, we can also write this last equality as
\begin{equation*}
 {\sf d}^{\otimes 2} \, \prod_{k=1}^m w(z_k)  ^{\alpha_k} =  
\left(
     \left(
      \prod_{i=1}^2 \sum_{k=1}^{m}  \alpha_k \frac {z_k}{1-h^{(i)} z_k} \right)
    +
 \PP_{1,2}
    \sum_{k=1}^{m}\alpha_k \prod_{i=1}^2%
      \frac {z_k}{1-h^{(i)} z_k}%
\right)
 \,\,\,\prod_{k=1}^m
    w(z_k)^{\alpha_k}.
  \end{equation*}
  In what follows we will often use this notation. For instance, it is convenient for the following generalization of (\ref{eq:26}):
\beq
\label{eq:27}
\quad {\sf d}^{\otimes n} \, \prod_{k=1}^m w(z_k)  ^{\alpha_k} 
= \left(\sum_{\sigma\in S_n} \PP_\sigma%
  \prod_{c\in
  \mathcal{C}(\sigma)}
\sum_{k=1}^m \alpha_k \prod_{i\in c} \frac {z_k} {1-h^{(i)} z_k} \right) \prod_{k=1}^m 
w(z_k) ^{\alpha_k},%
\eeq
where $\mathcal{C}(\sigma)$ denotes the set of the
cycles\footnote{Rigorously, what we call cycle should actually be
  called orbits, because we view them as non-ordered set. For instance
  if $n=3$, there are two cyclic permutations (sometimes denoted as
  (1,2,3) and (1,3,2) in the literature); in the present notation we
  say that both of them have one single cycle $\{1,2,3\}$, which does
  not distinguish these two permutations.} of the
permutation $\sigma$: for instance if $n=2$, then the identity
permutation has two cycles $\mathcal{C}(1)=\{\{1\},\{2\}\}$, whereas the
cyclic permutation has only one cycle
$\mathcal{C}(\tau)=\{\{1,2\}\}$.
The expression \eqref{eq:27} can be proven by recurrence by the same
argument as \eqref{eq:25}.

Thus, \eqref{eq:27} allows one to express $Q(z, \zeta)$ as 
\begin{equation}
\label{eq:3}
Q(z, \zeta)=\frac { w(z)/w(\zeta)} {z-\zeta}\sum_\sigma \PP_\sigma%
\prod_{c\in \mathcal{C}(\sigma)} \left[\left(
\prod_{i\in c} %
  \frac z {1-h ^{(i)} z}%
\right)- \left(
\prod_{i\in c} %
  \frac \zeta {1-h ^{(i)}
    \zeta}%
\right)\right]\,.
\end{equation}

As an example of manipulations with this operator, one can check that 
for arbitrary $k,l \in\{1,2,\ldots n\}$ we have
$Q(z,\zeta)= \PP_{k,l} Q(z,\zeta)\PP_{k,l}$: for each $\sigma$, we have $\PP_{k,l}
\PP_{\sigma} \PP_{k,l}=\PP_{\tau_{(k,l)}\circ\sigma\circ\tau_{(k,l)}}$, where $\tau_{(k,l)}$
is defined above as the transposition which exchanges $k$ and $l$, and
where $\circ$ denotes the composition of permutations. Moreover, 
$\sigma\mapsto {\tau_{(k,l)}\circ\sigma\circ\tau_{(k,l)}}$ is a bijection of
$S_n$,  and the set $\{i_1,i_2,\cdots,i_m\}$ is
 a cycle of $\tau_{(k,l)}\circ\sigma\circ\tau_{(k,l)}$ if and only if
 $\{\tau_{(k,l)}(i_1),\tau_{(k,l)}(i_2),\cdots,\tau_{(k,l)}(i_k)\}$ is a cycle of
 $\sigma$. Hence the relation $Q(z,\zeta)= \PP_{k,l} Q(z,\zeta)\PP_{k,l}$ is
 clear as soon as we notice that $h^{(i)}
\PP_{k,l}= \PP_{k,l} h^{(\tau_{(k,l)}(i))}$.
This relation, which can also be obtained directly from the commutation
relations $ {\sf d}_i{\sf d}_j={\sf d}_j{\sf d}_i$, also implies by recurrence
that
\begin{gather}
  \label{eq:6}
  \forall \sigma_0\in S_n,\qquad  \PP_{\sigma_0}^{-1} ~%
  Q(z,\zeta) ~%
 \PP_{\sigma_0}=Q(z,\zeta).
\end{gather}

For notational simplicity, we introduce the operators 
\begin{align}
  \label{eq:28}
\mathcal{O}_m\equiv & \mathcal{O}(z_m, \zeta_m) \equiv Q(z_m, \zeta_m)\frac{((1-h\,z_m)(1-h\,\zeta_m))^{\otimes n}}{w(z_m)/w(\zeta_m)}
\\=&
\frac {1} {z_m-\zeta_m}\sum_\sigma \PP_\sigma%
\prod_{c\in \mathcal{C}(\sigma)} \left(\left[
\prod_{i\in c} z_m (1-h ^{(i)} \zeta_m) \right]- \left[
\prod_{i\in c} \zeta_m (1-h ^{(i)}
    z_m)\right]\right)\,,
\label{eq:30}
\end{align}
Relation \eqref{eq:23} is then equivalent to 
\begin{align}
  \label{eq:29}
  \mathcal{S}(\mathcal{O}_1~%
 \mathcal{O}_2)=0\,,
\end{align}
where $\mathcal{S}(\mathcal{O}_1~%
 \mathcal{O}_2)$ is a polynomial
of degree at most two in each $h^{(i)}$.
Of course, one can note that due to relations like $h^{(i)} \PP_{i,j}=
\PP_{i,j} h^{(j)}$, the degree in each individual $h^{(i)}$ is
well defined only if the position of the operators $\PP_{\sigma}$ is
specified. We will define degrees by putting all
permutation operators to the left of all operators $h^{(i)}$.

 We will first see that the
term with degree $0$ in all of the $h^{(i)}$'s vanishes, then that the terms
having degree $2$ in any of the  $h^{(i)}$'s vanish, and finally
that the terms having degree $1$ in any of the  $h^{(i)}$'s also
vanish. To this end we will proceed by recurrence assuming that $n\geq
2$ and that for all $1\leq n'<n$, we have
$\mathcal{S}(\mathcal{O}^{[n']}_1~%
 \mathcal{O}^{[n']}_2)=0$, where
$\mathcal{O}_m^{[n]}$ denotes the operator defined in \eqref{eq:28} for
a spin chain of length\footnote{If there is no indication of 
the length of spin chain, 
we assume that $\mathcal{O}_m$ always denote the operators
for the length $n$ spin chain.} 
$n$. In particular, one can see that
$\mathcal{O}^{[1]}_m=1$, which makes the statement
$\mathcal{S}(\mathcal{O}^{[1]}_1~%
 \mathcal{O}^{[1]}_2)=0$ obvious.

\paragraph{Terms of degree 0.}
\label{sec:terms-with-degree-0}
 
The terms with degree $0$ in all of the $h^{(i)}$'s are obtained by
substituting $h= 0$ in \eqref{eq:30}, to get 
\begin{equation}
\label{eq:4}
\left.\mathcal{O}_m\right|_{h= 0}=
\frac {1} {z_{m}-\zeta_{m}}\sum_\sigma
\PP_\sigma%
\prod_{c\in \mathcal{C}(\sigma)} 
\left[ z_m ^{\# c} - \zeta_m ^{\#c}\right]\,,
\end{equation}
where $\# c$ denotes the number of elements in $c$.

Interestingly enough, one can show from the expression \cite{1937,nakayama1940some} of
the characters of the representations of $S_n$, 
that \eqref{eq:4} can be rewritten as
\begin{equation}
\label{eq:7}
\left.\mathcal{O}_m\right|_{h=
  0}= 
  \sum_{k=1}^{n} 
  z_m ^{n-k} (-  \zeta_m )^{k-1}
\sum_\sigma \PP_\sigma\,\, \chi_{n-k,k-1}(\sigma)\,,
\end{equation}
where $\chi_{\alpha,\beta}(\sigma)$ denotes the character of the
permutation $\sigma$ in the hook representation
$(\alpha+1,1^{\beta})$ of the symmetric group $S_{\alpha+\beta+1}$. 
Indeed, the Murnaghan-Nakayama rule \cite{1937,nakayama1940some} says that for $\sigma\in S_n$, the
character $\chi_{\alpha,\beta}(\sigma)$ is obtained by summing
contributions from each Young tableau like the following one (which
contributes to $\chi_{\alpha,\beta}(\sigma)$ where $\alpha=7$,
$\beta=4$, and $\sigma$ is for instance the permutation
$1,2,3,4,5,6,7,8,9,10,11,12\mapsto 1,4,2,3,5,7,8,9,10,6,12,11$)
\begin{equation}
  \label{eq:12}
\includegraphics{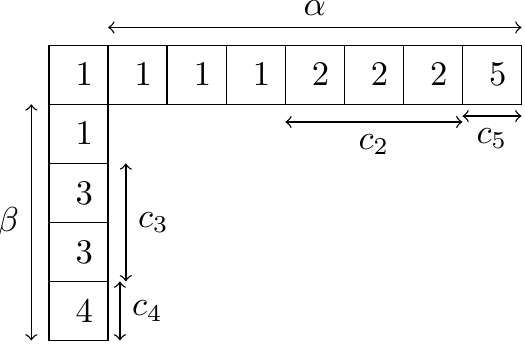}
\end{equation}
i.e. for each tableau such that each number $i$ appears $c_i$ times 
(where the permutation $\sigma$ has $l$ cycles of respective size $c_1,
c_2, \dots,c_l$),
in such a way the these numbers increase in
each row or column, and that for each number $i$, the set of boxes
having the label $i$ is convex\footnote{For a more general representation than the one
  associated to the hook-diagram $(\alpha+1,1^{\beta})$, this
  Murnaghan-Nakayama rule also requires an additional constraint that
  there is no two-by-two-square of boxes 
  labeled by the same number.}.
The contribution of each such tableau is $\prod_{i=1}^l (-1)^{h_i-1}$ where
  $h_i$ is the number of rows where the label $i$ is present at least once.

Hence (by denoting as $I$ the set of all labels $i\ge 2$ present in
the first column) one gets
\begin{align}
  \label{eq:21}
  \chi_{n-k,k-1}(\sigma)=&\sum_{\substack{I\subset \{2,3,\dots,l\}\\
k-c_1\leq \sum_{i\in I}c_i\leq k-1}} (-1)^{(k-\sum_{i\in I}c_i)-1} \prod_{i\in
  I} (-1)^{c_i -1}
=
\sum_{\substack{I\subset \{2,3,\dots,l\}\\
k-c_1\leq \sum_{i\in I}c_i\leq k-1}} 
(-1)^{k+1+|I|}\,,
\end{align}
where $|I|$ denotes the number of elements in $I$.

By comparison, the coefficient of $ z_m ^{n-k}   %
\zeta_m %
^{k-1}
\PP_\sigma
$ in \eqref{eq:4} (i.e. in 
$\sum_\sigma \PP_\sigma \sum_{j\geq 0}
(\tfrac{\zeta_m}{z_m})^j \linebreak[1] \prod_{i=1}^l 
[z_m ^{c_i} \linebreak[1]  - \zeta_m ^{c_i}]
$) is equal to
\begin{align}
  \label{eq:39}
\sum_{\substack{I\subset\{1,2,\dots,l\}
\\
\sum_{I\in I}c_i\leq k-1}}
(-1)^{|I|}
=%
\sum_{\substack{I\subset\{2,\dots,l\}
\\
k-c_1\leq\sum_{I\in I}c_i\leq k-1}}
(-1)^{|I|}\,,
\end{align}
where the equality is obtained by noticing that in the l.h.s., each term where $I\ni 1$ cancels with the term where $I$
is substituted with $I\setminus\{1\}$. This proves that the relations
\eqref{eq:4} and \eqref{eq:7} are identical.

One can then note that up to a numerical 
factor $a_{\alpha,\beta}$, each combination
$\sum_\sigma \PP_\sigma \chi_{\alpha,\beta}(\sigma)$ is a projector
onto the space associated to the irreducible representation 
 $(\alpha+1,1^{\beta})$ of  $S_{\alpha+\beta+1}$. Hence the
orthogonality of these spaces implies the relation\footnote{
One way to show \eqref{eq:5} is by introducing the Young symmetrizer
$c_\lambda$ associated to a representation $\lambda$,
which obey $\epsilon_\lambda\equiv\sum_\sigma \PP_\sigma \chi_{\lambda}(\sigma) \propto
\sum_\sigma \PP_\sigma c_\lambda \PP_{\sigma^-1}$ (see for instance
(3.11) on page 126 of \cite{Weylgroups}). In the case $\lambda\neq
\lambda'$, the Lemma (4.3.H) on page 126 of \cite{Weylgroups} says
that $\epsilon_{\lambda} \epsilon_{\lambda'}=0$, whereas the relation
written as $\epsilon\epsilon=\epsilon$ right before (3.11) on page 126
of \cite{Weylgroups} is the statement that if $\lambda =\lambda'$,
then  $\epsilon_{\lambda} \epsilon_{\lambda'}\propto\epsilon_\lambda$.
}
\begin{align}
  \sum_\sigma \PP_\sigma \chi_{\alpha,\beta}(\sigma) \sum_{\sigma'}
  \PP_{\sigma'} \chi_{\alpha',\beta'}(\sigma')\label{eq:5}
  =a_{\alpha,\beta} \delta_{\alpha,\alpha'}\delta_{\beta,\beta'} \sum_\sigma \PP_\sigma \chi_{\alpha,\beta}(\sigma) \,.
\end{align}

This allows us to compute 
\begin{equation}
\left.\mathcal{O}_1~%
 \mathcal{O}_2\right|_{h= 0}
= \sum_{k=1}^{n}  a_{n-k,k-1} \left(  {z_1 z_2} \right)^{n-k} \left(
  {\zeta_1\zeta_2} \right)^{k-1}
\sum_\sigma \PP_\sigma \chi_{n-k,k-1}(\sigma)\,,
\end{equation}
which clearly shows that $\mathcal{S}(\left.\mathcal{O}_1~%
 \mathcal{O}_2\right|_{h= 0})=0$, which is the vanishing of the
term with degree $0$ in all the $h^{(i)}$'s.

\paragraph{Terms of degree 2.}
\label{sec:terms-with-degree-2}

In order to understand the terms of higher degree in $h^{(i)}$ of the
product $\mathcal{O}_1~%
 \mathcal{O}_2$, we
will first investigate the terms of higher degree in $h^{(i)}$ of the
operator $\mathcal{O}_m$:
for instance, if we denote by $(\mathcal{O}_m)_{(h^{(n)})^1}$
the coefficient\footnote{In this notation,
  $(\dots)_{(h^{(k)})^1}$ denotes the coefficient of degree
  one in $h^{(k)}$, where the ambiguity in the degree of each
    individual $h^{(k)}$ (due to $\PP_{i,j}h^{(i)}=h^{(j)}\PP_{i,j}$)
      is fixed as explained above: by moving all permutation operators to the left of all $h^{(i)}$.
} of $h^{(n)}$ in $\mathcal{O}_m$, we get
\begin{multline}
(\mathcal{O}_m)_{(h^{(n)})^1}=
\label{eq:13}\frac {-\zeta_m z_m} {z_m-\zeta_m}\sum_{\substack{\sigma\in
    S_n\\\sigma(n)\neq n}} \PP_\sigma%
\prod_{c\in \mathcal{C}(\sigma)} \left(\left[
\prod_{\substack{i\in c\\i\neq n}}  %
  {z_m}(1-h^{(i)} \zeta_m ) %
\right]- \left[
\prod_{\substack{i\in c\\i\neq n}}  %
  \zeta_m (1-h^{(i)} z_m) %
\right]\right)  %
\\=
\frac {-\zeta_m z_m} {z_m-\zeta_m}\!\!\sum_{\substack{
1\le k\le n-1\\
\hat\sigma\in
    S_{n-1}}} \!\!\!\PP_{k,n} ~%
 \PP_{\hat\sigma} ~%
 \!\! \prod_{c\in
  \mathcal{C}\left(\hat\sigma\right)} \!\left(\left[ 
\prod_{\substack{i\in c}} %
  {z_m}(1-h^{(i)} \zeta_m) %
\right]- \left[
\prod_{\substack{i\in c}}  %
  \zeta_m (1-h^{(i)} z_m) %
\right]\right) %
\end{multline}
where $\hat\sigma\in S_{n-1}$ is defined by
\begin{align}
  \label{eq:8}
  \hat\sigma(i)=&
  \begin{cases}
    \sigma(i)&\textrm{if } \sigma(i)\neq n\\
    \sigma(n)&\textrm{if } \sigma(i)= n\,,
  \end{cases}& \textrm{hence }\sigma=&\tau_{(n,\sigma(n))} \circ   \hat\sigma\,,
\end{align}
where $\tau_{(i,j)}$ is the transposition of $i \leftrightarrow j$
defined previously. The first line in \eqref{eq:13} comes from the fact that if
$\sigma(n)=n$, then $\{n\}$ is a cycle of $\sigma$ and $h^{(n)}$ can
only appear through the factor $\prod_{i\in\{n\}} \left( {z_m}(1-h^{(i)}
\zeta_m ) - \zeta_m (1-h^{(i)} z_m)\right) = z_n-\zeta_n$, hence $h^{(n)}$ is
actually absent from the terms where $\sigma(n)=n$. The second line in \eqref{eq:13} is a simple rewriting of the first line.

From there we see that %
\begin{equation}
(\mathcal{O}_m)_{(h^{(n)})^1}={\left(-\zeta_m z_m %
  \sum_{k=1}^{n-1}
\PP_{k,n}\right)~%
 (%
  \mathcal{O}^{[n-1]}_m%
\otimes
\II)}\label{eq:17}\,%
.
\end{equation}

Analogously, one can
see that for arbitrary $j\le n$, the coefficient of
degree 1 in $h^{(j)}$ is %
$(\mathcal{O}_m)_{(h^{(j)})^1}={(-\zeta_m z_m 
\PP_{j,n}   \sum_{k=1}^{n-1}  \PP_{k,n})~%
 (%
  \mathcal{O}^{[n-1]}_m
\otimes
\II)~%
 \PP_{j,n}}$.

Now we are ready to compute the coefficient of $\left(h^{(n)}\right)^2$ in
the product $\mathcal{O}^{[n]}_1~%
 \mathcal{O}^{[n]}_2$: it is given by the
coefficient of  $h^{(n)}$ in 
$ -\zeta_2 z_2\,\, 
\mathcal{O}_1^{[n]} ~%
{(\sum_{k=1}^{n-1}
\PP_{k,n})~%
 (  \mathcal{O}^{[n-1]}_2\otimes
\II)}=-\zeta_2 z_2\,\, 
(\sum_{k=1}^{n-1}
\PP_{k,n})\linebreak  %
\mathcal{O}_1^{[n]} ~%
{ (  \mathcal{O}^{[n-1]}_2\otimes
\II)} $, 
which
is equal to
\begin{equation}
  \label{eq:18}
z_1z_2  \zeta_1\zeta_2 \left(\sum_{k=1}^{n-1}
\PP_{k,n}\right)^2 ~%
 ((\mathcal{O}^{[n-1]}_1~%
 \mathcal{O}^{[n-1]}_2)\otimes
\II)\, ,
\end{equation}
where we used \eqref{eq:6} and \eqref{eq:17}. 
Using the recurrence hypothesis, which contains the statement that
$\mathcal{S}(\mathcal{O}^{[n-1]}_1~%
 \mathcal{O}^{[n-1]}_2)=0$, we see that the coefficient of
$\left(h^{(n)}\right)^2$ in $\mathcal{S}(\mathcal{O}^{[n]}_1~%
\mathcal{O}^{[n]}_2)$ (which we will denote as $(\mathcal{S}(\mathcal{O}^{[n]}_1~%
\mathcal{O}^{[n]}_2))_{(h^{(n)})^2}$) does vanish.

\paragraph{Some notations.}
\label{sec:notations}

The analysis of the terms with degree $1$ requires a couple of notations
that we will now introduce. We have already seen that the coefficient of
$\mathcal{O}_m$ with degree $1$ in $h^{(j)}$ is
 $(\mathcal{O}_m)_{(h^{(j)})^1}={(-\zeta_m z_m \PP_{j,n} %
  \sum_{k=1}^{n-1}  \PP_{k,n})~%
 (
  \mathcal{O}^{[n-1]}_m
\otimes
\II)~%
 \PP_{j,n}} %
$, where one can notice
that $\sum_{k=1}^{n-1}
\PP_{j,n}~%
 \PP_{k,n}=  \sum_{k\in\{1,2,\ldots,n\}\setminus
  j}\PP_{k,j}~%
 \PP_{j,n}$. This allows to rewrite  this
expression as $(-\zeta_m z_m 
  \sum_{k\neq j}  \PP_{k,j})~%
(
  \mathcal{O}^{[n-1]}_m
\triangleleft\,
\II^{(j)}) %
$, where  
\begin{align}
  \label{eq:22}
    \mathcal{O}^{[n-1]}_m
\triangleleft\,
\II^{(j)}&\textrm{ denotes }\PP_{j,n}~%
 (
  \mathcal{O}^{[n-1]}_m 
\otimes
\II)~%
 \PP_{j,n}\,.
\end{align}
One can also notice, %
using the same
arguments as in the proof of \eqref{eq:6}, that it is equal to $ \PP_{j,n}~%
 (
  \mathcal{O}^{[n-1]}_m 
\otimes
\II)~%
 \PP_{j,n}=\frac {\prod_{k\neq j}
  (1-h^{(k)}z_m)(1-h^{(k)}\zeta_m)} {w(z_m)/w(\zeta_m)} \prod_{k\neq
  j}
{\sf d}_k \frac {w(z_m)}{w(\zeta_m)}
$. We will also use natural generalizations of this notation, for instance
\begin{align}
  \label{eq:32}
  \mathcal{O}^{[n-1]}_m
\triangleleft\,
h^{(j)}&\equiv \PP_{j,n}~%
 (
  \mathcal{O}^{[n-1]}_m 
\otimes
h)~%
 \PP_{j,n}\,,\\  \mathcal{O}^{[n-2]}_m
\triangleleft\,
h^{(j)}\triangleleft\,
\II^{(k)}&\equiv \PP_{j,n-1}\PP_{k,n}~%
 (
  \mathcal{O}^{[n-2]}_m 
\otimes
h\otimes \II)~%
 \PP_{k,n}\PP_{j,n-1}\,&\textrm{when }j\neq& k,n.
\end{align}

Another important remark is that all the operators that we deal
with are of the form $\sum_\sigma \PP_{\sigma} c_\sigma$, where $c_\sigma$
is a coefficient which only contains some scalars and the operators
$h^{(j)}$. We denote this coefficients $c_\sigma$ as
$\left[\cdots\right]_\sigma$. For instance, we have
\begin{align}
\mathcal{O}_m=&\sum_{\sigma\in S_n}\PP_\sigma
\left[\mathcal{O}_m\right]_{\sigma}\,,\hspace{2cm}
\mathcal{O}_1~%
 \mathcal{O}_2=\sum_{\sigma\in S_n}\PP_\sigma
\left[\mathcal{O}_1~%
 \mathcal{O}_2\right]_{\sigma}\,,\\
  \label{eq:31}
\textrm{where }\left[\mathcal{O}_m\right]_{\sigma}=&
\frac 1 {z_m-\zeta_m}\prod_{c\in \mathcal{C}(\sigma)} \left(\left[
\prod_{i\in c} z_m (1-h ^{(i)} \zeta_m) \right]- \left[
\prod_{i\in c} \zeta_m (1-h ^{(i)}
    z_m)\right]\right)\,,
\\%
  \left[\mathcal{O}_1~%
 \mathcal{O}_2\right]_{\sigma}=&\sum_{\tilde
    \sigma\in S_n} \PP_{{\tilde \sigma}^{-1}}
\left[\mathcal{O}_1\right]_{\sigma\circ {\tilde
    \sigma}^{-1}} \PP_{\tilde \sigma}
\left[\mathcal{O}_2\right]_{{\tilde \sigma}}\,,
\label{eq:16}
\end{align}
where 
$ \PP_{{\tilde \sigma}^{-1}}
\left[\mathcal{O}_1\right]_{\sigma\circ {\tilde
    \sigma}^{-1}} \PP_{\tilde \sigma}$ coincides with $\left[\mathcal{O}_1\right]_{\sigma\circ {\tilde
    \sigma}^{-1}}$ up to the substitution $h^{(j)}\to h^{(\tilde
  \sigma^{-1}(j))}$.

We also have 
\begin{align*}
  \mathcal{O}^{[n-1]}_m\otimes \II=&\sum_{\sigma\in S_n}\PP_\sigma
\left[\mathcal{O}^{[n-1]}_m\otimes \II \right]_{\sigma}& \textrm{where  }
\left[\mathcal{O}^{[n-1]}_m\otimes \II \right]_{\sigma}=&
\begin{cases}
  0&\textrm{ if }\sigma(n)\neq n \\
  [\mathcal{O}^{[n-1]}_m%
]_{\sigma}\otimes \II &\textrm{ if }\sigma(n)= n, 
\end{cases}
\end{align*}
where we identify the permutations $\sigma\in S_{n-1}$ with the
permutations $\sigma\in S_n$ such that $\sigma(n)=n$.
For any operator $\mathcal{O}$ on 
$(\CC^{N})^{\otimes n}$ and any $\sigma, \tau \in S_{n}$ 
with $\tau^2=1$, we also have\footnote{Multiplying $\PP_{\tau}$ from the right of the expansion 
$\mathcal{O} \PP_{\tau}=\sum_{\sigma} \PP_{\sigma} \left[\mathcal{O} \PP_{\tau} \right]_{\sigma}$, we obtain 
$\mathcal{O}=\sum_{\sigma} \PP_{\sigma} \left[\mathcal{O} \PP_{\tau} \right]_{\sigma}\PP_{\tau}
=\sum_{\sigma} \PP_{\sigma \circ \tau} \PP_{\tau}\left[\mathcal{O} \PP_{\tau} \right]_{\sigma}\PP_{\tau}
=\sum_{\sigma} \PP_{\sigma } \PP_{\tau}\left[\mathcal{O} \PP_{\tau} \right]_{\sigma \circ \tau }\PP_{\tau}$, where we used the fact that 
$ \PP_{\tau}^2=1$,  $ \PP_{\sigma}\PP_{\tau}=\PP_{\sigma \circ \tau}$ and $\tau^2=1$. Taking note on the fact that 
$ \PP_{\tau}\left[\mathcal{O} \PP_{\tau} \right]_{\sigma \circ \tau }\PP_{\tau}$ does not effectively contain the permutation operator, and comparing the above equation with an expansion 
$ \mathcal{O} =\sum_{\sigma} \PP_{\sigma} \left[\mathcal{O} \right]_{\sigma}$, we obtain \eqref{eq:op}. 
\eqref{eq:op2} follows from a similar argument.} 
\begin{align}
\left[\mathcal{O}\right]_{\sigma}&=
\PP_{\tau} \left[\mathcal{O} \PP_{\tau}\right]_{\sigma \circ \tau} \PP_{\tau} 
 \label{eq:op}
\\
&=  \left[\PP_{\tau} \mathcal{O} \right]_{\tau \circ \sigma } 
 \label{eq:op2}
\end{align}

Using this notation (and at intermediate steps the convention $\PP_{n,n}=\II$), one can see that expression \eqref{eq:17} can
be rewritten as 
\begin{align}
\label{eq:2}
{\left(-\zeta_m z_m 
  \sum_{k=1}^{n-1}
\PP_{k,n}\right)~%
 (
  \mathcal{O}^{[n-1]}_m
\otimes
\II)}
=&\zeta_m z_m \left(\II-
  \sum_{k=1}^{n}
\PP_{k,n}\right)~%
 (
  \mathcal{O}^{[n-1]}_m
\otimes
\II)\nonumber\\\nonumber
=&\sum_{\sigma \in S_n} \PP_{\sigma} \left[ \zeta_m z_m
  (\II-\PP_{n,\sigma(n)}) ~%
  (
  \mathcal{O}^{[n-1]}_m
\otimes
\II) \right]_{\sigma}\\\nonumber
=&\sum_{\sigma \in S_n} \PP_{\sigma} \left[\zeta_m z_m
 (
  \mathcal{O}^{[n-1]}_m
\triangleleft\,
\II^{(\sigma(n))}) ~%
  (\II-\PP_{n,\sigma(n)}) \right]_{\sigma}\!\\=&
\sum_{\sigma \in S_n} \PP_{\sigma} \left[  -\zeta_m z_m
( \mathcal{O}^{[n-1]}_m \triangleleft\, \II^{(\sigma(n))})
 ~%
 \sum_{\!\!\!\!\!\!\!\!\!\!\!k\in\{1,2,\ldots,n\}\setminus\{\sigma(n)\}\!\!\!\!\!\!\!\!\!\!\!}
\PP_{k,\sigma(n)}\right]_{\sigma}\!\!. %
\end{align}
Here the first equality %
comes from
the fact that $ (
  \mathcal{O}^{[n-1]}_m
\otimes
\II)$ only contains $\PP_{\tilde \sigma}$ where $\tilde \sigma
(n)=n$. Hence, one can only get $\PP_{k,n}~%
 \PP_{\tilde
  \sigma}=\PP_{\sigma}$ if $k=\sigma(n)$. The next line is an elementary 
manipulation with the permutation operators (using the fact that in
$\II-\PP_{n,\sigma(n)}$ the term $\II$ contributes only if $\sigma(n)=n$), and the last equality comes
out by the same argument as the first one.

For simplicity, the sequence of equalities \eqref{eq:2}
will be written as 
$-\zeta_m z_m 
  \sum_{k=1}^{n-1} \linebreak[1]
\langle \PP_{k,n} \linebreak[1] (
  \mathcal{O}^{[n-1]}_m
\otimes
\II)\rangle
=\zeta_m z_m 
\langle 
(\II-  \sum_{k=1}^{n} 
\PP_{k,n}) (
  \mathcal{O}^{[n-1]}_m
\otimes
\II)\rangle
=\zeta_m z_m  \langle (\II-\PP_{n,\sigma(n)}) (
  \mathcal{O}^{[n-1]}_m
\otimes
\II) \rangle
=\zeta_m z_m
\langle (
  \mathcal{O}^{[n-1]}_m
\triangleleft\, %
\II^{(\sigma(n))})  (\II-\PP_{n,\sigma(n)}) \rangle=
-\zeta_m z_m 
\langle( \mathcal{O}^{[n-1]}_m \triangleleft\, \II^{(\sigma(n))})
\sum_{k\neq \sigma(n)}
\PP_{k,\sigma(n)}\rangle$, where\footnote{If $F(\sigma)$ is
  independent of
$\sigma \in S_{n}$, then 
$\langle F(\sigma)\rangle =\langle F\rangle =
\sum_{\sigma}\PP_\sigma [F]_\sigma =F$. In general $\langle F(\sigma)\rangle\neq F(\sigma)$.}
$\langle F(\sigma)\rangle$ for the operator $F(\sigma)$ 
of $\sigma \in S_{n}$ 
by definition equals to
$\sum_{\sigma}\PP_\sigma [F(\sigma)]_\sigma$.

As a first example of using the above notations, let us note that for any $i<n$,
$\sum_{\sigma\in S_n} \PP_{\sigma}  \PP_{i,n}
[\mathcal{O}_1\mathcal{O}_2]_{\sigma}\PP_{i,n}$, which coincides with
$\mathcal{O}_1 \mathcal{O}_2$ up to the substitution
$h^{(i)}\leftrightarrow h^{(n)}$, can also be written as $\langle\PP_{\sigma}
\PP_{i,n} \PP_{\sigma} ^{-1} \mathcal{O}_1\mathcal{O}_2\PP_{i,n}\rangle$, which
is also equal to $\langle\PP_{\sigma}
\PP_{i,n} \PP_{\sigma} ^{-1} \PP_{i,n} \mathcal{O}_1\mathcal{O}_2\rangle$ (due
to \eqref{eq:6}). A
consequence is that, since we have already proven that in
$\mathcal{O}_1\mathcal{O}_2$, the coefficient of $(h^{(n)})^2$
vanishes, we deduce that it also vanishes in $\langle\PP_{\sigma}
\PP_{i,n} \PP_{\sigma} ^{-1} \PP_{i,n} \mathcal{O}_1\mathcal{O}_2\rangle$. It
means that in $\mathcal{O}_1 \mathcal{O}_2$, the coefficient of
$(h^{(i)})^2$ vanishes just like the coefficient of  $(h^{(n)})^2$.

An important remark about the equalities \eqref{eq:2} is
that the left-hand side is the expression \eqref{eq:17} of 
$(\mathcal{O}_m)_{(h^{(n)})^1}%
=\sum_\sigma \PP_\sigma
([\mathcal{O}_m]_\sigma)_{(h^{(n)})^1}$. This expression allows to conveniently find that
for instance
$(\PP_{i,j} \mathcal{O}_m)_{(h^{(n)})^1}=\sum_\sigma \PP_{i,j}  \PP_\sigma
([\mathcal{O}_m]_\sigma)_{(h^{(n)})^1}=-\zeta_m z_m   \PP_{i,j}
\sum_{k=1}^{n-1} \linebreak[1]
\PP_{k,n}~%
 (
  \mathcal{O}^{[n-1]}_m
\otimes
\II)$. On the other hand, the right-hand side $-\zeta_m z_m 
\langle( \mathcal{O}^{[n-1]}_m \triangleleft\, \II^{(\sigma(n))})
\sum_{k\neq \sigma(n)} \linebreak[1]
\PP_{k,\sigma(n)}\rangle$ is convenient in order to express 
$(\mathcal{O}_m \PP_{i,j})_{(h^{(n)})^1}$: for instance one sees that 
\begin{align}
  \label{eq:14}
  (\mathcal{O}_m
  \PP_{1,n})_{(h^{(n)})^1}=&\textstyle(\mathcal{O}_m)_{(h^{(1)})^1}\PP_{1,n}=-\zeta_m
  z_m \langle 
( \mathcal{O}^{[n-1]}_m \triangleleft\, \II^{(\sigma(1))})
\sum_{k\neq \sigma(1)}
\PP_{k,\sigma(1)}\rangle \PP_{1,n}\nonumber{}\\=&\textstyle
-\zeta_m
  z_m  \sum_\sigma \PP_\sigma 
[( \mathcal{O}^{[n-1]}_m \triangleleft\, \II^{(\sigma(1))})
\sum_{k\neq \sigma(1)}
\PP_{k,\sigma(1)}]_{\sigma} \PP_{1,n}\nonumber{}\\=&\textstyle
-\zeta_m
  z_m  \sum_\sigma \PP_\sigma \PP_{1,n}
[( \mathcal{O}^{[n-1]}_m \triangleleft\, \II^{(\sigma(1))})
{\textstyle \sum_{k\neq \sigma(1)}}
\PP_{k,\sigma(1)} \PP_{1,n} ]_{\sigma\circ \tau_{(1,n)}}\nonumber{}\\=&\textstyle
-\zeta_m
  z_m  \sum_{\tilde \sigma} \PP_{\tilde \sigma}
[( \mathcal{O}^{[n-1]}_m \triangleleft\, \II^{(\tilde \sigma(n))})
{\textstyle \sum_{k\neq \tilde \sigma(n)}}
\PP_{k,\tilde \sigma(n)} \PP_{1,n} ]_{\tilde\sigma}\nonumber{}\\
=&\textstyle
-\zeta_m
  z_m \langle 
( \mathcal{O}^{[n-1]}_m \triangleleft\, \II^{(\sigma(n))})
\sum_{k\neq \sigma(n)}
\PP_{k,\sigma(n)}\PP_{1,n}\rangle\,,
\end{align}
where \eqref{eq:op} is used. 
More generally we get $(\mathcal{O}_m
  \PP_{i,j})_{(h^{(n)})^1}=-\zeta_m
  z_m \langle 
( \mathcal{O}^{[n-1]}_m \triangleleft\, \II^{(\sigma(n))}) \linebreak[1]
\sum_{k\neq \sigma(n)}
\PP_{k,\sigma(n)}\PP_{i,j}\rangle$, for arbitrary $i, j$.

\paragraph{Terms of degree 1.}
\label{sec:terms-with-degree-1}

Finally, let us investigate
$(\mathcal{O}^{[n]}_1~%
 \mathcal{O}^{[n]}_2)_{(h^{(n)})^1}$, i.e.
 the coefficient of degree $1$ in $h^{(n)}$
in the expansion of $\mathcal{O}^{[n]}_1~%
 \mathcal{O}^{[n]}_2$. Using the
arguments above (cf.\ eqs.\ \eqref{eq:6}, \eqref{eq:17}, 
\eqref{eq:2}, and the last paragraph above), we see that this term is
equal to 
\begin{multline}
 \!\!\!\!\!  \left
    (-\zeta_2 z_2\,\,  \sum_{k=1}^{n-1}
\PP_{k,n}~%
\mathcal{O}_1^{[n]} 
~%
 (%
  \mathcal{O}^{[n-1]}_2
\otimes
\II)
-\zeta_1 z_1\,\,  \langle(%
  \mathcal{O}^{[n-1]}_1
\triangleleft\,
\II^{(\sigma(n))})%
~%
\mathcal{O}_2^{[n]} ~%
 \!\sum_{k\neq \sigma(n)}
\PP_{k,\sigma(n)}\rangle %
 \right
)_{(h^{(n)})^0}
\\
=
  -
\left.\left\langle\left(%
\zeta_2 z_2%
\sum_{k=1}^{n-1}
\PP_{k,n}~%
\mathcal{O}_1^{[n]} 
~%
 (%
  \mathcal{O}^{[n-1]}_2
\otimes
\II)
+\zeta_1 z_1\,%
  \mathcal{O}^{[n-1]}_1
\triangleleft\,
\II^{(\sigma(n))}\right)%
~%
\mathcal{O}_2^{[n]} ~%
 \sum_{k\neq \sigma(n)}
\PP_{k,\sigma(n)}\right\rangle \right|_{h^{(n)}= 0}\!\!\!\!
\label{degree1}
\end{multline}
where we used
the notations defined above\footnote{For simplicity, we denote by
  $(X)|_{h^{(n)}=0}$ the coefficient $(X)|_{(h^{(n)})^0}$ of
  $(h^{(n)})^0$ in $X$.}. 
In what follows we will show, with the necessary details, how this term vanishes (when $\mathcal{S}$ is applied) for permutations $\sigma$ such that
$\sigma(n)\neq n$. We will also discuss very shortly how the 
argument can be transposed to the simpler case of the
permutations such that $\sigma(n)=n$.

In order to prove that it vanishes (after the functions $\mathcal{S}$
is applied), we can use the recurrence hypothesis
$\mathcal{S}\left( Q^{[n-1]}(z_1, \zeta_1)~%
 Q^{[n-1]}(z_2,
    \zeta_2)\right)=0$,
 where $Q^{[n-1]}$ denotes $\frac{1}{(z -\zeta)}\linebreak[1]
{\sf d}^{\otimes (n-1)}\! \left ( \frac{w(z)}{w(\zeta )}\right )
$. If we take the derivative 
(i.e. apply ${\sf d} \otimes $ from the left) 
of this relation and use the definition
\eqref{eq:28} to rewrite $Q(z_m,\zeta_m)$ in terms of $\mathcal{O}_m$,
we get
\begin{gather}
  \label{eq:33}
\mathcal{S}\left(\mathcal{A}\right)=0\\
\textrm{where } \mathcal{A}\equiv
  \mathcal{O}^{[n]}_1~%
 (\mathcal{O}^{[n-1]}_2 \otimes ((1-h
  z_2)(1-h \zeta_2)))
+ (\mathcal{O}^{[n-1]}_1 \otimes ((1-h
  z_1)(1-h \zeta_1))) ~%
 \mathcal{O}^{[n]}_2. \nonumber
\end{gather}

Hence, we also have $\mathcal{S}[f(\mathcal{A})]=0$ for any linear
transformation $f$ which commutes with $\mathcal{S}$. In particular we have\footnote{We will use the notation 
$\mathcal{S}\left[\dots \right]_{\sigma}$ for $\mathcal{S}(\left[\dots
\right]_{\sigma})$.}
\begin{equation}
  \label{eq:15}
  \left(\mathcal{S}\left[\sum_{k=1}^{n-1}\PP_{k,n}\mathcal{A}+\PP_{n,\sigma(n)}\,
  \mathcal{A}\,  \PP_{n,\sigma(n)}
\sum_{k\neq \sigma(n)}\PP_{k,\sigma(n)}
\right]_{\sigma}\right)_{(h^{(n)})^2}=0\,.
\end{equation}
Assuming that
$\sigma(n)\neq n$, this identity reads %
\begin{align}
\label{eq:34}
\sum_{k=1}^{8} \mathcal{S}\left[{\mathcal B}_{k}\right]_\sigma=0,
\end{align}
where 
\begin{align}
{\mathcal B}_{1}=&
\left.
\vphantom{\left({\textstyle \sum_{k\neq \sigma(n)}}\PP_{k,\sigma(n)}\right)^2}
z_2 \zeta_2 {\textstyle \sum_{k=1}^{n-1}}\PP_{k,n} \mathcal{O}^{[n]}_1~%
(\mathcal{O}^{[n-1]}_2 \otimes \II) 
\right|_{h^{(n)}= 0} ,
\nonumber\\
{\mathcal B}_{2}= &
z_1\zeta_1(z_2+\zeta_2)
\left({\textstyle \sum_{k=1}^{n-1}}\PP_{k,n}\right)^2
(\mathcal{O}^{[n-1]}_1\otimes \II)~%
 (\mathcal{O}^{[n-1]}_2\otimes
\II) 
 ,
\nonumber\\
{\mathcal B}_{3}=&
 z_1 \zeta_1 (z_2-\zeta_2)\PP_{n,\sigma(n)}
(\mathcal{O}_1^{[n-1]}\otimes \II)
~%
(\mathcal{O}_2^{[n-1]}\otimes \II)
, \nonumber\\
{\mathcal B}_{4}=&
z_1\zeta_1z_2\zeta_2  \textstyle 
 \sum_{%
     k\neq n,
      \sigma(n)}%
\PP_{k,n}%
(\mathcal{O}^{[n-2]}_1 \triangleleft\, \II^{(\sigma(n))}\otimes ((1-h
   z_1)(1-h \zeta_1))) 
\nonumber\\
&\hspace{.4\textwidth}
\textstyle
(\mathcal{O}^{[n-1]}_2\triangleleft\, \II^{(\sigma(n))})
\sum_{%
     k'\neq n,
      \sigma(n)}%
\PP_{k',n}
\PP_{n,\sigma(n)}
 ,
\nonumber\\
{\mathcal B}_{5}=&
(z_1-\zeta_1)z_2\zeta_2
\PP_{n,\sigma(n)}(\mathcal{O}_1^{[n-1]}\otimes
\II)~%
(\mathcal{O}_2^{[n-1]} \otimes \II)
, \nonumber\\
{\mathcal B}_{6}=&
\left.z_1\zeta_1z_2\zeta_2 \textstyle
\sum_{\substack{1\le k\le n\\k\neq n,
      \sigma(n)}}\right(
(\mathcal{O}_1^{[n-1]}\triangleleft\,
\II^{(\sigma(n))})
~%
(\mathcal{O}_2^{[n-2]}
 \triangleleft\, \II^{(\sigma(n))}
\triangleleft\, ((1-h z_2)(1-h \zeta_2))^{(k)})
\nonumber\\
&\hspace{.4\textwidth}%
\PP_{n,\sigma(n)}\PP_{\sigma(n),k}
\left. {\textstyle \sum_{\substack{1\le k'\le n\\k'\neq n,
      \sigma(n)}}}\PP_{k',\sigma(n)}\right)
,
\nonumber\\
{\mathcal B}_{7}=&
\left. 
z_1\zeta_1 \left(\mathcal{O}_1^{[n-1]}\triangleleft\,
  \II^{(\sigma(n))}\right)\mathcal{O}_2^{[n]} {\textstyle \sum_{k\neq
  \sigma(n)}}\PP_{k,\sigma(n)}
\right|_{h^{(n)}= 0} ,
\nonumber\\
{\mathcal B}_{8}=&
(z_1+\zeta_1)z_2 \zeta_2 (\mathcal{O}_1^{[n-1]}\triangleleft\, \II^{(\sigma(n))})~%
(\mathcal{O}_2^{[n-1]}\triangleleft\, \II^{(\sigma(n))})~%
\left({\textstyle \sum_{k\neq
      \sigma(n)}}\PP_{k,\sigma(n)}\right)^2
\,.
\end{align}

The identity \eqref{eq:34} is just an explicit reformulation of
\eqref{eq:15}, as discussed in more details in the very end of the
present appendix.

We can rewrite the terms $\mathcal{S}\left[\mathcal{B}_k\right]_\sigma$ by manipulating the permutation
operators: for instance we can rewrite
$\mathcal{S}\left[\mathcal{B}_8\right]_\sigma$ using
\begin{align}
  \label{eq:19}
\lefteqn{[ 
(\mathcal{O}_1^{[n-1]}\triangleleft\, \II^{(\sigma(n))})~%
(\mathcal{O}_2^{[n-1]}\triangleleft\, \II^{(\sigma(n))})~%
{\textstyle \sum_{k'\neq
      \sigma(n)}}\PP_{k',\sigma(n)}
{\textstyle \sum_{k\neq
      \sigma(n)}}\PP_{k,\sigma(n)}
]_{\sigma}}\nonumber\hspace{3cm}\\
 =&\textstyle
\left[ 
(\mathcal{O}_1^{[n-1]}\triangleleft\, \II^{(\sigma(n))})~%
(\mathcal{O}_2^{[n-1]}\triangleleft\, \II^{(\sigma(n))})~%
\PP_{n,\sigma(n)}~%
\sum_{k\neq n,\sigma(n)} \PP_{k,\sigma(n)}
\right]_{\sigma} \nonumber\\ =&
\textstyle
\left[ \PP_{n,\sigma(n)}~%
\sum_{k\neq n,\sigma(n)} \PP_{k,\sigma(n)}~%
(\mathcal{O}_1^{[n-1]}\triangleleft\, \II^{(n)})~%
(\mathcal{O}_2^{[n-1]}\triangleleft\, \II^{(n)})
\right]_{\sigma} \nonumber\\ =&
\textstyle \left[ 
\sum_{k\neq n,\sigma(n)} \PP_{k,n} ~%
\PP_{n,\sigma(n)}~%
(\mathcal{O}_1^{[n-1]}\triangleleft\, \II^{(n)})~%
(\mathcal{O}_2^{[n-1]}\triangleleft\, \II^{(n)})
\right]_{\sigma} \,,
\end{align}
where the first equality comes from the fact that we take the
coefficient $[\dots]_\sigma$, which implies that in $({\textstyle \sum_{k\neq
      \sigma(n)}}\PP_{k,\sigma(n)})^2=\sum_{k,k^{\prime}\neq
    \sigma(n)}\PP_{\tau_{(k',\sigma(n))}\circ \tau_{(k,\sigma(n))}}$,
  we should only keep the terms where $(\tau_{(k',\sigma(n))}\circ
  \tau_{(k,\sigma(n))})(n)=\sigma(n)$, i.e. the terms where $k'=n$ and
  $k\neq n$. From  \eqref{eq:19}, 
we see that $\mathcal{S}\left[\mathcal{B}_1\right]_\sigma$ and $\mathcal{S}\left[\mathcal{B}_8\right]_\sigma$ sum up to
\begin{multline}
  \label{eq:35}
\mathcal{S}\left[\mathcal{B}_2+\mathcal{B}_8\right]_\sigma
=\\
  \mathcal{S}%
    \left[z_1\zeta_1z_2\zeta_2(\tfrac 1 {z_1}+\tfrac 1
    {z_2}+\tfrac 1 {\zeta_1}+\tfrac 1 {\zeta_2})
{\textstyle\sum_{k\neq n, \sigma(n)}}\PP_{k,n}\PP_{n,\sigma(n)}
(\mathcal{O}^{[n-1]}_1\otimes \II)~%
 (\mathcal{O}^{[n-1]}_2\otimes
\II)
\right]_\sigma %
\,,
\end{multline}
which is zero due to the recurrence hypothesis
$\mathcal{S}(\mathcal{O}^{[n-1]}_1~%
\mathcal{O}^{[n-1]}_2)=0$. 
Similarly, %
we get
\begin{equation}
  \label{eq:36}
\mathcal{S}\left[{\mathcal B}_{3}+{\mathcal B}_{5}\right]_\sigma=
\mathcal{S}%
  \left[   z_1 \zeta_1 z_2\zeta_2(-\tfrac 1 {z_1}-\tfrac 1 {z_2}+\tfrac 1 {\zeta_1}+\tfrac 1 {\zeta_2})\PP_{n,\sigma(n)}
(\mathcal{O}_1^{[n-1]}\otimes \II)~%
(\mathcal{O}_2^{[n-1]}\otimes \II) \right]_\sigma %
\,,
\end{equation}
which is zero by the same argument. 
We can also express the sum $\mathcal{S}\left[{\mathcal B}_{4}+{\mathcal B}_{6}\right]_\sigma$ as
\begin{multline}
  \label{eq:37}
\mathcal{S}\left[{\mathcal B}_{4}+{\mathcal B}_{6}\right]_\sigma= \textstyle 
\mathcal{S}\left[
z_1\zeta_1z_2\zeta_2  \textstyle 
 \sum_{
     k\neq n,
      \sigma(n)}
\PP_{k,n}\right.\\\textstyle
\left(
(\mathcal{O}^{[n-2]}_1 \triangleleft\, \II^{(\sigma(n))}\otimes ((1-h
   z_1)(1-h \zeta_1)))
~%
 (\mathcal{O}^{[n-1]}_2\triangleleft\, \II^{(\sigma(n))})\right.
\\ \textstyle
+
\left.
(\mathcal{O}^{[n-1]}_1 \triangleleft\, \II^{(\sigma(n))})
~%
 (\mathcal{O}^{[n-2]}_2\triangleleft\, \II^{(\sigma(n))}
\otimes ((1-h
   z_1)(1-h \zeta_1)))\right)\\\textstyle
\left.
\sum_{%
     k'\neq n,
      \sigma(n)}%
\PP_{k',n}
\PP_{n,\sigma(n)}
\right]_\sigma
\end{multline}
which contains the same quantity as \eqref{eq:33} up to the
transformation $n\to n-1$. Hence $\mathcal{S}\left[{\mathcal B}_{4}+{\mathcal B}_{6}\right]_\sigma$ vanishes if $n\ge 3$,
because the recurrence hypothesis contains $\mathcal{S}(\mathcal{O}^{[n-2]}_1~%
\mathcal{O}^{[n-2]}_2)\linebreak[1]=0$, which allows to obtain \eqref{eq:33} even
for $n-1$ spins. On the other hand, if $n=2$, then ${\mathcal B}_{4}+{\mathcal B}_{6}$ vanishes %
because the sum over $k\neq n,
      \sigma(n)$ has no terms.

Finally, the above discussion 
(\eqref{eq:34} and  
\eqref{eq:35}$=$\eqref{eq:36}$=$\eqref{eq:37}$=0$) implies that 
the remaining terms $\mathcal{S}\left[\mathcal{B}_1\right]_\sigma$ and $\mathcal{S}\left[\mathcal{B}_7\right]_\sigma$ sum up to zero, which gives 
\begin{multline}
  \label{eq:38}
\mathcal S\left[{\mathcal B}_{1}+{\mathcal B}_{7}\right]_\sigma=
\\
  \left.\mathcal{S}\left[
z_2 \zeta_2 {\textstyle 
\sum\limits_{k=1}^{n-1}}\PP_{k,n} \mathcal{O}^{[n]}_1  ~%
(\mathcal{O}^{[n-1]}_2 \otimes \II)  + z_1\zeta_1 \left(\mathcal{O}_1^{[n-1]}\triangleleft\,
  \II^{(\sigma(n))}\right)\mathcal{O}_2^{[n]} 
   {\textstyle \sum\limits_{k\neq
  \sigma(n)}} \PP_{k,\sigma(n)}
\right]_\sigma \right|_{h^{(n)}=
0}
\\
 = 0\,.
\end{multline}
For simplicity, the argument was here written for permutations
$\sigma$ such that $\sigma(n)\neq n$. On the other hand, if $\sigma(n)=
n$ then one obtains the same outcome \eqref{eq:38}, by writing the
condition $\left(\mathcal{S}\left[\sum_{k=1}^{n-1}\PP_{k,n}\mathcal{A}+
  \mathcal{A} 
\sum_{k\neq \sigma(n)}\PP_{k,\sigma(n)}
\right]_{\sigma}\right)_{(h^{(n)})^2}=0$, and following the same steps
as above, the main difference
being that ${\mathcal B}_{3}$ and 
${\mathcal B}_{5}$ are then absent from \eqref{eq:34}\footnote{Indeed,
  for $\sigma(n)\neq
n$, $\mathcal{B}_3$ was the term where $k=\sigma(n)$ in
$(\mathcal{S}[\sum_{k\neq
    n} \PP_{k,n} (\mathcal{O}^{[n-1]}_1 \otimes ((1-h
  z_1)(1-h \zeta_1))) ~%
 \mathcal{O}^{[n]}_2]_\sigma)
  _{(h^{(n)})^2}$. In the case when $\sigma(n)=n$, there is no term
  $k=\sigma(n)$ in this sum.}.

By summing up over all permutations, we hence obtain that
\begin{equation}
 \left. \mathcal{S}\left(\left\langle
z_2 \zeta_2 {\textstyle 
\sum\limits_{k=1}^{n-1}}\PP_{k,n} \mathcal{O}^{[n]}_1  %
(\mathcal{O}^{[n-1]}_2 \otimes \II)  + z_1\zeta_1 \left(\mathcal{O}_1^{[n-1]}\triangleleft\,
  \II^{(\sigma(n))}\right)\mathcal{O}_2^{[n]} 
   {\textstyle \sum\limits_{k\neq
  \sigma(n)}} \PP_{k,\sigma(n)} \right\rangle
\right) \right|_{h^{(n)}=
0}\!\!= 0\,,
\end{equation}
which exactly shows that
in $\mathcal{S}(\mathcal{O}^{[n]}_1~%
 \mathcal{O}^{[n]}_2)$
 the term of degree one in $h^{(n)}$ 
(cf.\ \eqref{degree1}) 
does vanish
as we wished to prove.

To conclude the proof, we have seen that in
$\mathcal{S}(\mathcal{O}^{[n]}_1~%
 \mathcal{O}^{[n]}_2)$, the terms
of degree $2$ in any $h^{(i)}$ vanish, as well as the terms of degree
$1$ in $h^{(n)}$. Multiplying by permutations, we deduce that the terms
of degree $1$ in any $h^{(i)}$ vanish, and we have also shown that
the term of degree $0$ in all the $h^{(i)}$'s vanishes. Hence we
proved (by recurrence) that $\mathcal{S}(\mathcal{O}^{[n]}_1~%
\mathcal{O}^{[n]}_2)=0$.

\paragraph{More details for equation \eqref{eq:34}.}
\label{sec:more-deta-equat}

In this paragraph we will show that \eqref{eq:34} is equivalent to \eqref{eq:15}.
First, we can see that the first two terms in this expression are 
$%
  {\mathcal B}_{1}+{\mathcal B}_{2}%
=
(%
\sum_{k=1}^{n-1}\PP_{k,n}  
 \mathcal{O}^{[n]}_1~%
 (\mathcal{O}^{[n-1]}_2 \otimes ((1-h
  z_2)(1-h
  \zeta_2)))
) _{(h^{(n)})^2}
$, where we used \eqref{eq:17}. 
Next, we can see that 
$%
\left[{\mathcal B}_{3}\right]_\sigma=(%
[\PP_{n,\sigma(n)}  (\mathcal{O}^{[n-1]}_1 \otimes ((1-h
  z_1)(1-h \zeta_1))) ~%
  \mathcal{O}^{[n]}_2]_\sigma)_{(h^{(n)})^2}$: 
namely, if we introduce $\mathcal{C}=[\PP_{n,\sigma(n)}  (\mathcal{O}^{[n-1]}_1 \otimes ((1-h
  z_1)(1-h \zeta_1))) ~%
  \mathcal{O}^{[n]}_2]_\sigma=[(\mathcal{O}^{[n-1]}_1 \otimes ((1-h
  z_1)(1-h \zeta_1))) ~%
  \mathcal{O}^{[n]}_2]_{\hat \sigma}$, where $\hat \sigma =
  \tau_{(n,\sigma(n))} \circ \sigma$ obeys $\hat\sigma
  (n)=n$, and we decompose it as in \eqref{eq:16} then we see that
  $\mathcal{C}$ only involves $\left[\mathcal{O}_2\right]_{{\sigma'}}$
  where $\sigma'(n)=n$. But from \eqref{eq:31} one sees that for such
  permutations, one has $[\mathcal{O}^{[n]}_2
]_{\sigma'}=(z_2-\zeta_2)
  [\mathcal{O}^{[n-1]}_2\otimes \II
]_{\sigma'}$, hence $\mathcal{C}=(z_2-\zeta_2)[\PP_{n,\sigma(n)}  (\mathcal{O}^{[n-1]}_1 \otimes ((1-h
  z_1)(1-h \zeta_1))) ~%
  (\mathcal{O}^{[n-1]}_2\otimes \II)]_\sigma$, which gives $%
\left[{\mathcal B}_{3}\right]_\sigma=(%
[\PP_{n,\sigma(n)}  (\mathcal{O}^{[n-1]}_1 \otimes ((1-h
  z_1)(1-h \zeta_1))) ~%
  \mathcal{O}^{[n]}_2]_\sigma)_{(h^{(n)})^2}$.

Let us also show that $
\left[{\mathcal B}_{4}\right]_\sigma
=([\sum_{k\neq
    n,\sigma(n)} \PP_{k,n} (\mathcal{O}^{[n-1]}_1 \otimes ((1-h
  z_1)(1-h \zeta_1))) ~%
 \mathcal{O}^{[n]}_2]_\sigma)
  _{(h^{(n)})^2}$, i.e. that $
\left[\mathcal{B}_3+{\mathcal B}_{4}\right]_\sigma
=([\sum_{k\neq
    n} \PP_{k,n} (\mathcal{O}^{[n-1]}_1 \otimes ((1-h
  z_1)(1-h \zeta_1))) ~%
 \mathcal{O}^{[n]}_2]_\sigma)
  _{(h^{(n)})^2}$: we have
\begin{align}
    \label{eq:20}
    \lefteqn{\textstyle    ([\sum_{k\neq
     n,\sigma(n)} \PP_{k,n} (\mathcal{O}^{[n-1]}_1 \otimes ((1-h
   z_1)(1-h \zeta_1))) ~%
 \mathcal{O}^{[n]}_2]_\sigma)
   _{(h^{(n)})^2}}\hspace{2cm}\nonumber\\=&\textstyle
([\sum_{k\neq
     n,\sigma(n)} \PP_{k,n} (\mathcal{O}^{[n-1]}_1 \otimes ((1-h
   z_1)(1-h \zeta_1))) ~%
 (\mathcal{O}^{[n]}_2)  _{(h^{(n)})^1}]_\sigma)
   _{(h^{(n)})^1}\nonumber\\
=&-
\textstyle
([z_2\zeta_2\sum_{k\neq
     n,\sigma(n)} \PP_{k,n} (\mathcal{O}^{[n-1]}_1 \otimes ((1-h
   z_1)(1-h \zeta_1))) 
\nonumber\\*&\hspace{4cm} \textstyle
\sum_{k'\neq n} \PP_{k',n}
(\mathcal{O}^{[n-1]}_2\otimes \II)]_\sigma)  _{(h^{(n)})^1}
\qquad \text{[by \eqref{eq:17}]}
\nonumber\\
=&-
\textstyle
[z_2\zeta_2\sum_{k\neq
     n,\sigma(n)} \PP_{k,n} \sum_{k'\neq n} \PP_{k',n}
 (\mathcal{O}^{[n-1]}_1 \triangleleft\, ((1-h
   z_1)(1-h \zeta_1))^{(k')}) _{(h^{(n)})^1} 
\nonumber\\*&\hspace{7cm} \textstyle
(\mathcal{O}^{[n-1]}_2\otimes \II)]_\sigma 
\nonumber\\
=&\textstyle
z_1\zeta_1 z_2\zeta_2[\sum_{k\neq
     n,\sigma(n)} \sum_{k'\neq n}  \sum_{\tilde k\neq n,k'}  
\PP_{k,n}  \PP_{k',n}  \PP_{\tilde k,n
}\nonumber\\*&\hspace{3cm}\textstyle
(\mathcal{O}^{[n-2]}_1 \triangleleft\, ((1-h
   z_1)(1-h \zeta_1))^{(k')}\otimes \II) 
~%
 (\mathcal{O}^{[n-1]}_2\otimes \II)]_\sigma
\nonumber\\=
&\textstyle
z_1\zeta_1 z_2\zeta_2 [\sum_{k\neq
     n,\sigma(n)} \sum_{k'\neq
     n,\sigma(n)} 
\PP_{k,n}  \PP_{k',n}  \PP_{n,\sigma(n)
}\nonumber\\*&\hspace{3cm}\textstyle
(\mathcal{O}^{[n-2]}_1 \triangleleft\, ((1-h
   z_1)(1-h \zeta_1))^{(k')}\otimes \II) 
~%
 (\mathcal{O}^{[n-1]}_2\otimes \II)]_\sigma
\nonumber\\=&\left[\mathcal{B}_4\right]_\sigma
  \end{align}
where the first equality relies on the condition $\sigma(n)\ne n$ and the fourth equality rewrites $(\mathcal{O}^{[n-1]}_1 \triangleleft\, ((1-h
   z_1)(1-h \zeta_1))^{(k')}) _{(h^{(n)})^1}=\PP_{n,k'}
(-z_1\zeta_1 \linebreak[1]\sum_{\tilde k \neq k',n}
\PP_{\tilde k ,k'} \mathcal{O}^{[n-2]}_1
\triangleleft\,I^{(k')}\otimes ((1-h
   z_1)(1-h \zeta_1))
)
\PP_{n,k'}
 $ (which is obtained from
   the definition \eqref{eq:22} and the relation $(\mathcal{O}_m)_{(h^{(j)})^1}\linebreak[1]=(-\zeta_m z_m 
  \sum_{k\neq j}  \PP_{k,j})~%
(
  \mathcal{O}^{[n-1]}_m
\triangleleft\,
\II^{(j)}) %
$ obtained before) as  $-z_1\zeta_1 \sum_{\tilde k\ne k',n}\PP_{\tilde k,n
}
(\mathcal{O}^{[n-2]}_1 \triangleleft\, ((1-h
   z_1)(1-h \zeta_1))^{(k')}\otimes \II)$. The fifth equality in
   \eqref{eq:20} uses the same argument as in the derivation of 
\eqref{eq:2}, namely that when we pick the coefficient
$[\dots]_{\sigma}$ all terms with $\tilde k\neq \sigma(n)$ vanish. 
Then we find 
$\sum_{k=1}^{4}\left[{\mathcal B}_{k}\right]_\sigma=
([ 
\sum_{k=1}^{n-1} \PP_{k,n} \mathcal{A} 
]_\sigma) _{(h^{(n)})^2}$.

In order to conclude the proof that \eqref{eq:34} is an explicit
rewriting of \eqref{eq:15}, we will show that we also have $\sum_{k=5}^{8}\left[{\mathcal B}_{k}\right]_\sigma=(%
[\PP_{n,\sigma(n)}\,
  \mathcal{A}\,  \PP_{n,\sigma(n)}
\sum_{k\neq \sigma(n)}\PP_{k,\sigma(n)}
]_\sigma) _{(h^{(n)})^2}$, which will require to use a slightly different
notation for the degrees in the variables $h^{(i)}$. So far, we have been
using the notation $X_{(h^{(k)})^p}$ to denote the coefficient in front of the term which has degree $p$ in $h^{(k)}$ in $X$
. In this definition, the ambiguity
which arises  in the definition of
the degree in each individual $h^{(i)}$ (due to relations like
$\PP_{i,j}h^{(i)}=h^{(j)}\PP_{i,j}$) was fixed by conventionally
moving all permutation operators to the left of the operators
$h^{(i)}$. We will now also use the notation  $X^{(h^{(k)})^p}$ to
denote the coefficient in front of the term which has degree $p$ in $h^{(k)}$ in $X$  when
the ambiguity in the degree is fixed by moving all permutation
operators to the right. For instance, when $i\neq j$, we have
{\belowdisplayshortskip=0pt
\belowdisplayskip=0pt
\begin{align}
 \left( \PP_{i,j} h^{(i)} \right)_{(h^{(i)})^1}=&\PP_{i,j}&
 \left( \PP_{i,j} h^{(i)} \right)_{(h^{(j)})^1}=&0\\
 \left( \PP_{i,j} h^{(i)} \right)^{(h^{(i)})^1}=&0&
 \left( \PP_{i,j} h^{(i)} \right)^{(h^{(j)})^1}=&\PP_{i,j}\,,
\end{align}}{\abovedisplayshortskip=0pt
\abovedisplayskip=5pt
\begin{align}
  \label{eq:41}
  (\mathcal{O}_m)_{(h^{(i)})^1}=&{\left(-\zeta_m z_m %
  \sum_{k\neq i}
\PP_{k,i}\right)~%
 (%
  \mathcal{O}^{[n-1]}_m%
\triangleleft
\II^{(i)})}\,,\\
  (\mathcal{O}_m)^{(h^{(i)})^1}=& (
  \mathcal{O}^{[n-1]}_m
\triangleleft
\II^{(i)})
~%
{\left(-\zeta_m z_m %
  \sum_{k\neq i}
\PP_{k,i}\right)}\,.
\end{align}}

Let us now show that $\sum_{k=5}^{8}\left[{\mathcal B}_{k}\right]_\sigma=([\PP_{n,\sigma(n)}\,
  \mathcal{A}\,  \PP_{n,\sigma(n)}
\sum_{k\neq \sigma(n)}\PP_{k,\sigma(n)}
]_\sigma) _{(h^{(n)})^2}$, by splitting the right-hand side into
pieces. The first piece is ${(%
  [}\PP_{n,\sigma(n)}\,
   \mathcal{O}^{[n]}_1~%
 (\mathcal{O}^{[n-1]}_2 \otimes ((1-h
  z_2)(1-h \zeta_2)))\,  \PP_{n,\sigma(n)}
\sum_{k\neq \sigma(n)}\PP_{k,\sigma(n)}
]_\sigma) _{(h^{(n)})^2}$, which splits into the term $k=n$ on the one
hand, and the other terms in the other hand. The term where
$k=n$ is given by:
\begin{align}
  \label{eq:45}
 \lefteqn{ %
   \left(%
    \left[\PP_{n,\sigma(n)}\,
   \mathcal{O}^{[n]}_1~%
 (\mathcal{O}^{[n-1]}_2 \otimes ((1-h
  z_2)(1-h \zeta_2)))\,  \PP_{n,\sigma(n)}
\PP_{n,\sigma(n)}
\right]_{\sigma}\right)_{(h^{(n)})^2}}\qquad\qquad\qquad\nonumber\\
=&%
\left[\PP_{n,\sigma(n)}
\left(\mathcal{O}^{[n]}_1\right)_{(h^{(n)})^1} ~%
 (\mathcal{O}^{[n-1]}_2 \otimes ((1-h
  z_2)(1-h \zeta_2))) _{(h^{(n)})^1}
\right.\nonumber\\*&\hspace{1cm}
\left.
+
\PP_{n,\sigma(n)} \left(\mathcal{O}^{[n]}_1\right)_{(h^{(n)})^0} ~%
 (\mathcal{O}^{[n-1]}_2 \otimes ((1-h
  z_2)(1-h \zeta_2)))_{(h^{(n)})^2}\right]_{\sigma}\nonumber{}
\\=&0+\left[\mathcal{B}_5\right]_\sigma\,,
\end{align}
by the same argument as in the derivation of $\mathcal{B}_3$. Next the
terms with $k\ne n$ are given by
\begin{align}
\lefteqn{ \left(%
    \left[\PP_{n,\sigma(n)}\,
   \mathcal{O}^{[n]}_1~%
 (\mathcal{O}^{[n-1]}_2 \otimes ((1-h
  z_2)(1-h \zeta_2)))\,  \PP_{n,\sigma(n)}
\sum_{k\neq n,\sigma(n)}\PP_{k,\sigma(n)}
\right]_{\sigma}\right)_{(h^{(n)})^2}}\qquad\nonumber\\
=&
\left[
  \left(
    \left(
\mathcal{O}^{[n]}_1
    \right)^{(h^{(\sigma(n))})^1}~%
 \left(\mathcal{O}^{[n-1]}_2 \triangleleft ((1-h
  z_2)(1-h \zeta_2))^{(\sigma(n))}\right)  \right)^{(h^{(\sigma(n))})^1}
\sum_{k'\neq n,\sigma(n)}\PP_{k',\sigma(n)}
\right]_{\sigma}%
\\
=&
\left[-z_1\zeta_1
(\mathcal{O}^{[n-1]}_1\triangleleft \II^{(\sigma(n))})~%
\sum_{k\ne \sigma(n)}\PP_{\sigma(n),k}
   ~%
 \left(\mathcal{O}^{[n-1]}_2 \triangleleft ((1-h
  z_2)(1-h \zeta_2))^{(\sigma(n))}\right)^{(h^{(k)})^1}\right.\nonumber\\*&\hspace{10cm}
\left.
\sum_{k'\neq n,\sigma(n)}\PP_{k',\sigma(n)}
\right]_{\sigma}%
\\
=&
z_1\zeta_1z_2\zeta_2\,%
\left[
(\mathcal{O}^{[n-1]}_1\triangleleft \II^{(\sigma(n))})~%
\sum_{k\ne \sigma(n)}\PP_{\sigma(n),k}
(\mathcal{O}^{[n-1]}_2\triangleleft \II^{(k)}\triangleleft ((1-h
  z_2)(1-h \zeta_2))^{(\sigma(n))})
\right.\nonumber\\*&\hspace{6cm}
\left.
\sum_{\tilde k\ne k,\sigma(n)}\PP_{\tilde k,k} %
\sum_{k'\neq n,\sigma(n)}\PP_{k',\sigma(n)}
\right]_{\sigma}\label{eq:44}\\=&
z_1\zeta_1z_2\zeta_2\,%
\left[
\sum_{k\ne \sigma(n),n}
(\mathcal{O}^{[n-1]}_1\triangleleft \II^{(\sigma(n))})   
(\mathcal{O}^{[n-1]}_2\triangleleft \II^{(\sigma(n))}\triangleleft ((1-h
  z_2)(1-h \zeta_2))^{(k)}) %
\right.\nonumber\\*&\hspace{6cm}
\left.
\PP_{\sigma(n),k}  ~%
 \PP_{n,k}~%
\sum_{k'\neq n,\sigma(n)}\PP_{k',\sigma(n)}
\right]_{\sigma}\label{eq:43}\\=&\left[\mathcal{B}_6\right]_\sigma\,,
\end{align}
where the equality \eqref{eq:43} arises because only the terms with $\tilde k=n$
contribute to \eqref{eq:44}.
Finally, the second piece in $([\PP_{n,\sigma(n)}\,
  \mathcal{A}\,  \PP_{n,\sigma(n)}
\sum_{k\neq \sigma(n)}\PP_{k,\sigma(n)}
]_\sigma) _{(h^{(n)})^2}$ is given by $([\PP_{n,\sigma(n)}%
 \linebreak[1]  (\mathcal{O}^{[n-1]}_1 \otimes ((1-h
  z_1)(1-h \zeta_1))) ~%
 \mathcal{O}^{[n]}_2\,  \PP_{n,\sigma(n)}
\sum_{k\neq \sigma(n)}\PP_{k,\sigma(n)}
]_\sigma) _{(h^{(n)})^2}$, which is given by
\begin{align}
  \nonumber
    \lefteqn{\left(%
        \left[\PP_{n,\sigma(n)}\,
(\mathcal{O}^{[n-1]}_1 \otimes ((1-h
  z_1)(1-h \zeta_1))) ~%
 \mathcal{O}^{[n]}_2
 \PP_{n,\sigma(n)}
\sum_{k\neq \sigma(n)}\PP_{k,\sigma(n)}
\right]_{\sigma}\right)_{(h^{(n)})^2}}\qquad\qquad\\=&\left(
\left[
\mathcal{O}_1^{[n-1]}\triangleleft \omhz[1]^{(\sigma(n))}~%
 \mathcal{O}_2 \sum_{k\neq \sigma(n)}
\PP_{k,\sigma(n)}
  \right]_{\sigma}
  \right)_{(h^{(n)})^2}
\\
=&
\left[
 \left(
\mathcal{O}_1^{[n-1]}\triangleleft \omhz[1]^{(\sigma(n))}~%
 \mathcal{O}_2 
  \right)^{(h^{(\sigma(n))})^2}
\sum_{k\neq \sigma(n)}
\PP_{k,\sigma(n)}
  \right]_{\sigma}
\\=&
\left[\vphantom{\sum_{k\neq \sigma(n)}
\PP_{k,\sigma(n)}}
\left(
 \left(
\mathcal{O}_1^{[n-1]}\triangleleft \omhz[1]^{(\sigma(n))}
  \right)^{(h^{(\sigma(n))})^1}
~%
 \left(\mathcal{O}_2 
  \right)^{(h^{(\sigma(n))})^1}
\right.\right.\nonumber \\*
&\hspace{1cm}+\left.
 \left(
\mathcal{O}_1^{[n-1]}\triangleleft \omhz[1]^{(\sigma(n))}
  \right)^{(h^{(\sigma(n))})^2}
~%
 \left(\mathcal{O}_2 
  \right)^{(h^{(\sigma(n))})^0}\right)\nonumber\\*&\hspace{10cm}
\left.
\sum_{k\neq \sigma(n)}
\PP_{k,\sigma(n)}
  \right]_{\sigma}
\label{eq:40}\\=&\left[\mathcal{B}_8+\mathcal{B}_7\right]_\sigma\,,
\end{align}
where the equality \eqref{eq:40} uses the fact that $[\mathcal{O}_1^{[n-1]}\triangleleft \omhz[1]^{(\sigma(n))}]_{\tilde \sigma}$ vanishes unless $\tilde\sigma(\sigma(n))=\sigma(n)$.

Hence we
have seen that 
$\sum_{k=1}^{4}[{\mathcal B}_{k}]_\sigma=
(%
[ 
\sum_{k=1}^{n-1} \PP_{k,n} \mathcal{A} 
]_\sigma) _{(h^{(n)})^2}$ and $\sum_{k=5}^{8}[{\mathcal B}_{k}]_\sigma
={(%
[\PP_{n,\sigma(n)}}\linebreak[2]
  \mathcal{A}\,  \PP_{n,\sigma(n)}
\sum_{k\neq \sigma(n)}\PP_{k,\sigma(n)}
]_\sigma) _{(h^{(n)})^2}$, which shows that \eqref{eq:34} is nothing
but an explicit
rewriting of \eqref{eq:15}.

\def\thesubsection{D}
\subsection{The bilinear identity and matrix integrals}
\label{Hirpr}
\setcounter{equation}{0}

In this appendix we give an independent 
proof of the fact that the master $T$-operator satisfies the 
bilinear identity for the 
KP hierarchy. We use the technique of matrix derivatives 
(see section \ref{sec:matrix})
and matrix integrals (see \cite{mehta}).
As it follows from that section,
it is enough to prove the bilinear identity in the form
(\ref{Hir}) with an auxiliary matrix $A$. Here we assume that 
the matrix $A$ is positively defined 
and Hermitian\footnote{The type of the matrix 
$A$ is actually 
not very important for us. We take it to be Hermitian for simplicity 
reasons. For our purpose it is sufficient to assume 
that this matrix has
enough number (namely, $N^2$), of independent entries,
so that one could take derivatives with respect to them in (\ref{phi}).}.

\subsubsection*{A complex matrix integral}

Let $A$ be a positively defined (Hermitian) matrix and 
$B,C$ be arbitrary $N\times N$ matrices.
We need the following Gaussian matrix integral
over $N\times N$ complex matrices $W$:
\beq\label{NMM}
\int \left[d^2 W \right]\exp\left(-\tr \left(A^{-1} W W^\dagger -BW-CW^\dagger\right) \right) = \left(\det A\right)^N\exp(\tr A C B ),
\eeq
with the flat measure $\displaystyle{\left[d^2 W \right] \propto 
\prod_{i,j=1}^N d^2 W_{ij}}$, where
$d^2 W_{ij}\equiv d{\rm Re}W_{ij}\, d{\rm Im}W_{ij}$,
normalized in such a way that 
$
\int \left[d^2 W \right]\exp\left(-\tr W W^\dagger \right)=1.
$
This formula follows from eq. (14) of \cite{Orlov} (which is
equivalent to (\ref{NMM}) with $A=\II$) after the change of
integration variables $W \to A^{-\frac{1}{2}}W$, 
$W^\dagger \to W^\dagger A^{-\frac{1}{2}}$.
Note that the integral (\ref{NMM}) is invariant under
the Hermitian conjugation of the matrix integration variable
$W \to W^\dagger$:
\begin{multline}\label{symm}
\int \left[d^2 W \right]\exp\left(-\tr \left(A^{-1} W W^\dagger -BW-CW^\dagger\right) \right) \\
=\int \left[d^2 W \right]\exp\left(-\tr \left(A^{-1} W^\dagger W -BW^\dagger-CW\right) \right)
\end{multline}

\subsubsection*{Proof of the bilinear identity
in the special case $x=x_1= \ldots =x_n$}

Let us start with the special case 
of the master $T$-operator at $x=x_1= \ldots =x_n$ 
given by (\ref{mtrest}). 
The matrix integral representation (\ref{NMM}) suggests 
to consider the modified generating function
$$
\widetilde{\Phi}^G({\bf t}, {\bf t'};A)=
\left(\det A\right)^{N}\Phi^G({\bf t}, {\bf t'};A)
$$
instead of (\ref{bic}).
Obviously, this does not affect the bilinear identity (\ref{Hir})
and so it is enough to prove it for $\widetilde{\Phi}^G({\bf t}, {\bf t'};A)$.
The combination of (\ref{phi}) and (\ref{NMM}) gives:
\begin{equation}\label{mint1}
\begin{array}{c}
\displaystyle{\widetilde{\Phi}^G({\bf t}, {\bf t'};A)=\int \left[d^2 W \right]  \exp\left(- \tr \left(A^{-1} W W^\dagger -W\frac{\p}{\p g^{t}} - W^\dagger\frac{\p}{\p h^{t}} \right)\right)\times}  \\ \\
\displaystyle{\times\left.\exp \left (\sum_{k\geq 1}\left(t_k \, \mbox{tr}\, h^k+t'_k \, \mbox{tr}\, g^k\right)\right )\right|_{g=h} }\\ \\
\displaystyle{= \int \left[d^2 W \right]\exp{\left(-\tr A^{-1} W W^\dagger+\tr\sum_{k\geq 1}\left(t_k \,  (h+W^\dagger)^k+t'_k \,  (h+W)^k\right)\right)}}\\ \\ 
\displaystyle{=   \int \left[d^2 W \right]\exp{\left(-\tr A^{-1} (W-h) (W^\dagger-h)+\tr\sum_{k\geq 1}\left(t_k \, W^{\dagger k}+t'_k \, W^k\right)\right). }}
\end{array}
\end{equation}
Here we have used invariance of the complex matrix 
integral with respect to the shifts of the integration variables 
$W \to W + B$, $W^\dagger \to W^\dagger +C$ with
arbitrary complex matrices $B$ and $C$. 
(This invariance clearly holds at $N=1$ while 
the $N>1$ case is reduced 
to a multiple integral of the same type.) 

\noindent
{\bf Remark.} Note that in general the matrix integrals in the last two
lines of (\ref{mint1}) diverge, so they make sense only as formal 
series in ${\bf t}_k$ and ${\bf t'}_k$, each coefficient of this series
being well-defined. This is precisely the meaning that we need from
the generating function, so one should not worry about convergence
of the integrals of this type 
((\ref{mint1}) and below) since their integrands have to be 
understood as power series in ${\bf t}_k$ and ${\bf t'}_k$.

The commutativity of the master $T$-operators 
implies that
$
\Phi^G({\bf t}, {\bf t'};A)=\Phi^G({\bf t'}, {\bf t};A).
$
Therefore, we have two different 
expressions for the same generating function: 
\beq\label{sym}
\Phi^G({\bf t}, {\bf t'};A)=\left.\exp\left(\tr A\frac{\p}{\p h^{t}}
\frac{\p}{\p g^{t}}\right)\exp \left (\sum_{k\geq 1}t_k \, \mbox{tr}
\, h^k\right )\exp \left (\sum_{k\geq 1}t'_k \, \mbox{tr}\, g^k\right )
\right|_{g=h}
\eeq
and
\beq\label{sym111}
\Phi^G({\bf t}, {\bf t'};A)=\left.\exp\left(\tr A\frac{\p}{\p g^{t}}
\frac{\p}{\p h^{t}}\right)\exp \left (\sum_{k\geq 1}t_k \, \mbox{tr}
\, h^k\right )\exp \left (\sum_{k\geq 1}t'_k \, \mbox{tr}\, g^k\right )
\right|_{g=h}.
\eeq
Then, using (\ref{symm}), we obtain another integral representation
of the same generating function:
\beq\label{mint2}
\widetilde{\Phi}^G({\bf t}, {\bf t'};A)=  \int \left[d^2 W \right]
\exp\left({-\tr A^{-1} (W^\dagger-h) (W-h)+\tr\sum_{k\geq 1}\left(t_k \, W^{\dagger k}+t'_k \, W^k\right) }\right).
\eeq

A complex matrix $W$ can be decomposed as 
\beq\label{dec1}
W=U (w+R) U^\dagger
\eeq
with unitary $U$, diagonal $w$ and strictly upper-triangular $R$. 
The elements of the matrix $w$ are eigenvalues of $W$: $w=\diag (w_1,\ldots, w_N)$. 
The flat measure on the space of complex matrices is
\beq
\left[d^2 W \right]=c_N\left[d U\right] \prod_{i<j} d^2 R_{ij}\, |\Delta(w)|^2 \prod_{i=1}^N d^2 w_i,
\eeq 
where $\left[d U\right]$ is the Haar measure for the unitary group, 
$
\displaystyle{\Delta(z)=\prod_{i>j}(z_i-z_j)}
$
is the Vandermonde determinant and $c_{N}$ is an $N$-dependent constant 
(see, e.g., \cite{mehta,measure}).
There is an equivalent but different decomposition: \beq\label{dec2}
W={V} (w+{Q}^\dagger){V}^\dagger ,
\eeq
where ${V}$ is unitary and ${Q}$ is strictly upper-triangular 
so that $Q^\dagger$ is strictly lower-triangular. For this decomposition the measure is
\beq
\left[d^2 W \right]=c_N\left[d {V}\right] \prod_{i<j} d^2 Q_{ij}\, 
|\Delta(w)|^2 \prod_{i=1}^N d^2 w_i.
\eeq 

Let us use decomposition (\ref{dec1}) for the integral from the 
last line of (\ref{mint1}),
\begin{equation}\label{mi1}
\begin{array}{c}
\displaystyle{\widetilde{\Phi}^G({\bf t}, 
{\bf t'};A)=c_N\!\! \int \left[d U\right]  
\prod_{i<j} d^2 R_{ij}\, |\Delta(w)|^2 
\prod_{i=1}^N d^2 w_i} \\ \\ 
\displaystyle{\times \exp\left({-\tr A^{-1} (U (w+R) U^\dagger-h) (U (\bar{w}+R^\dagger) U^\dagger-h)+\tr\sum_{k\geq 1}\left(t_k \, 
\bar{w}^{ k}+t'_k \, w^k\right)} \right),}
\end{array}
\end{equation}
and decomposition (\ref{dec2}) for the integral (\ref{mint2}):
\begin{equation}\label{mi2}
\begin{array}{c}
\displaystyle{\widetilde{\Phi}^G({\bf t}, {\bf t'};A)=c_N \!\!
\int \left[d V\right] \prod_{i<j} d^2 Q_{ij}\, |\Delta(w)|^2 \prod_{i=1}^N d^2 w_i } \\ \\ 
\displaystyle{\times \exp\left({-\tr A^{-1} (V (\bar{w}+Q) V^\dagger-h) (V (w+Q^\dagger) V^\dagger-h)+\tr\sum_{k\geq 1}\left(t_k \, \bar{w}^{ k}+t'_k \, w^k\right) }\right).}
\end{array}
\end{equation}
Let us re-denote the integration variables 
in the last integral as
$V\to U$ (a unitary matrix) and
$Q\to R$ (an upper-triangular matrix). 
Then the half-sum of (\ref{mi1}) 
and (\ref{mi2}) gives
\begin{equation}\label{eqHir}
\begin{array}{c}
\displaystyle{ \widetilde{\Phi}^G({\bf t} - \left[z^{-1}\right], {\bf t'} + \left[ z^{-1}\right];A)}\\ \\ 
\displaystyle{ =\, \frac{c_N}{2}\!\!\!
\int \left[d U\right] \prod_{i<j} d^2 R_{ij}\, |\Delta(w)|^2 \prod_{i=1}^N d^2 w_i \prod_{j=1}^N \left(1\! +\! \frac{w_j\! -\!
\bar{w}_j}{z} \sum_{k\geq 0}\left(\frac{w_j}{z}\right)^k\right) 
e^{\tr\sum_{k\geq 1}\left(t_k \, \bar{w}^{ k}+t'_k \, w^k\right) }}\\ \\ 
\displaystyle{ \times \left(e^{-\tr A^{-1} (U(w+R)U^\dagger-h) (U(\bar{w}+R^\dagger)U^\dagger-h)}+e^{-\tr A^{-1} (U(\bar{w}+R)U^\dagger-h) (U(w+R^\dagger)U^\dagger-h)}\right).}
\end{array}
\end{equation}
which is again well-defined as
a formal series in $\bf{t}$, $\bf{t}'$. Then, for each term 
of this series, one should take the residue at $z=\infty$ 
in the l.h.s. of the bilinear identity (\ref{Hir}).
This means that if we treat the integral as a formal series, 
the $z$-integration should be performed first:
\beq
\frac{1}{2\pi i}\!
\oint_{\infty}\!\!\! dz \, e^{\xi({\bf t-t'},z)}
\! \! \prod_{j=1}^N \Bigl (\! 1\! +\! \frac{w_j\! -\! \bar w_j}{z}
\sum_{k\geq 0}\left (\frac{w_j}{z}\right)^k\Bigr ) 
\! = \!\! \sum_{k=1}^N  e^{\xi({\bf t-t'},w_k)} 
(w_k\! -\! \bar{w}_k) \prod_{j\neq k} 
\frac{w_k\! -\! \bar{w}_j}{w_k\! \! -w_j}
\eeq
This function is regular at $w_i=w_j$ for all $i$ and $j$. 
Thus the bilinear identity (\ref{Hir}) is equivalent to the following
relation:
\begin{equation}
\begin{array}{c}
\displaystyle{\sum_{k=1}^N \int \left[d U\right] \prod_{i<j} d^2 R_{ij}\, |\Delta(w)|^2 \prod_{i=1}^N d^2 w_i (w_k-\bar{w}_k) \prod_{j\neq k} \frac{w_k-\bar{w}_j}{w_k-w_j}} \\ \\
\displaystyle{\times \exp{\left(\sum_{m\geq 1}\sum_{j\neq k}\left(t_m \bar{w}_j^k+t'_m w_j^k\right) + \sum_{m\geq 1}t_m\left(w_k^m+\bar{w}_k^m\right)\right)}} \\ \\
\displaystyle{ \times \left(e^{-\tr A^{-1} (U(w+R)U^\dagger-h) (U(\bar{w}+R^\dagger)U^\dagger-h)}+e^{-\tr A^{-1} (U(\bar{w}+R)U^\dagger-h) (U(w+R^\dagger)U^\dagger-h)}\right)=0}
\end{array}
\end{equation}

In fact each term in the sum over $k$ is equal to zero. 
To see this, let us consider only the dependence of the integrand on the corresponding variable $w_k$. 
It is easy to see that the integrand is antisymmetric 
with respect to the interchange 
$w_k \leftrightarrow \bar{w}_k$ (complex conjugation), hence 
the integral over the complex plane vanishes. 
Indeed, the second and the third lines are obviously symmetric
while in the first line we have
\beq
|\Delta(w)|^2  (w_k-\bar{w}_k) \prod_{j\neq k} \frac{w_k-\bar{w}_j}{w_k-w_j}=(w_k-\bar{w}_k)\prod_{j\neq k} (w_k-\bar{w}_j)(\bar{w}_k-\bar{w}_j) 
\cdot N (w),
\eeq
where $N(w)$ does not depend on $w_k$. This expression 
is obviously antisymmetric w.r.t. the 
conjugation $w_k \leftrightarrow \bar{w}_k$, so the integral vanishes 
and the bilinear identity is thus proved.

{\bf Example: $N=1$.}
Let us consider the simplest possible example $N=1$. 
Then $A \in  \RR_+$ is just a real positive number. Then
\beq
\widetilde{\Phi}^G({\bf t}, {\bf t'};A)=  
\sum_{i,j=1}^\infty h_i ({\bf t}) h_j ({\bf t'})  P_{ij}
\eeq
where the $h_i$'s are the elementary 
Schur polynomials (\ref{Schur-p}) and
\beq
P_{ij} =\int_{\CCC} d^2 w \, \exp\left({- \,\frac{(w-h) 
(\bar{w}-h)}{A} }\right) \bar{w}^i w^j
\eeq
 Then
 \beq
 \widetilde{\Phi}^G({\bf t}-\left[z^{-1}\right], 
 {\bf t'}+\left[z^{-1}\right];A) =  
 \sum_{i,j=1}^\infty h_i ({\bf t}) h_j ({\bf t'})  \tilde{P}_{i,j}(z)
  \eeq
where 
\beq
\tilde{P}_{ij}(z) = \int_{\CCC} d^2 w \, 
\exp\left ({- \frac{(w-h) (\bar{w}-h)}{A} }\right) \bar{w}^i w^j
\left(1 + \frac{w-\bar{w}}{z} \sum_{k=1}^\infty \left( \frac{w}{z}\right)^k\right)  
\eeq
is a series in $z^{-1}$. Then the right-hand side of (\ref{eqHir}) reduces to
\begin{equation}
\begin{array}{c}
\displaystyle{
 \sum_{i,j=1}^\infty h_i ({\bf t}) h_j ({\bf t'})  \oint_{\infty} dz\, e^{\xi({\bf t-t'},z)} \tilde{P}_{i,j}(z)}
\\
\displaystyle{=  \sum_{i,j,k,l=1}^\infty h_i ({\bf t}) h_j ({\bf t'}) h_k({\bf t}) h_l({\bf -t'}) \int_{\CCC} d^2 w \, (w-\bar{w}) \exp\left({- \frac{(w-h) (\bar{w}-h)}{A}}\right) \bar{w}^{i} w^{j+k+l}
}
\\
\displaystyle{=  \sum_{i,k=1}^\infty h_i ({\bf t})  h_k({\bf t})  
\int_{\CCC} d^2 w \, (w-\bar{w}) \exp\left({- \frac{(w-h) 
(\bar{w}-h)}{A}}\right) \bar{w}^{i} w^{k}
}
\end{array}
\end{equation}
where the third line follows from the second one and the relation (\ref{hhs}). The integral in the last line is anti-symmetric with respect to the change $i \leftrightarrow k$, thus the whole sum vanishes.

As it follows from the arguments given in section 4.2, the 
special case of the bilinear identity proven above is already 
enough for the proof of the general case. However,
a sketch of the direct 
proof by means of the matrix integrals
is given below for completeness.

\subsubsection*{Proof of the bilinear identity
in the general case}

To prove the bilinear identity for arbitrary $x-x_i$ let us consider a generalization of (\ref{phi}).
\beq\label{phi1}
\Phi^G({\bf t}, {\bf t'};A_1,\ldots,A_n)=\left.\exp\left(\tr Q(A)\right)\exp \left (\sum_{k\geq 1}t_k \, \mbox{tr}\, h^k\right )
\exp \left (\sum_{k\geq 1}t'_k \, \mbox{tr}\, g^k\right )\right|_{g=h}.
\eeq
This generating function depends on $n$ external matrices 
$A_1,\ldots,A_n$ (all of which are assumed to be positively defined) through
\beq
Q(A)=\sum_{\alpha=1}^n A_\alpha \left(x-x_\alpha+\frac{\p}{\p h^{t}}\right)\left(x-x_\alpha+\frac{\p}{\p g^{t}}\right).
\eeq
The master $T$-operator (\ref{master1}) is a coefficient of the expansion of (\ref{phi1}) in front of the term linear in all $A_\alpha$, thus 
for the proof of (\ref{hir1}) it is enough to show that
\beq\label{Hir1}
\oint_{\infty}e^{\xi({\bf t-t'},z)}\Phi^G({\bf t}-\left[z^{-1}\right], {\bf t'}+\left[z^{-1}\right];A_1,\ldots,A_n) dz =0.
\eeq

The prove is a generalization of the one 
given above. 
First of all, let us simplify the notation and denote $x_\alpha -x$ by $x_\alpha$. As above, we also slightly modify the definition of the generating function:
\beq
\widetilde{\Phi}^G({\bf t}, {\bf t'};A_1,\ldots,A_n)=\left(\det A_1\ldots A_n\right)^N \,{\Phi}^G({\bf t}, {\bf t'};A_1,\ldots,A_n).
\eeq
Then
\begin{equation}\label{int21}
\begin{array}{c}
\displaystyle{\widetilde{\Phi}^G({\bf t}, {\bf t'};A_1,\ldots,A_n)}\\ \\
\displaystyle{=\int \prod_{\alpha=1}^n \left[d^2 W_\alpha \right]e^{-\tr \sum_{\alpha=1}^n \left(A_\alpha ^{-1} W_\alpha W_\alpha^\dagger - W_\alpha\left(\frac{\p}{\p g^{t}}-x_\alpha\right)- W_\alpha^\dagger\left(\frac{\p}{\p h^{t}}-x_\alpha\right)\right)}\left.e^{ \sum\left(t_k \, \mbox{tr}\, h^k+t'_k \, \mbox{tr}\, 
g^k\right)}\right|_{g=h} }\\ \\
\displaystyle{=\int \prod_{\alpha=1}^n \left[d^2 W_\alpha \right]e^{-\tr \sum_{\alpha=1}^n \left(A_\alpha ^{-1} W_\alpha W_\alpha^\dagger +x_\alpha \left(W_\alpha+W_\alpha^\dagger\right) \right)+\tr\sum_{k\geq 1}\left(t_k \,  (h+\sum_{\beta=1}^nW^\dagger_\beta)^k+t'_k \,  (h+\sum_{\beta=1}^n W_\beta)^k\right)}} \\ \\ 
\displaystyle{=\int \prod_{\alpha=1}^n \left[d^2 W_\alpha \right]e^{-\tr  A_1^{-1} \left( W_1-h-\sum_{\beta=2}^nW_\beta\right) \left(W_1^\dagger -h-\sum_{\beta=2}^nW_\beta^\dagger\right)
-\tr \sum_{\alpha=2}^n 
\left(A_\alpha ^{-1} W_\alpha W_\alpha^\dagger \right)}}\\ \\ 
\displaystyle{\times \,\, 
e^{\tr \left(\sum_{k\geq 1}\left(t_k \,  W^{\dagger k}_1+t'_k \,   
W_1^k\right)-x_1(W_1+W_1^\dagger -2h)- \sum_{\alpha=2}^n (x_\alpha-x_1)\left(W_\alpha+W_\alpha^\dagger\right)\right),}}
\end{array}
\end{equation}
where we have used the change of variables 
$$\displaystyle{
W_1 \to W_1-h-\sum_{\beta=2 }^n W_\beta}, \quad
\displaystyle{W_1^\dagger \to W_1^\dagger -h-
\sum_{\beta= 2}^nW_\beta^\dagger}.$$ 
Again, using the commutativity of the master $T$-operators and symmetry (\ref{symm}) of the complex matrix integral, we can obtain another matrix integral representation for the same generating function:
\begin{equation}\label{int22} 
\begin{array}{c}
\displaystyle{\widetilde{\Phi}^G({\bf t}, {\bf t'};A_1,\ldots,A_n)}\\ \\
\displaystyle{=\int \prod_{\alpha=1}^n 
\left[d^2 W_\alpha \right] e^{-\tr \left(A_1^{-1} W_1^\dagger W_1-W_1\left(\frac{\p}{\p g^{t}}-x_1\right)- 
W_1^\dagger\left(\frac{\p}{\p h^{t}}-x_1\right)\right) }}\\ \\ 
\displaystyle{\times \,\, 
e^{-\tr \sum_{\alpha=2}^n\left(A_\alpha^{-1} W_\alpha W_\alpha^\dagger-W_\alpha\left(\frac{\p}{\p h^{t}}-x_\alpha\right)- W_\alpha^\dagger\left(\frac{\p}{\p g^{t}}-x_\alpha\right)\right)}\left.e^{ \sum\left(t_k \, \tr h^k+t'_k \, \tr g^k\right)}\right|_{g=h}}\\ \\ 
\displaystyle{=\int \prod_{\alpha=1}^n \left[d^2 W_\alpha \right]e^{-\tr  A_1^{-1} \left( W_1^\dagger-h-\sum_{\beta=2}W_\beta\right) \left(W_1 -h-\sum_{\beta=2}W_\beta^\dagger\right)
-\tr \sum_{\alpha=2}^n \left(A_\alpha ^{-1} 
W_\alpha W_\alpha^\dagger \right)}}\\ \\ 
\displaystyle{\times \,\,
e^{\tr \left(\sum_{k\geq 1}\left(t_k \,  W^{\dagger k}_1+t'_k \,   
W_1^k\right)-x_1(W_1+W_1^\dagger -2h)- \sum_{\alpha=2}^n (x_\alpha-x_1)\left(W_\alpha+W_\alpha^\dagger\right)\right).}}
\end{array}
\end{equation}
Now we apply the same 
argument as in the previous proof
to the integral over the matrix $W_1$. Namely, we take the sum 
of the last lines of (\ref{int21}) and (\ref{int22}) with different parametrization of complex matrix $W_1$ ((\ref{dec1}) and (\ref{dec2}) respectively). After taking the residue in $z$ in (\ref{Hir1}) 
the same argument as before goes through.

\def\thesubsection{E}
\subsection{More about commutation relations}
\label{sec:more-about-comm}
\setcounter{equation}{0}

As stated in the main text, the commutation of the $T$-operators can be
derived from the Yang-Baxter relation, or can as well be obtained
from the commutation of $T$-operators of the spin chain, by taking the
limit $\eta\to 0$. The derivation of this commutation is a bit
technical, and in \cite{Kazakov2007na} it was first obtained for
$T$-operators corresponding to symmetric representations (using the
Yang-Baxter equation), and then for general representations using the
Cherednik-Bazhanov-Reshetikhin formula.

In this appendix, we give another proof of this
commutation, based on the results of Appendix
\ref{sec:alternative-proof}, where the commutation relation 
\begin{equation}
  \label{eq:1}
  [\,Q(z_1,\zeta_1)\,,\,Q(z_2,\zeta_2)\,]=0
\end{equation}
was proved. The present proof uses the polynomial 
normalization \eqref{g1}, but the normalization does not affect
the commutation relations, hence the result holds in the normalization
\eqref{d2} as well.

Let us deduce from \eqref{eq:1} that the $T$-operators
corresponding to different hook-rep\-re\-sen\-ta\-ti\-ons, 
but to the same value
of $x$, do commute with each
other. At the level of generating functions, we should prove that 
\begin{equation}
  \label{eq:10}
  \mathcal{S}_{z,\zeta}\left(\left[(y_n +{\sf d}_n)\, \ldots \,
(y_1 +{\sf d}_1) \frac {w(z_1)}{w(\zeta_1)} \right]\cdot
\left[(y_n +{\sf d}_n)\, \ldots \,
(y_1 +{\sf d}_1) \frac {w(z_2)}{w(\zeta_2)} \right]\right)=0\,,
\end{equation}
where $y_i=x-x_i$ and the notation $\mathcal{S}_{z,\zeta}$ was introduced in Appendix
\ref{sec:alternative-proof}. 
This relation \eqref{eq:10} is easily proven by recurrence over the
number $n$ of spins, if we notice that its l.h.s. is
polynomial of degree at most two in each of the variables $y_i$. The
term with degree $2$ in $y_n$ is exactly the relation with one less
spins while the
term with degree $1$ in $y_n$ is its derivative. Hence the
l.h.s. is independent of $y_n$. Using the relation ${\sf d}_i
{\sf d}_j={\sf d}_j {\sf d}_i$, we
deduce that it is also independent of all the
$y_i$'s. But when all the $y_i$'s are set to zero, \eqref{eq:10}
reduces to the relation \eqref{eq:1}, which was already proven. This
proves the relation \eqref{eq:10}, by recurrence over the number $n$
of spins.

We have proven that the $T$-operators corresponding to
hook-representations commute with each other. Then the Giambelli formula
\eqref{det1a} allows us to deduce that all the operators 
$(x-x_n +{\sf d}_n)\, \ldots \,
(x-x_1 +{\sf d}_1) f(h)$ commute with
each other, for all
functions of $h$ which are arbitrary linear combinations of
characters, if they have the same value of $x$. Noticing that
\begin{equation}
  \label{eq:11}
  \begin{array}{c}
  \partial_x (x-x_n +{\sf d}_n)\, \ldots \,
(x-x_1 +{\sf d}_1) f(h) \\ \\
\hspace{2cm}=\, \Bigl [ (x-x_n +{\sf d}_n)\, \ldots \,
(x-x_1 +{\sf d}_1) f(h) \tr h\Bigr ] 
\\ \\
\hspace{4cm}-\, \tr h \Bigl [ (x-x_n +{\sf d}_n)\, \ldots \,
(x-x_1 +{\sf d}_1) f(h) \Bigr ],
\end{array}
\end{equation}
one deduces that the successive derivatives 
$\partial _x ^k (x-x_n +{\sf d}_n)\, \ldots \,
(x-x_1 +{\sf d}_1) f(h)$ also commute with all operators $(x-x_n +{\sf d}_n)\, \ldots \,
(x-x_1 +{\sf d}_1) \tilde f(h)$, which allows one to conclude that the
commutation also holds for two operators corresponding to different
values of $x$.

\mciteSetMidEndSepPunct{\newline}{}{}
\bibliographystyle{hutphysp}
\bibliography{biblio}

\end{document}